\documentclass[aps,reprint,10pt,superscriptaddress,showpacs,longbibliography]{revtex4-1}

\usepackage{amsmath,amssymb,amsthm,amscd,latexsym,epsfig,bm,subfigure}
\usepackage[usenames,dvipsnames]{color}
\usepackage[colorlinks,bookmarks=false,citecolor=blue,linkcolor=red,urlcolor=blue,breaklinks=true]{hyperref}
\usepackage{verbatim}
\usepackage{bbm}
\usepackage{array}
\usepackage[normalem]{ulem}
\usepackage{amsmath,amssymb}
\usepackage{hyperref}
\usepackage{bm}
\usepackage{graphicx}
\usepackage{color}
\usepackage{longtable}

\newcommand{\PreserveBackslash}[1]{\let\temp=\\#1\let\\=\temp}
\newcolumntype{C}[1]{>{\PreserveBackslash\centering}p{#1}}

\newcommand{\uone}{\ensuremath{\mathrm U(1)}}
\DeclareMathOperator{\tr}{tr}
\DeclareMathOperator{\img}{img}

\begin{document}
\title{Real-space recipes for general topological crystalline states}
\author{Zhida Song}
\altaffiliation{Current address: Department of Physics, Princeton University, Princeton, New Jersey 08544, USA.}
\affiliation{Beijing National Research Center for Condensed Matter Physics,
and Institute of Physics, Chinese Academy of Sciences, Beijing 100190, China}
\affiliation{University of Chinese Academy of Sciences, Beijing 100049, China}
\author{Chen Fang}
\email{cfang@iphy.ac.cn}
\affiliation{Beijing National Research Center for Condensed Matter Physics,
and Institute of Physics, Chinese Academy of Sciences, Beijing 100190, China}
\affiliation{Kavli Institute for Theoretical Sciences, Chinese Academy of Sciences, Beijing 100190, China}
\affiliation{CAS Center for Excellence in Topological Quantum Computation, Beijing, China}
\affiliation{Songshan Lake Laboratory For Materials Science, Guangdong 523808, China}
\author{Yang Qi}
\email{qiyang@fudan.edu.cn}
\affiliation{Center for Field Theory and Particle Physics, Department of Physics, Fudan University, Shanghai 200433, China}
\affiliation{State Key Laboratory of Surface Physics, Fudan University, Shanghai 200433, China}
\affiliation{Collaborative Innovation Center of Advanced Microstructures, Nanjing 210093, China}
\date{\today}

\begin{abstract}
Topological crystalline states are short-range entangled states jointly protected by onsite and crystalline symmetries. While the non-interacting limit of these states, \emph{e.g.}, the topological crystalline insulators, have been intensively studied in band theory and have been experimentally discovered, the classification and diagnosis of their strongly interacting counterparts are relatively less well understood. Here we present a unified scheme for constructing all topological crystalline states, bosonic and fermionic, free and interacting, from real-space ``building blocks'' and ``connectors''. Building blocks are finite-size pieces of lower dimensional topological states protected by onsite symmetries alone, and connectors are ``glue'' that complete the open edges shared by two or multiple pieces of building blocks. The resulted assemblies are selected against two physical criteria we call the ``no-open-edge condition'' and the "bubble equivalence", which, respectively, ensure that each selected assembly is gapped in the bulk and cannot be deformed to a product state. The scheme is then applied to obtaining the full classification of bosonic topological crystalline states protected by several onsite symmetry groups and each of the 17 wallpaper groups in two dimensions and 230 space groups in three dimensions. We claim that our real-space construction gives the complete set of topological crystalline states for bosons and fermions, and prove the boson case analytically using a spectral sequence expansion of group cohomology.
\end{abstract}
\maketitle

\section{Introduction}
\label{sec:intro}
Symmetry-protected topological states (SPT)~\cite{Gu2009,ChenCZX,ChenSPTScience,ChenSPTPRB} are gapped states that do not have topological orders~\cite{WenTO1990,Wen1991a} (fractional excitations) but cannot be deformed into product states of localized wave functions without either symmetry breaking or gap closing.
The constituent particles of SPT can either be bosonic or fermionic.
They are probably the most well-understood topological states, and the famous examples are AKLT-like states \cite{Affleck1987} (bosonic), topological insulators and topological superconductors \cite{Hasan2010,Qi2011} (both fermionic).
Specially, SPT protected by onsite symmetries (only acting on internal degrees of freedom) have been studied for years, and we now know that bosonic SPT are classified by group cohomology of the symmetry group~\cite{Gu2009,ChenCZX,ChenSPTScience,ChenSPTPRB,Chen2014} (with the exception of the so-called ``beyond-group-cohomology'' states~\cite{VishwanathBGC,WangBGC,BurnellBGC,WenSOinf}), and SPT of free fermions are classified by the K-theory in the ``tenfold way'' \cite{Schnyder2008,Kitaev2009}.
The classification of interacting fermions are much harder.
Progresses in recent years~\cite{GuSuH,Kapustin2015,Gaiotto2016,Kapustin2017,Putrov2017,MChengPRX2018,QRWangBSuH,JuvenWang2018,Cheng2018,QRWang2018X,LanFSPT2018X,JuvenWang2018} have provided mathematical frameworks to describe the classification, but the detailed computation is still challenging for general symmetry groups.
In contrast to SPT protected by onsite symmetries are crystalline symmetry-protected topological states, or simply topological crystalline states (TCS).

As suggested by name, TCS have their nontrivial topology protected by both onsite symmetries and crystalline symmetries.
Crystalline symmetries are the symmetry groups of periodic lattices in various dimensions (restricted, for simplicity, to two and three in this paper), and the study of crystalline symmetries as groups has been complete since the end of the last century \cite{Bradley2010}.
All crystalline symmetries are classified into 17 wallpaper groups in two dimensions (2D) and 230 space groups in three dimensions (3D).
Among TCS, those constituted of non-interacting fermions with charge conservation have so far attracted most theoretical and experimental effort.
These states are also known as the topological crystalline insulators \cite{Fu2011,Chiu2016}, the classification and diagnosis of which have only recently been completed \cite{Bradlyn2017,Po2017,Slager2017,Song2018a,Khalaf2018,Song2018}.
Interacting TCS, especially the fermionic ones, are far less understood.
On one hand, the framework of group cohomology for onsite-symmetry bosonic SPT cannot be directly applied; on the other hand, there is not an obvious way of adapting the K-theory, which is key to solving the classification problem of free fermions \cite{Kitaev2009,Shiozaki2014,Shiozaki2018}, to the task of classifying interaction fermions.
A recent work by \citet{ThorngrenElse2018} provides a mathematical connection between TCS and onsite-symmetry SPT states.
Another way to understand TCS is the process of dimensional reduction~\cite{Isobe2015,Song2017,SJHuang2017}, which constructs TCS by decorating high-symmetry points, lines and planes with lower-dimensional onsite-symmetry SPT states.
Comparing to the method of~\citet{ThorngrenElse2018}, these dimensional-reduction constructions are easier to compute in practice (because both the dimensionality and symmetry groups are reduced), and offer extra physical insight on the nature of these TCS states.
For interacting bosonic and fermionic SPT states, a large class of TCS states have been constructed this way, but the previous works have not been systematically applied to arbitrary symmetry groups, and it has not been shown whether the real-space construction is complete.

We in this paper show that all TCS, bosonic and fermionic, free and interacting, can be built up in real space from two types of elementary ingredients, the building blocks (or simply blocks) and the connectors.
Building blocks are finite-size pieces of lower dimensional SPT that are protected by the respective little symmetry group alone.
The little symmetry group includes all symmetry operations, onsite or spatial, that leave each point in the lower dimensional subspace invariant: they can be considered as enlarged onsite groups by the spatial symmetries that do not change the spatial coordinates on specific subspaces of the lattice.
A building block defined on a $p$-dimensional subspace is called a $p$-block.
For three dimensional TCS, one considers $p=3,2,1,0$-blocks~\footnote{3-block TCSs are simply 3D SPT states protected by the onsite symmetry alone, which are compatible with the crystalline symmetries. This is discussed in more details in Appendix~\ref{app:trivial}}.
To construct a gapped TCS, we arrange $p$-blocks in such a way that the space group of the TCS is preserved, including translation symmetries, point group symmetries and nonsymmorphic symmetries.
The $p$-blocks in general have open boundaries, and, being SPT themselves, gapless boundary states (or more precisely speaking, boundary anomalies) on their ($p-1$)-dimensional boundaries.
Therefore a symmetric construction with $p$-blocks alone cannot be gapped in the bulk, and in order to build a gapped state, one needs ``glue'' to close the open edges in the assembly.
The glue is the connectors, which are technically speaking torsors defined on ($p-1$)-dimensions.
($p-1$)-connectors are inserted where multiple $p$-blocks share one ($p-1$)-dimensional open edge, and should hybridize the gapless states contributed from the joining $p$-blocks so that the edge becomes gapped.
When all open edges are completed, that is, when the ``no-open-edge condition'' is met, we obtain assemblies that are (i) symmetric under onsite and spatial symmetries and (ii) gapped.
However, this does not mean that the crystal is topologically nontrivial, and for TCS, we require that it cannot be deformed into a product state.
Obviously, this implies that there is at least one building block with $p>0$ that is a nontrivial SPT, but this alone is insufficient: there are constructions from nontrivial ($p>0$)-building blocks that can still be trivialized.
We show that the trivialization can be understood as a ``bubbling process'', in which constructions that can be canceled by the ``bubbles'' are excluded, considered trivial.
Two TCS are hence topologically equivalent if the decorations can be related by a bubbling process, and this is called the ``bubble equivalence''.
The space of all TCS is hence the space of symmetric assemblies of building blocks satisfying the no-open-edge condition quotient the bubble equivalence.

One should be aware that both the no-open-edge condition and the bubbling equivalence, simple enough in appearance, have their subtleties.
While it is obvious that one may use a ($p-1$)-connector to complete the open edges at the meeting of two or multiple $p$-blocks, after all necessary ($p-1$)-connectors are added, at the ($p-2$)-joints where these ($p-1$)-connectors meet, there may be ($p-2$)-dimensional open edges.
Similarly, while it is natural that bubbles in $p+1$ dimensions can be used to annihilate $p$-blocks, there are cases where ($p+2$)-bubbles, leaving all ($p+1$)-blocks intact, annihilate $p$-blocks.
A third and related subtlety, called the group extension problem, is about the relations between TCS constructed from $p$-blocks and those constructed from ($p^\prime<p$)-blocks.
All three subtleties have to do with constructions that have nontrivial connectors, or torsors.
Torsors are not SPT (but may be understood as fractions of SPT), and their topological properties should be separately considered.

The real-space construction scheme given above allows an automated generation of all inequivalent TCS for arbitrary spatial and onsite symmetry groups in any dimension $D$, defined in the following steps: (a) make a symmetric arrangement of $p$-blocks; (b) add ($p-1$)-connectors to complete open edges; (c) use ($p+1$)-bubbles to ``modulo out'' trivial constructions; (d) add ($p-2$)-connectors to complete open edges of ($p-1$)-connectors; (e) use $p+2$ bubbles to ``modulo out'' trivial constructions and (f) repeat until the connectors are zero-dimensional and the bubbles are $d$-dimensional, where $d$ is the spatial dimension.
This process naturally fits the construction of TCS into a \emph{spectral sequence}, a successive approximation originally designed for computing homology (cohomology) groups of a topological space~\cite{GuideSpSeq,Brown}.
We adapt the terminology in mathematics and refer to this successive approximation as different \emph{pages} in the spectral sequence.
Each page, being worked out from the previous page, can be roughly understood as a certain level of approximation to the exact classification, more accurate than its previous page, and less accurate than the next.
Following this observation, we can analytically prove that the real-space construction process as defined above gives exactly the same classification of general bosonic TCS, leading to same results as derived from the gauging-spatial-symmetry argument.
This proof can be found in Appendix~\ref{app:ss}.
We develop an automated code and generate all bosonic TCS having typical onsite symmetries (such as unitary and antiunitary $Z_n$ symmetries) and any of the 17 wallpaper groups in 2D and 230 space groups in 3D.
On the fermionic side, the completeness of the construction scheme has been demonstrated for free fermions with charge conservation, time-reversal symmetry, and any of the 230 space group symmetries in Ref.~\cite{Song2018}.
We also point out while the general real-space construction still holds, the difficulty in classifying interacting fermionic TCS lies in the nontrivial superposition of states due to fermion statistics, and the lack of a unified bulk and boundary description for fermionic SPT.

Compared with the group cohomology formula in Ref.~\cite{ThorngrenElse2018}, our method for classifying TCS is not only easier to compute, but also more concrete so that for each TCS we have a real-space, piecewise construction.
In particular, it allows us to distinguish TCS made of building blocks in different dimensions, which is an additional structure in the TCS classification.
Additionally, the paginated structure of the spectral sequence also has physical interpretations: the different levels of approximation can be used to construct variants of the Hasting-Oshikawa-Lieb-Schultz-Mattis (HOLSM) Theorem~\cite{Lieb1961,OshikawaLSM,Arun2004,Hastings2004,Hastings2005,ZaletelLSM2015,ChengLSM2016,SJHuang2017,QiLSM2017,DyonLSM,Lu2017a,Lu2017b}.

The rest of the paper are divided into five sections. In Section~\ref{sec:simplified}, we introduce the decomposition of a three dimensional space into a chain complex of cells of difference dimensions, describe in detail an intuitive picture of how a topological crystalline state can be piecewise constructed in real space, with specific examples illustrating the physical meaning of the ``no-open-edge condition'' and the ``bubbling equivalence''.
In Section~\ref{sec:full}, the subtle issues in the process related to nontrivial connectors, such as the ($p-2$)-open-edges, the ($p+2$)-bubbles and the group extension problem, are addressed,  illustrated using both bosonic and fermionic examples. In Section~\ref{sec:examples}, we compute the full classification of bosonic TCS protected by typical onsite symmetries and any of the 17 wallpaper groups and 230 space groups.
We demonstrate the steps of computation using two examples, and the complete list of results for all groups is provided in Appendix~\ref{app:wgsg}.
In Section~\ref{sec:lsm}, we apply our real-space construction to study the Hasting-Oshikawa-Lieb-Schultz-Mattis (HOLSM) Theorem~\cite{Lieb1961,OshikawaLSM,Arun2004,Hastings2004,Hastings2005} and its generalizations~\cite{ZaletelLSM2015,ChengLSM2016,SJHuang2017,QiLSM2017}, including the recent one which enforcing a nontrivial SPT state~\cite{DyonLSM,Lu2017a,Lu2017b}.
In Section~\ref{sec:conclusion}, we summarize our results and possible discuss future directions.

\section{Real-space construction}
\label{sec:simplified}

In this section, we introduce an intuitive picture for constructing TCS in two and three dimensions (2D, 3D).
The first step is the decomposition of $d$ dimensional Euclidian space into ``cells'' of various dimensions with respect to the lattice symmetry group, building a \emph{chain complex} for later use.
Next, we decorate the $p<d$ cells with $p$-blocks in a symmetric way, obtaining a collection of assemblies as candidates for TCS.
Then we check each candidate against the condition that the open ($p-1$)-dimensional edges of the $p$-blocks are closed by glueing together adjacent $p$-blocks.
Candidates meeting the no-open-edge condition do not have gapless modes on the ($p-1$)-cells in the bulk.
Finally, we discard from the candidates assemblies that equal vacuum (product states) under the bubble equivalence defined by ($p+1$)-bubbles, so that the rest of the candidates are better approximations of TCS.
In this simplified version of real-space construction, we have not explicitly calculate the connectors, or the glue, at the ($p-1$)-cells, which may lead to subtle issues treated in the next section.

\subsection{Cell decomposition and chain complex}
\label{sec:blocks}
Following Ref.[\onlinecite{Song2018}], for each space group, there is a well-defined cell decomposition process that a $d$-dimensional Euclidean space ($R^d$) is decomposed into the union of $p=0,1,2,...,d$-dimensional ``cells'' with zero overlap.
Here a $p$-cell is topologically equivalent to $R^p$, or an $p$ dimensional disk minus its boundary, and we denote the collection of all $p$-cells in a decomposition as $Y_p$.
We use greek letters $\sigma$, $\tau$, $\mu$ and $\gamma$ to denote $p=d,d-1,d-2,d-3$-cells, respectively, and also use $\sigma$ to denote a general cell with unspecified dimension.
We require that any space-group operation maps one $p$-cell to another $p$-cell.
We also require that the union of all $Y_p$ is $R^d$ itself.
Mathematically, the collection of all cells under these conditions form a topological space, $Y$, called a $G$-complex.
Furthermore, for any $\sigma\in{Y}_p$, we consider the subgroup $G_\sigma$ that maps $\sigma$ to itself.
We require that any $g\in G_\sigma$ has a \emph{pointwise} action on $\sigma$, which in general include onsite symmetries and crystalline symmetries.
In other words, $G_\sigma$ acts as an onsite symmetry group locally on $\sigma$.
Additionally, unlike in Ref.[\onlinecite{Song2018}], the cells are oriented.
The orientation is arbitrary subject to the condition that the orientations of $\sigma$ and $g\cdot\sigma$ are also related by $g$.
\begin{figure}
\begin{centering}
\includegraphics[width=1.0\linewidth]{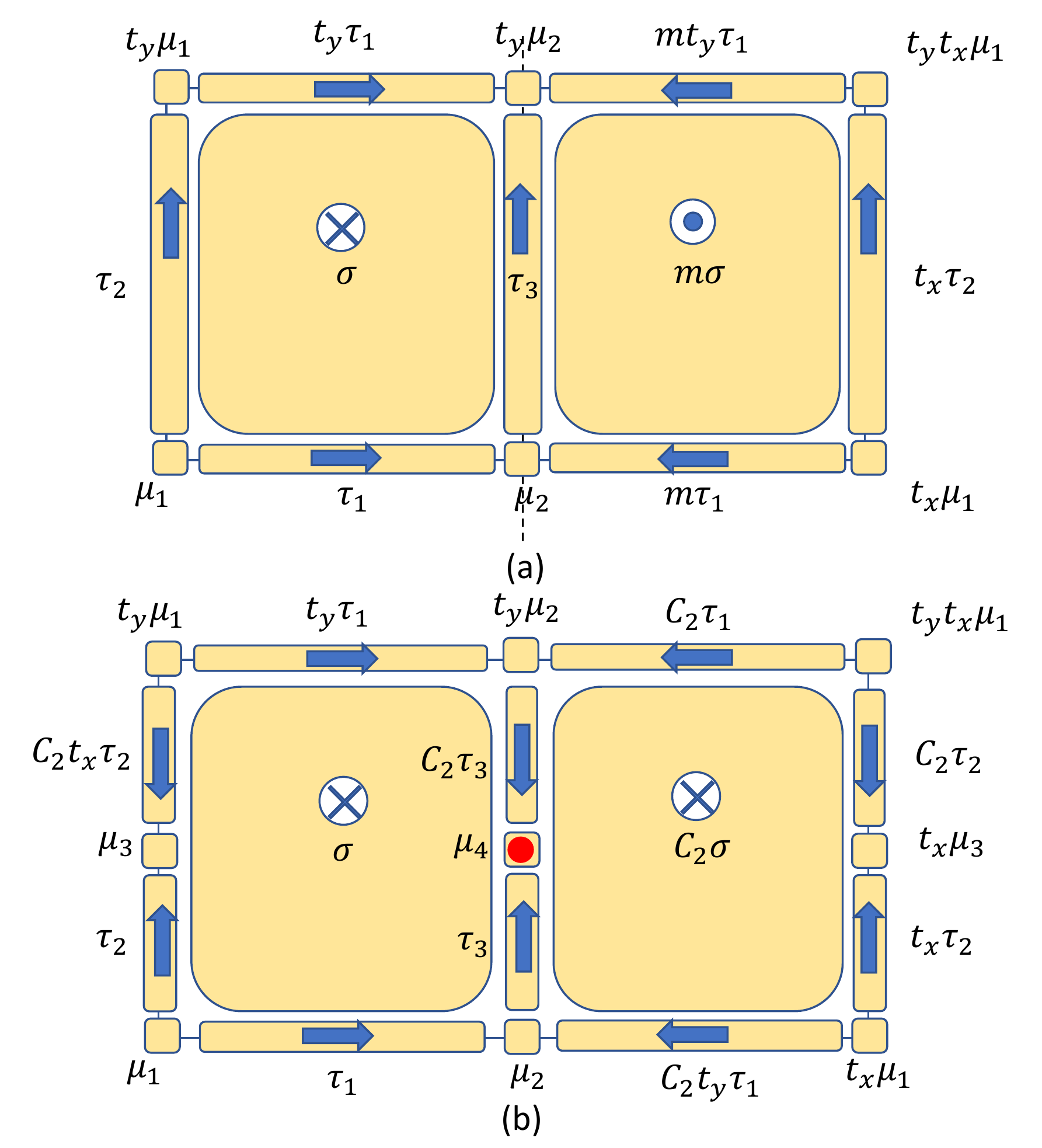}
\par\end{centering}
\caption{(a) The cell decomposition for 2D wallpaper group $Pm$. The mirror line is the vertical dashed line, and the $2, 1, 0$-cells within one unit cell are labeled, with their respective orientations marked by arrows. (b) The cell decomposition for 2D wallpaper group $P2$. The rotation center is marked in red, and the $2, 1, 0$-cells within one unit cell are labeled, with their respective orientations marked by arrows.}\label{fig:1}
\end{figure}
We illustrate the cell decomposition with simple examples.
In Fig.\ref{fig:1}(a), we show the cell decomposition within one unit cell of 2D space group $pm$, \emph{i.e.}, orthogonal with one mirror line mapping $x$ to $-x$.
The independent cells are labeled by $\sigma_{1,2,3,\ldots}$ ($\tau$ for 1-cells and $\mu$ for 0-cells), and the other cells are labeled as $g\sigma$ with $g\in{G}$.
Their orientations are marked by arrows.
In Fig.\ref{fig:1}(b), there is another example of decomposition for 2D space group $p2$, \emph{i.e.}, orthogonal with a twofold rotation center $C_2$.
For this decomposition, we comment that the 1D segment passing the rotation center, instead of being a 1-cell by itself, is decomposed into two 1-cells ($\tau_3$ and $C_2\tau_3$) and one 0-cell ($\mu_4$).
This is in contrast to the example in Fig.\ref{fig:1}(a), where the segment coincident with the mirror line needs no additional decomposition.
The difference is because we require that symmetry group of a $p$-cell, $G_\sigma$, should act pointwise on $\sigma$, and while each point in $\tau_3$ is invariant under mirror in Fig.\ref{fig:1}(a), only $\mu_4$ is invariant under $C_2$, and $\tau_{3}$ and $C_2\tau_3$ are mapped to each other.

After decomposition, we are ready to define the operator $\partial$ acting on a $p$-cell $\sigma_p$: $\partial\sigma_p$ gives the ($p-1$)-cells that are at the boundary at $\sigma$, \emph{i.e.},
\begin{equation}
\partial\sigma_p=\sum_{\sigma_{p-1}\subset\bar\sigma_p}\langle\sigma_{p-1}|\partial\sigma_p\rangle\sigma_{p-1}.
\end{equation}
The coefficient $\langle\sigma_{p-1}|\partial\sigma_p\rangle=0$ if $\sigma_{p-1}$ is not part of the boundary of $\sigma_p$; $\langle\sigma_{p-1}|\partial\sigma_p\rangle=\pm1$ if $\sigma_{p-1}$ is part of the boundary, and the sign depends on the relative orientations of $\sigma_p$ and $\sigma_{p-1}$.
For example, in Fig.\ref{fig:1}(a), the orientation of $\sigma$ gives, by right-hand rule, natural orientations for all adjacent 1-cells, and if the designated orientation of $\tau_i$ is parallel (opposite) to it, $\langle\sigma_{p-1}|\sigma_p\rangle=+1$ ($-1$): $\partial\sigma=\tau_1+\tau_3-t_y\tau_1-\tau_2$.
For 1-cell $\tau_i$, the arrow naturally gives the coefficients $\langle\mu_i|\partial\tau_j\rangle$: the zero cell at the head and at the tail of the arrow have $+1$ and $-1$ coefficients, respectively.
For example, $\partial\tau_1=\mu_2-\mu_1$.
The $\partial$-operation hence defines a chain complex from $d$-cells all the way down to $0$-cells:
\begin{equation}
  \label{eq:Ycc}
Y_d\xrightarrow{\partial}Y_{d-1}\xrightarrow{\partial}\cdots\xrightarrow{\partial}Y_0.
\end{equation}

Here, we notice that the homology groups of this chain complex are all trivial, because $Y$ is a decomposition of $R^d$, which is topologically trivial.
As a result, the chain complex in Eq.~\eqref{eq:Ycc} must be an exact sequence, {\it i.e.}, $\partial^2=0$.
This fact is important to the proof in Appendix~\ref{app:ss}.

\subsection{Building blocks and connectors}
\label{sec:block}

We start by reviewing topological states we can decorate on each cell.
This forms the basis of our real-space construction:
Each cell $\sigma$ of dimension $p$ can be decorated with a topological state, which is protected by the local onsite symmetry group $G_\sigma$.
On an isolated cell, such topological states are classified by the $p$-dimensional SPT states protected by $G_\sigma$, which we denote by $\Phi^p(G_\sigma)$.
We will refer to such decorations as ``$p$-dimensional building blocks", or $p$-blocks for short.
However, in our real-space construction, we need to consider a more general case, where $\sigma$ is part of the boundary of a $(p+1)$-dimensional cell $\tau$, which is in turn decorated by another topological state.
Hence, we need to extend the scope of topological states to also include boundary topological states.
In general, we consider a set of representative fixed-point topological states, which we denote by $\Psi^p(G)$.

There are two important operations on $\Psi^p(G)$, which are the mathematical fundation for the computation of our real-space constructions.
First, we can stack two states $\psi_{1,2}$ in $\Psi^p(G)$, and get a new state, which we denote by $\psi_1\boxplus\psi_2$.
Second, we can compute the coboundary of $\psi\in\Psi^p(G)$, which should be a state in $\Psi^{p+1}(G)$.
The physical meaning of $d\psi$ is that $\psi$ can be realized on a $p$-dimensional boundary of $d\psi$.

The structure of such sets $\Psi^p(G)$ and the rules of these two operations depend on the nature of the system: being a free-fermion, interacting-boson or interacting-fermion system.
In particular, for bosonic systems, $\Psi^p(G)$ is simply the cochain space of $G$.
In the main text, we assume that $\Psi^p(G)$, $\boxplus$ and $d$ are known, and use them to compute real-space constructions, while leaving the details to Appendix~\ref{app:spt-review}.
In particular, the SPT phases $\Phi^p(G)$ can be computed from $\Psi^p(G)$ and the coboundary maps $d$:
$\Phi^p(G)$ is given by the subgroup of cocycles in $\Psi^p(G)$ satisfying $d\alpha=0$, quotient the coboundries given by the image of the coboundary map $d[\Psi^{p-1}(G)]$.

\subsection{``First page'' (first-order approximation) candidates for TCS}
In general, a $d$-dimensional TCS can be constructed by assumbling $p$-blocks ($p<d$) and gluing them with lower-dimensional connectors, while leaving all higher-dimensional cells empty.
The state after this construction is termed a $p$-block \emph{topological crystal}~\cite{Song2017,SJHuang2017,Song2018}.
In this way, all TCSs are organized into $p$-block topological crystals, where $p=0,\ldots,d$:
in particular, one can argue that a TCS with trivial (but not necessarily empty) SPT states decorationed on $d>p$ cells can be continuously deformed to a topological crystal with all ($d>p$)-cells empty.
The classification and construction of TCS then amounts to enumerating all inequivalent topological crystals for a given symmetry group $G$.
Here, when enumerating all $p$-block topological crystals, we do not distinguish different choices of lower-dimensional connectors.
This is because the difference~\footnote{The difference between two states $A$ and $B$ is defined as the stacking of $A$ and the inverted state of $B$, denoted by $-B$.} between two such choices on $q$-cell connectors can be viewed as a $q$-block topological crystals and described by the $q$-block classification.

We now consider $p$-block topological crystals.
They are labeled by difference choices of $p$-blocks, which are $p$-dimensional SPT states at $p$ cells $\sigma\in Y_p$.
Their classification is determined by the local onsite group $G_\sigma$, and given by $\Phi^p(G_\sigma)$.
For bosonic systems, there is $\Phi^p(G_\sigma)=H^{p+1}[G_\sigma,U(1)]$.
However, for interacting fermionic systems, the calculation of $\Phi^p(G_{\sigma})$ for arbitrary $G_{\sigma}$ can be much more complicated.

Naturally, if some $p$-cell $\sigma$ is decorated with some $p$-block, $[\alpha]$, then symmetry requires that the $p$-cell $g\sigma$ must be decorated by the $p$-block $g\cdot[\alpha]_\sigma\in\Phi_{g\sigma}$ for $g\in{G}$.
This is possible because of the isomorphism $G_{g\sigma}=gG_\sigma g^{-1}\simeq G_\sigma$.
In Appendix~\ref{app:gaction}, we show the explicit definition of $g\cdot[\alpha]$, and for now we intuitively understand it as copying the $p$-block at $p$-cell $\sigma$ to another $p$-cell $g\sigma$.
Therefore, only the cells in the $G$-orbits of $Y_p$, $Y_p/G$, may have independent $p$-blocks, and the $p$-blocks at all the other cells are determined by symmetry.
Physically, $\sigma\in{Y_p}/G$ are the $p$-cells that are not related by any $g\in{G}$ and $g\neq{e}$.
Once the $p$-blocks for all cells in $Y_p/G$ are specified, we obtain a symmetric assembly denoted by $[\psi_p]$, such that
\begin{equation}
[\psi_p|_{\sigma}]\in\Phi^p(G_{\sigma})
\end{equation}
and
\begin{equation}
  \label{eq:a'=ga}
[\psi_p|_{g\sigma}]=g\cdot[\psi_p|_{\sigma}].
\end{equation}
Throughout the paper, we use $\psi_p$ to denote the state, or wave function, on all $p$-cells, use $|_\sigma$ to denote the same wave function restricted to a certain cell $\sigma$.
We use $[\psi]$ for the phase the representative state of which is $\psi$.
The collection of all possible symmetric assemblies from $p$-blocks is denoted by $E^p_{p,1}$
\begin{equation}
  \label{eq:E1}
E^p_{p,1}\equiv\{[\psi_p]\}\equiv\bigoplus_{\sigma\in{Y}_p/G}\Phi^p(G_\sigma).
\end{equation}
For reasons to be revealed in later sections, these assemblies are called the \emph{first-page} candidates for TCS, which can also be taken as first-order approximations to TCS.
The exact meaning of $E^q_{p,r}$ will also be introduced and put to use in due course.

\subsection{No-open-edge condition}
\label{sec:gluing}

Being SPT, $[\psi_\sigma]$ necessarily has gapless modes at $\tau\in\partial\sigma$.
In fact, in the $G$-complex, each $(p-1)$-cell is in the boundary of at least two $p$-cells.
(For example, $\tau_3\in\partial\sigma\cap\partial{m}\sigma$ in Fig.\ref{fig:1}(a).)
Therefore, for a gapped TCS, we require that the gapless modes at $\tau$ contributed from all adjacent $\tau$ can gap out each other, so that there is no open edge.
More precisely speaking, the gapless boundary modes reflect the quantum anomaly of an SPT boundary.
The necessary condition for a gapped $(p-1)$-cell $\tau$ is that the anomalies contributed by all adjacent $[\psi_p|_\sigma]$ cancel each other.

Suppose a ($p-1$)-cell $\tau$ is the common edge of $n$ pieces of $p$-cells $\sigma_{1,\ldots,n}$, each decorated by a $p$-cell $[\psi_p|_{\sigma_i}]\in\Phi^p(G_{\sigma_i})$.
The no-open-edge condition requires that when we direct sum the bordering $p$-blocks, the resulted total state is a trivial SPT.
However, as each $p$-cell has its own local onsite symmetry group $G_{\sigma_i}$, and as the ``trivialness'' or ``nontrivialness'' of SPT is only well-defined with respect to some symmetry group, we have to make clear what is the symmetry for the direct sum of the $p$-blocks.
The answer is $G_\tau$, the local onsite symmetry group of $\tau$, which is the shared edge of $\sigma_{1,\ldots,n}$.
Physically, $\tau$ is a high-symmetry line, and hence have higher symmetry than the bordering high-symmetry planes: $G_{\sigma_i}\subset{G}_\tau$.
Suppose for a pair of $\tau$ and $\sigma$ satisfying $\langle\tau|\partial\sigma\rangle\neq0$, there is $g\in{G}_\tau$ and $g\notin{G}_\sigma$, then $g\sigma$ must be another $p$-cell bordering $\tau$, $\langle\tau|\partial{g}\sigma\rangle=\langle\tau|\partial\sigma\rangle$.
(For better understanding, use Fig.\ref{fig:1}(a), where $m\in{G}_{\tau_3}$ and $m\notin{G}_\sigma$, and $\langle\tau_3|\partial_\sigma\rangle=\langle\tau_3|\partial{m}\sigma\rangle=1$.)
The direct sum $[\psi_p|_\sigma]\oplus[\psi_p|_{g\sigma}]$ is hence symmetric under not only under $G_\sigma$ and $G_{g\sigma}$, but also under $g$.
Therefore, the total direct sum of all bordering $p$-blocks
\begin{equation}\label{eq:defd1}
[d_1\psi_p|_\tau]\equiv\bigoplus_{\sigma\in{Y_p}}\langle\tau|\partial\sigma\rangle[\psi_p|_\sigma]\in\Phi^p(G_\tau).
\end{equation}
We remark that in Eq.(\ref{eq:defd1}), if $\langle\tau|\partial\sigma\rangle=-1$, it means that we need to invert the $p$-block at $\sigma$, before stacking it.
We also note while $[d_1\psi^p_\tau]$ has the symmetry group of $\tau$, a $(p-1)$-cell, $[d_1\psi^p_\tau]$ is still an SPT (trivial or nontrivial) in $p$ dimensions.
Or one can say that $[d_1\psi^p_\tau]$ is a $p$-SPT associated with a ($p-1$)-cell, protected by the symmetries of the ($p-1$)-cell.
The collection of these $p$-SPT attached to ($p-1$)-cells are denoted by $E^{p}_{p-1,1}$.
If $[d_1\psi_p|_\tau]$ is a nontrivial SPT with respect to $G_\tau$, then on $\tau$, which is the common edge of the stacked layers, there must be gapless modes, meaning that this particular $\psi_p$ fails the no-open-edge condition.
For a given $[\psi_p]\in{E}^p_{p,1}$, the collection of all $[d_1\psi_p|_\tau]$ is given by $d_1[\psi_p]=\oplus_{\tau\in{Y}_{p-1}}[d_1\psi_p|_\tau]$.
This defines $d_1$ as a linear mapping between $E^p_{p,1}$ to $E^p_{p-1,1}$.
Therefore, the no-open-edge condition states that only the kernel of $d_1$ may be candidates for TCS, \emph{i.e.}, $d_1[\psi_p]=0$.
Here, $E^p_{p-1,1}$ has a similar definition as \eqref{eq:E1},
\begin{equation}
  \label{eq:E1a}
E^p_{p-1,1}\equiv\bigoplus_{\sigma\in{Y}_{p-1}/G}\Phi^p(G_\sigma).
\end{equation}
Physically, it describes attaching elements of $\Phi^p$ on ($p-1$)-cells, where they are interpreted as symmetry anomalies on the boundary of $p$-cells.
Hence, we refer to elements of $E^p_{p-1,1}$ as \emph{anomaly patterns}.

On the other hand, if $[d_1\psi_p|_\tau]$ is a trivial $p$-SPT with symmetry group $G_\tau$, on $\tau$ we can place a ``mass term'' that gaps out the edge modes contributed by $\sigma_{1,\ldots,n}$.
The resulted gapped state on the ($p-1$)-cell $\tau$ is called a \emph{connector}, as in real space it acts as the nexus of the bordering $p$-cells $\sigma_{1,\ldots,n}$.
Connectors are not SPT in general, but are \emph{torsors} the definition and properties of which we defer to later sections.
For now, we only need to know that any $[\psi_p]\in{E}^p_{p,1}$ that satisfies the no-open-edge condition can be glued by ($p-1$)-connectors such that any $\tau\in{Y}_{p-1}$ is also gapped.
\begin{figure*}
\begin{centering}
\includegraphics[width=1.0\linewidth]{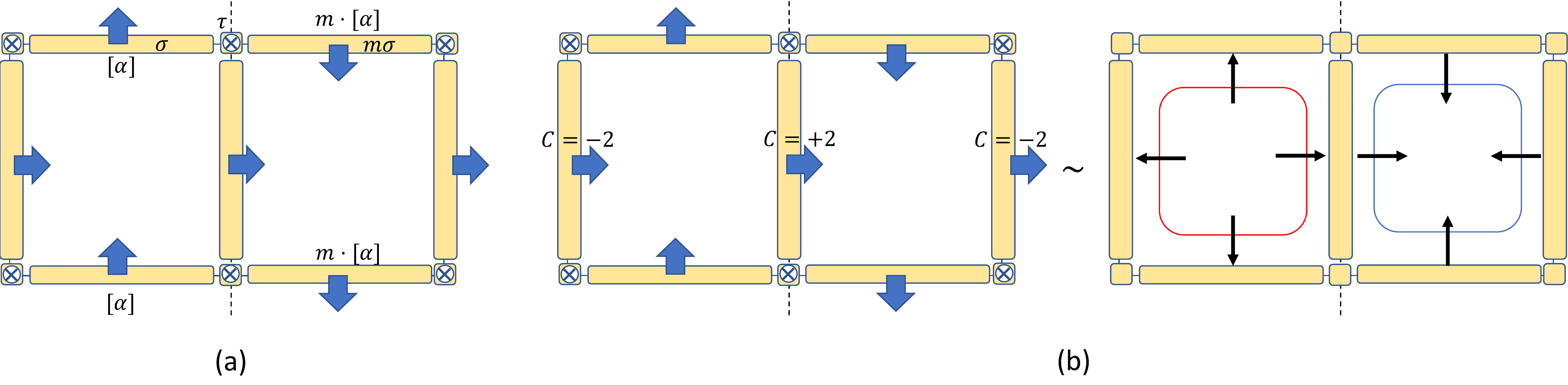}
\par\end{centering}
\caption{(a) One symmetric assembly in $E^2_{2,1}$ for space group $Pm$ in one unit cell. The figure shows the cross section of a unit cell that is perpendicular to the $z$-direction. The arrows show the orientation of $2$- and $1$-cells. This assembly fails the no-open-edge condition. (b) The evolution of assembly in $\mathrm{ker}d_1$ for space group $Pm$ in a fermionic system. The $2$-cells are decorated with Chern insulators, the Chern numbers of which are indicated. This state is equivalent to an assembly in $E^2_{3,1}$, where two $3$-bubbles are decorated to the $3$-cells. The outward (inward) thin arrow means that the Chern number on the associated boundary is $+1$ ($-1$).}\label{fig:2}
\end{figure*}
We use two examples in bosonic systems to show, respectively, that certain $p$-assemblies in $E^p_{p,1}$ satisfy and do not satisfy the no-open-edge condition.
The space group is that of a 3D orthogonal lattice having one mirror plane mapping $x$ to $-x$ plotted in Fig.\ref{fig:2}(a), and the onsite symmetry group is a unitary $Z_4$-symmetry.
We differentiate the cases where (i) the $Z_4$ generator $g_0$ commutes with mirror symmetry $m$: $g_0m=mg_0$, and (ii) they do not commute $g_0m=mg_0^{-1}$.
Consider an assembly $\psi_2$ generated from decorating $\sigma$ with a $2$-block which is the generator of $\Phi^2(G_{\sigma_1})=Z_4$, $[\alpha]$.
The decorations on the other orbits of $Y_2/G$ are set to be vacuum, and within a unit cell, the only decorated $2$-cells are $\sigma$ and $m\sigma$.
For case-(i), the mirror operation does not act on the local degrees of freedom, $m\cdot[\alpha]=[\alpha]$; but for case-(ii), we have $m\cdot[\alpha]=-[\alpha]$.
We check the assembly agains the no-open-edge condition at $\tau$ between $\sigma$ and $m\sigma$.
Following Eq.(\ref{eq:defd1}), in case-(i) there is
\begin{equation}
[d_1\psi_2|_\tau]=[\alpha]+[\alpha]=2[\alpha],
\end{equation}
where we have used $\langle\tau|\partial\sigma\rangle=\langle\tau|\partial{g}\sigma\rangle$.
Since $2[\alpha]\neq0$, $[\psi_2]$ in case-(i) fails the no-open-edge condition.
In case-(ii), we have
\begin{equation}
[d_1\psi_2|_\tau]=[\alpha]+(-[\alpha])=0.
\end{equation}
We conclude that $[\psi_2]$ in case-(ii) satisfies the no-open-edge condition.

\subsection{Bubble equivalence}
\label{sec:cobdry}

The kernel of $d_1:E^p_{p,1}\rightarrow{E}^p_{p-1,1}$ are gapped, symmetric assemblies of $p$-blocks (gapped at any $\tau\in{Y}_{p-1}$ by connectors).
But two different assemblies in $\mathrm{ker}d_1$ may be topologically equivalent to each other, and more importantly, some nontrivial (non-vacuum) state in $\mathrm{ker}d_1$ may even be equivalent to vacuum upon adiabatic deformation.
Following Ref. \cite{Song2018}, every adiabatic deformation is equivalent to the creation (annihilation) of \emph{bubbles} within some $p$-cell $(0<p\le{d})$.
A $p$-bubble is a $p$-dimensional disk inside some $p$-cell $\nu\in{Y_p}$, the inside of which is vacuum, and the boundary some $(p-1)$-SPT protected by $G_\nu$.
Therefore, it can be considered as some $(p-1)$-SPT attached to a $p$-cell.
It is straightforward to see that creation of bubbles cannot change the topology of the state.
Any bubble can shrink to a point and vanish, and, as the inside of any cell in our cell decomposition does not have any spatial symmetry, the shrinking and vanishing process does not break any symmetry.

Therefore, the topology of $[\psi_p]$ is unchanged after we decorate, in a $G$-symmetric way, the ($p+1$)-cells with some ($p+1$)-bubbles.
A $G$-symmetric assembly of ($p+1$)-bubbles is denoted $[\tilde\psi_{p+1}]$ and $[\tilde\psi_{p+1}]|_\nu$ is this assembly restricted to some $\nu\in{Y}_{p+1}$.
The collection of all ($p+1$)-bubbles are denoted by $E^p_{p+1,1}$, defined in a way similar to Eq.~\eqref{eq:E1}:
\begin{equation}
  \label{eq:E1b}
E^{p+1}_{p,1}\equiv\bigoplus_{\sigma\in{Y}_p/G}\Phi^{p+1}(G_\sigma).
\end{equation}
We refer to $[\tilde\psi_{p+1}]\in E^{p+1}_{p,1}$ as a \emph{bubbling pattern}.

To show the equivalence relations between assemblies $E^p_{p,1}$ induced by bubbles in $E^p_{p+1,1}$, we need to relate $[\tilde\psi_{p+1}]$ to an element in $E^p_{p,1}$.
We start with enlarging the bubbles, so that a bubble inside $\nu\in{Y}_{p+1}$ touches the boundary of $\nu$.
The $p$-SPT at the surface of the ($p+1$)-bubble $\nu$ then automatically attaches to all $\sigma\in{Y}_p$ at the boundary of $\nu$.
At the same time, we notice that any given $\sigma$ is the boundary of two or multiple $\nu\in{Y}_{p+1}$, so that the state induced at $\sigma$ comes from all bordering $\nu\in{Y}_{p+1}$,
\begin{equation}
  \label{eq:deftd1}
[\tilde{d}_1\tilde\psi_{p+1}|_\sigma]\equiv\bigoplus_{\nu\in{Y}_{p+1}}\langle\sigma|\partial\nu\rangle[\tilde\psi_{p+1}|_\nu].
\end{equation}
It is important to realize that although $[\tilde\psi_{p+1}|_\nu]\in\Phi^p(G_\nu)$ is a $p$-SPT protected by $G_\nu$, their sum is a $p$-SPT under a larger group $G_\sigma\supset{G}_\nu$ (after the orientation alignment is resolved by the coefficients $\langle\sigma|\partial\nu\rangle$ in the summation).
Therefore, we identify $[\tilde{d}_1\tilde\psi_{p+1}|_\sigma]$ as a $p$-SPT protected by $G_\sigma$, \emph{i.e.}, $[\tilde{d}_1\tilde\psi_{p+1}|_\sigma]\in\Phi^p(G_\sigma)\subset{E}^p_{p,1}$.
Eq.(\ref{eq:deftd1}) maps ($p+1$)-bubbles to $p$-blocks, establishing a linear map from $E^p_{p+1,1}$ to $E^p_{p,1}$, and since all elements in $E^p_{p+1,1}$ are topologically trivial, the image of the mapping is also trivial.
These trivial states in $\mathrm{img}\tilde{d}_1$ give the equivalence relations in $E^p_{p,1}$: $[\psi_p]\sim[\psi'_p]$ if and only if $[\psi_p]-[\psi'_p]\in\mathrm{img}\tilde{d}$.


We use the example in Fig.\ref{fig:2}(b) to illustrate the bubble equivalence.
It is a fermionic system with charge conservation symmetry and the lattice is 3D orthogonal with a mirror plane sending $x$ to $-x$.
We consider a certain $[\psi_2]$ where some $2$-cells are decorated with Chern insulators, the Chern numbers of which are shown in Fig.\ref{fig:2}(b).
It can be easily checked that the assembly in Fig.\ref{fig:2}(b) satisfies the no-open-edge condition.
But in Fig.\ref{fig:2}(b), we show that the state in $E^2_{2,1}$ is actually equivalent to a state made from $3$-bubbles only, that is, a state in  $\mathrm{img}\tilde{d}_1$.
There are two $3$-bubbles in one unit cell.
The boundary of the left bubble has Chern number $+1$ and that of the right boundary has Chern number $-1$.
Therefore this assembly is topologically trivial.

The no-open-edge condition requires that TCS be in the kernel of $d_1$ in Eq.(\ref{eq:defd1}), and the bubble equivalence states that two TCS related by an image of $\tilde{d}_1$ are topologically the same.
The first-order approximation to TCS, $E^p_{p,1}$, is refined by these two conditions into
\begin{equation}
  \label{eq:E2}
E^p_{p,2}\equiv\frac{\mathrm{ker}d^{p}_{p,1}}{\mathrm{img}d^p_{p+1,1}},
\end{equation}
where we introduce a general version of $d_1$ and $\tilde{d}_1$: $d^{p}_{q,r}: E^p_{q,r}\rightarrow{E}^{p+1-r}_{q-r,r}$.
$d_1$ and $\tilde{d}_1$ are the special cases $d_1=d^p_{p-1,1}$, $\tilde{d_1}=d^p_{p+1,1}$.

Like those in $E^p_{p,1}$, states in $E^p_{p,2}$ are also generated by decorating $p$-cells and are $G$-symmetric, but no-open-edge condition ensures that all cells in $Y_{p-1}$ are gapped, and the bubble equivalence relation ensures that they cannot be trivialized by ($p+1$)-bubbles.
$E^p_{p,2}$ is hence the second-page approximation to the set of TCS made from $p$-blocks ($p$-TCS for short).
Here, \emph{page} is the mathematical terminology referring to the order of approximation in the spectral sequence, and we denote the page number by $r$ in the notation $E^q_{p,r}$.
In further sections, we are to treat higher-page approximations $E^p_{p,3}, E^p_{p,4}...$, and in the end $E^p_{p,\infty}$ is the exact collection of $p$-TCS.
In~\ref{app:trivial}, however, we show that for bosonic systems, if $G$ is a direct product of the onsite and the space symmetry groups $G=SG\otimes{G}_0$, then the second page already contains the final answer: $E^p_{p,2}=E^p_{p,\infty}$.

\subsection{Further considerations}
\label{sec:further}
The discussion presented so far in this section provides an algorithm to enumerate possible topological crystals from lower-dimensional building blocks.
In this algorithm, we start by decorating $p$-dimensional cells with SPT states protected by the local symmetry groups.
We then compute whether the resulting anomalies cancel on the $(p-1)$-cells, and whether the state can be trivialized by a trivialization pattern on the $(p+1)$-cells.
However, a careful mathematical analysis in Sec.~\ref{sec:full} reveals that, for a general symmetry group $G$, such an intuitive algorithm is not complete as the answers in $E^{p}_{p,2}$ contains false entries that do not represent valid and nontrivial SPT states:

On one hand, $E^{p}_{p,2}$ may contain invalid entries:
Although we have checked that the building blocks can be glued together in a gapped way along the $(p-1)$-dimensional edges,
it is possible that doing so will always leave gapless modes on the $(p-2)$-dimensional cells.
In fact, two examples will be given in Sec.~\ref{sec:noe2}.
Therefore, to make sure that an SPT building block represents a gapped SPT state, one need to check that if the decoration can be extended to all lower-dimensional cells without anomaly.

On the other hand, $E^{p}_{p,2}$ may contain trivial entries which can be revealed by considering higher-dimensional bubbles. It is possible that bubbles on $(p+2)$-dimensional cells do not create any nontrivial SPT decorations on the $(p+1)$-cells, but creates such decorations on the $p$-cells.
Such a pattern of bubbles then indicates that the decoration on $p$-cells, $[\psi_p]\in{E}^p_{p,2}$ it creates is trivial.
In fact, such an example will be given in Sec.~\ref{sec:lsm}.
Consequently, to eliminate all trivial entries, one has to consider bubbles on all higher-dimensional cells.

There is one more subtlety we need to consider.
We have generally divided TCS according to the dimensionality of the building blocks.
For example, the 3D TCS can be divided into SPTs with plane-decorations and line-decorations, represented by $E^2_{2,\infty}$ and $E^1_{1,\infty}$, respectively.
(Point decorations correspond to atomic insulators and do not have boundary states in general, and so are excluded from the set of TCS.
Nevertheless, they can also be easily enumerated using our classification scheme.)
The complete classification of TCS is then a combination of all topological crystals with all possible $0<p<d$.
However, when recombining the classification of topological crystals with different building-block dimensions,
the result may be a nontrivial group extension of the two classification groups, instead of a direct sum.
For example, assume that for some $G$, $E^2_{2,\infty}$ and $E^1_{1,\infty}$ are both $\mathbb Z_2$.
Naively, the combined classification would be $\mathbb Z_2\times\mathbb Z_2$.
However, the combined result could also be $\mathbb Z_4$, which is a nontrivial extension of $\mathbb Z_2$ over $\mathbb Z_2$.
Intuitively, imagine stacking two copies of the nontrivial elements from $E^2_{2,\infty}$.
Since the classification is $\mathbb Z_2$, the resulting SPT state is trivial if viewed as a topological crystal with $p=2$, {\it i.e.}, the decoration on the 2-cells is trivial.
However, the resulting state can have nontrivial decorations on the 1-cells, and thus is the nontrivial $1$-TCS, or, the generator of $E^1_{1,\infty}$.
If this happens, the combined classification is then $\mathbb Z_4$, generated by the generator of $E^2_{2,\infty}$.
In general, stacking two or multiple $p$-TCS may produce trivial decorations on $p$-cells, but nontrivial decorations on $p^\prime$-cells where $p^\prime<p$, which is a nontrivial $p^\prime$-TCS.
Finding these relations in general is what we call a group extension problem.
From the arguments above, we can see that all the three subtleties outlined above involve the \emph{connectors} on lower-dimensional cells, the subject of the following section.

\subsection{Summary}
\label{eq:sum2nd}

In this section, we have seen that, for the simple cases of $G=SG\times G_0$, the classification of topological crystals with building-block dimension $d_b=p$ is given by $E^p_{p,2}=E^p_{p,\infty}$ in Eq.~\eqref{eq:E2}.

One advantage of the topological-crystal approach is that it allows us to consider topological crystals with different building-block dimensions separately.
In particular, it allows us to consider a more physical classification of crystalline SPTs, which ignores 0D building blocks.
The reason for considering this is that when considering the classification of topological states, we usually identify states that can be smoothly deformed to each other without breaking the symmetries.
Included in these smooth deformations are inserting and removal of local nondegenerate degrees of freedom, which in general can carry arbitrary 1D linear representations of the local symmetry group.
These degrees of freedom are precisely the content of 0D building blocks.
Therefore, the classification ignoring these 0D building blocks is the more physical one to consider, comparing to the full group-cohomology classification $H^{d+1}[G,\uone_{PT}]$~\cite{ThorngrenElse2018}.
Using the real-space construction, we can easily compute this classification, as
\begin{equation}
\label{eq:sum1}
\bigoplus_{p=1}^d E^p_{p,\infty}=\bigoplus_{p=1}^d E^p_{p,2}.
\end{equation}

\section{The Connectors}
\label{sec:full}

In this section, we explain how to solve the subtleties outlined in Sec.~\ref{sec:further}.
A key step in the computations is to determine the concrete content of the connectors decorated on the $(d_b-1)$-cells, which connect the SPT states on neighboring $d_b$-cells.
Using these connectors, we can compute the second-page no-open-edge conditions and bubbling equivalences, which together give the third-page result of the classification.
The connector also allows us to solve the group-extension problem arised in the process of combining classifications of TCSs with different $d_b$.

\subsection{Contents of the connectors}
\label{sec:connector}

We begin by reviewing the content of connectors.
Consider a $p$-block TCS $[\psi]\in E^p_{p,2}$, whose building blocks are $p$-dimensional SPT states decorated on the $p$-cell.
The connectors on the $(p-1)$-cells are then constrained by these building blocks through the bulk-boundary relation.
Previously, we studied the no-open-edge condition in Sec.~\ref{sec:gluing}, which ensures the existence of gapped symmetric connectors.
However, to further determine the concrete form of these connectors, we need not only the SPT phases $[\psi_\sigma]$ decorated on $\sigma\in Y_p$, but also the wave functions representing these phases.
Here, we use $\psi$ to denote a wave function representing a TCS phase $[\psi]$.
Just like $[\psi]$ is a collection of local SPT phases, $\psi$ is a collection of local SPT wave functions on all cells:
The local SPT phase on $\sigma\in Y_p$ is denoted by $\psi_\sigma\in\Psi^p(G_\sigma)$, where $\Psi^d(G)$ denotes the collection of $d$-dimensional $G$-symmetric wave functions, as reviewed in Appendix~\ref{app:spt-review}.
To form a symmetric wave function, we require that the local wave functions decorated to symmetry-related cells satisfy the following symmetry condition, similar to the one in Eq.~\eqref{eq:a'=ga},
\begin{equation}
\label{eq:a'=ga2}
\psi_{g\sigma} = g\cdot\psi_\sigma.
\end{equation}

As $\psi$ is made of $p$-dimensional building blocks, the decorations $\psi_\sigma=0$ on cells with dimensionality higher than $p$.
However, the connectors, which are decorations on cells in dimensions lower than $p$, are in general not vanishing.
Collectively, we denote the decorations on $d$-cells by $\psi_d$: $\psi_p$ is the building blocks on $p$-cells, $\psi_{p-1}$ is the connectors on $(p-1)$-cells, etc.

Now consider a $(p-1)$-cell $\tau\in Y_{p-1}$.
The connector decorated to $\tau$, which we denote by $\psi|_\tau$, satisfies the following bulk-boundary relation,
\begin{equation}
  \label{eq:psi-tau-sigma}
  d(\psi|_\tau) = \boxplus_{\sigma\in Y_p}\langle\tau|\partial\sigma\rangle\psi|_\sigma,
\end{equation}
where $\psi_1\boxplus\psi_2$ denotes the wave function obtained by stacking the two wave functions $\psi_1$ and $\psi_2$.
As reviewed in Appendix~\ref{app:spt-review}, for bosonic SPT states whose wave functions are represented by cochains, this stacking is just the normal addition between cochains.
However, for fermionic SPT states, this stacking operation is not commutative, $\psi_1\boxplus\psi_2\neq \psi_2\boxplus\psi_1$, because the statistical signs associated with reordering of fermionic operators.
Because of this subtlety, to unambiguously define the stacking in Eq.~\eqref{eq:psi-tau-sigma}, one must choose an ordering between the neighboring cells of $\tau$, and such ordering should be compatible with the crystal symmetries.

For simplicity, we introduce an operator $\partial$ to denote the operation on the right-hand-side of Eq.~\eqref{eq:psi-tau-sigma}:
$\partial$ transforms $\psi$ to an anomaly pattern $\partial\psi$, whose components on each cell are given by
\begin{equation}
  \label{eq:partial}
  (\partial\psi)|_\tau = \boxplus_{\sigma\in Y_p}\langle\tau|\partial\sigma\rangle\psi|_\sigma.
\end{equation}
Intuitively, the operator $\partial$ transfers the wave functions on $p$-cells to their boundary $(p-1)$-cells, where they are interpreted as boundary anomalies.
Using this operator, the relation in Eq.~\eqref{eq:psi-tau-sigma} is simplied as
\begin{equation}
  \label{eq:psi-tau-sigma2}
  d(\psi|_\tau) = (\partial\psi)|_\tau.
\end{equation}
Since such relation exists on every $(p-1)$-cell, it gives a relation between the $p$-blocks $\psi_p$ and the $(p-1)$-connectors $\psi_{p-1}$:
\begin{equation}
  \label{eq:psi-1-0}
  d\psi_{p-1} = \partial\psi_p.
\end{equation}
For bosonic SPT states, the detailed formula for computing the $\partial$ operator can be found in Appendix~\ref{app:transfer}.

Comparing the right-hand-side of Eq.~\eqref{eq:psi-tau-sigma} to Eq.~\eqref{eq:defd1}, it is easy to check that the SPT phase of $(\partial\psi)|_\tau$ is precisely $(\partial\psi)|_\tau\sim d_1[\psi]|_\tau$.
Hence, for a second-page TCS $[\psi]$ in $E^p_{p,2}$, the no-open-edge condition $d_1[\psi]=0$ ensures that Eq.~\eqref{eq:psi-tau-sigma2} has solutions for $\psi|_\tau$, representing possible choices of a connector bridging $p$-cells bordering $\tau$.
In the rest of this section, we shall use solutions of this equation to address the problems raised in Sec.~\ref{sec:further} and obtain a complete classification of TCSs.

\subsubsection{Bosonic example}
\label{sec:conn-b}

We use a simple bosonic example to demonstrate the process of determining the wave functions of connectors.
As shown in Appendix~\ref{app:trivial}, such examples can only be nontrivial when the symmetry group $G$ is not a direct product of $SG$ and $G_0$.
In fact, this example involves a magnetic translation symmetry group.
This example is adapted from the result of Ref.~\cite{DyonLSM}.
The connection to Ref.~\cite{DyonLSM} will be revealed in Sec.~\ref{sec:lsm}.

In this example, we consider 2D TCSs protected by the symmetry group $G=G^M\times\mathbb Z_2^T$, where $\mathbb Z_2^T$ is the usual (antiunitary) onsite time-reversal symmetry, and $G^M$ is a 2D magnetic translation symmetry group.
$G^M$ has three generators $t_x$, $t_y$, $x$, representing two translation symmetries and one onsite unitary $\mathbb Z_2$ symmetry, respectively.
Both $t_x$ and $t_y$ commutes with $x$.
However, $t_x$ and $t_y$ does not commute, and instead satisfies
\begin{equation}
  \label{eq:txty}
  t_xt_yt_x^{-1}t_y^{-1}=x.
\end{equation}
In this case, the onsite symmetry group $G_0=\mathbb Z_2^x\times\mathbb Z_2^T$, where $\mathbb Z_2^x$ denotes the $\mathbb Z_2$ group generated by $x$.
The space group is the quotient group $SG=G/G_0=\mathbb Z^2$, generated by the two translation operations.
\textcolor{red}{SZD: $SG$ or $SG$? }
However, $G$ is not a direct product of $G_0$ and $SG$, due to the nontrivial commutation relation in Eq.~\eqref{eq:txty}.

\begin{figure}
  \includegraphics{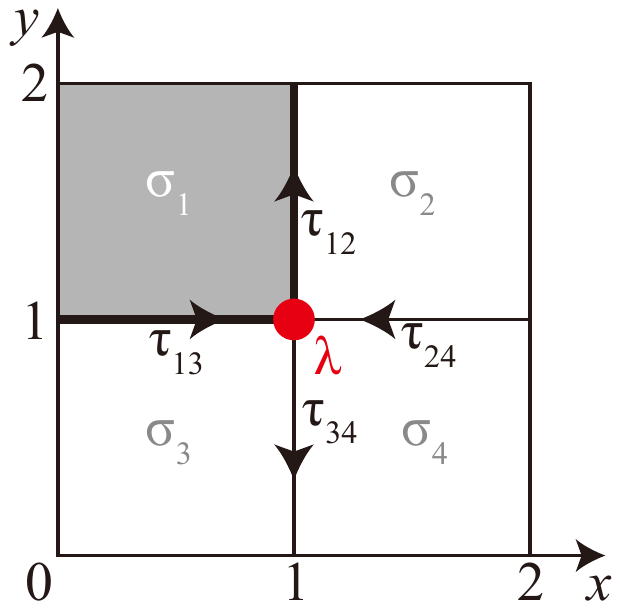}
  \caption{The cell decomposition for the magnetic translation symmetry group.}
  \label{fig:mag}
\end{figure}

We first decompose the 2D plane $\mathbb R^2$ into the $G$-complex $Y$ as outlined in Sec.~\ref{sec:blocks}.
Since $SG=\mathbb Z^2$ is the simplest wallpaper group, the result of the decomposition is simply a generic oblique lattice, as shown in Fig.~\ref{fig:mag}.
There is no point-group symmetries anywhere, and the local symmetry group of each cell is just $G_\sigma=G_0$.
Therefore, the SPT building blocks are obtained by attaching SPT states in $\Phi^p(G_0)=H^{p+1}[G_0,\uone_T]$ to $p$-cells.

In this example, we consider a particular 2-block assembly $[\psi]$ in $E^2_{2,1}$:
On each 2-cell, we decorate an SPT state represented by the following cocycle $[\alpha]\in\Phi^2(G_0)=H^3[G_0,\uone_T]$, represented by the following 3-cocycle:
\begin{equation}
\label{eq:alpha-z2z2t}
  \alpha(g_0,g_1,g_2,g_3)=[n_X(g_0)-n_X(g_1)]
  \beta(g_1,g_2,g_3)\pi,
\end{equation}
where the function $\beta$ is defined as
\begin{equation}
  \label{eq:beta-t2}
  \beta(g_0,g_1,g_2)=[n_T(g_1)-n_T(g_0)][n_T(g_2)-n_T(g_1)].
\end{equation}
Here, the $\mathbb Z_2$ variables $n_x$ and $n_T$ are obtained by writing the elements of $G_0$ as the following canonical form,
\begin{equation}
  \label{eq:nG0}
  g = x^{n_x(g)}T^{n_T(g)}.
\end{equation}
This cocycle represents a nontrivial 2D SPT state protected by both $x$ and $T$ symmetries.
The 2-blocks of $[\psi]$ is given by $[\psi|_\sigma]=[\alpha]$.

It is straightforward to check that this element satisfies the cocycle equation on the first page, $d_1[\psi]=0$, and remains a valid second-page SPT state in $E^2_{2,2}$.
To see this, we notice that $[\psi]$ decorates the same SPT state on every 2-cell.
Therefore, on each 1-cell, which borders two 2-cells, there are two counter-propagating anomalous edge modes, and they cancel each other. Hence, this decoration $[\psi]$ can be gapped out on 1-cells.

However, gapping out these 1-cells require nontrivial connectors.
In order to compute the connectors, we choose a wave function of $2$-blocks $\psi2$ representing $[\psi]$.
As reviewed in Appendix~\ref{app:spt-review}, on each 2-cell, the wave function is a $G$-valued-$G_0$-invariant 3-cococyle $\tilde\alpha$.
Without losing generality, we choose the following 3-cocycle $\tilde\alpha$:
\begin{equation}
\label{eq:ta-1}
\tilde\alpha(g_0,g_1,g_2,g_3)
=[n_x(g_0)-n_x(g_1)]\beta(g_1,g_2,g_3).
\end{equation}
This equation looks similar to Eq.~\eqref{eq:alpha-z2z2t}, but the group elements $g_i$ takes values in $G$ instead of $G_0$, and $\tilde n_X$ is extracted by writing the group elements in the following canonical form,
\begin{equation}
  \label{eq:nG}
  g = t_x^{n_{tx}(g)}t_y^{n_{ty}(g)}
  x^{n_x(g)}T^{n_T(g)}.
\end{equation}
Here, we emphasize that the $\tilde\alpha$ given in Eq.~\eqref{eq:ta-1} is not invariant under $G$-actions~\footnote{It is impossible to find a $G$-invariant $\tilde\alpha$ representing this SPT phase, due to the nontrivial structure of $G$.}.
In fact, using the commutation relation in Eq.~\eqref{eq:txty}, one can show that
  \begin{equation}
    \label{eq:ta-ty}
    \begin{split}
      &\tilde\alpha(t_yg_0, t_yg_1, t_yg_2, t_yg_3)=\\
      &[n_x(g_0)+n_{tx}(g_0)-n_x(g_1)-n_{tx}(g_1)]\beta(g_1,g_2,g_3),
    \end{split}
  \end{equation}
which is different from Eq.~\eqref{eq:ta-1}.

If we choose to decorate $\sigma_1$ with $\tilde\alpha$ and let $\psi|_{\sigma_1}=\tilde\alpha$,
the symmetry constraint will fix the decoration on other 2-cells.
In particular, the decoration on $\sigma_3$, which is related to $\sigma_1$ by the action of $t_y^{-1}$, is given by $\psi|_{\sigma_3}=t_y^{-1}\cdot\tilde\alpha$.
Using the explicit form of symmetry actions on cochains given in Appendix~\ref{app:gaction}, we get
\begin{equation}
  \label{eq:ta-3}
  \psi|_{\sigma_3}(g_0,g_1,g_2,g_3)
  =\tilde\alpha(t_yg_0,t_yg_1,t_yg_2,t_yg_3).
\end{equation}
Using the result of Eq.~\eqref{eq:ta-ty}, we see that the decorations on the two cells are actually different: $\psi|_{\sigma_1}\neq\psi|_{\sigma_3}$.
In fact, the two decorations still belong to the same cohomology class in $\Phi^2(G_0)=H^3[G_0,\uone_T]$.
However, to be compatible with the magnetic translation symmetry, different cochains (of the same cohomology class) have to be decorated to different 2-cells.
The difference between $\psi|_{\sigma_3}$ and $\psi|_{\sigma_1}$ then implies that one must decorate a nonvanishing 2-cochain to their boundary, $\tau_{13}$.
Using the explicit form of the cochains, one can derive the explicit form of the cocycle equation $d\psi|_{\tau_{13}}=\psi|_{\sigma_3}-\psi|_{\sigma_3}$:
\begin{equation}
  \label{eq:da-13}
    d\psi|_{\tau_{13}}(g_0,g_1,g_2,g_3)=
    [n_{tx}(g_0)-n_{tx}(g_1)]\beta(g_1,g_2,g_3).
\end{equation}
We now need to choose an arbitrary solution of this equation.
It is easy to check that the following 2-cocycle is a choice,
\begin{equation}
  \label{eq:ta-13}
  \psi|_{\tau_{13}}(g_0,g_1,g_2)=
n_{tx}(g_0)\beta(g_0,g_1,g_2).
\end{equation}

On the other hand, on the 1-cells along the $x$-direction, we are allowed to simply choose $\psi|_{\tau_{12}}=0$, because the cocycle $\tilde\alpha$ is invariant under the action of $T_x$ and therefore $\psi|_{\sigma_2}=\psi|_{\sigma_1}$.
Hence, we have to use nontrivial connectors on some 1-cells, because it is impossible to construct a wave function of $[\alpha]$ that is symmetric under all operations in $SG$.

\subsubsection{Free-fermion examples}
\label{sec:conn-ff}

\begin{figure}[t]
  \includegraphics[width=1.0\linewidth]{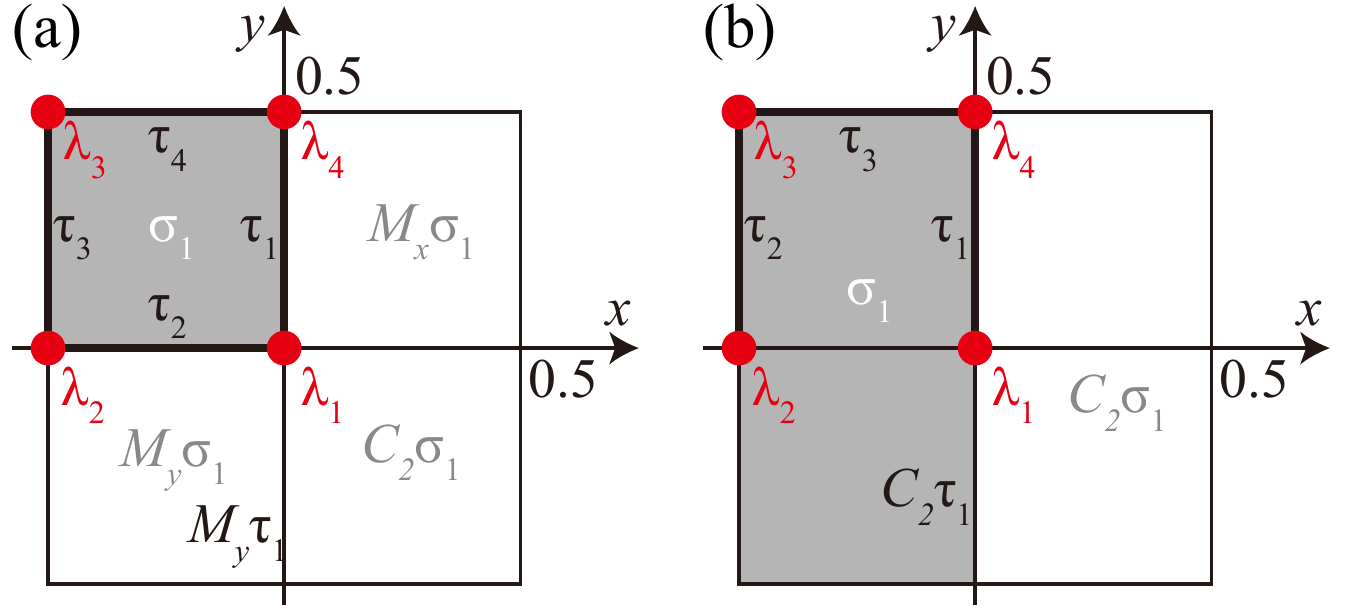}
  \caption{Cell decomposition for the $p2mm$ and $p2$ groups. }
  \label{fig:p2mm}
\end{figure}

Next, we present an example of determining nontrivial connectors in a 2D topological crystaline superconductor.
We consider that the 2D system has a wallpaper group $p2mm$, the generators of which are $\hat{M}_x$ and $\hat{M}_y$, a time-reversal symmetry, $\hat{T}$, and a particle-hole symmetry, $\hat{P}$.
The algebra relations of these generators are given by $\hat{T}^2=-1$, $\hat{P}^2=1$, $\hat{M}_x^2=-1$, $M_y^2=1$, $[\hat{T},\hat{P}]=0$, $[\hat{T},\hat{M}_x]=0$, $\{\hat{T},\hat{M}_y\}=0$, $[\hat{P},\hat{M}_x]=0$, and $[\hat{P},\hat{M}_y]=0$.
Such relations can be realized in a superconductor with significant spin-orbit coupling and an order parameter projectively representing the $\hat{M}_y$ symmetry.
To proceed, we represent these operators as $\hat{T}=is_2 K$, $\hat{P}=\mu_1 K$, $\hat{M}_x = i\mu_3 s_3$, $\hat{M}_y = \mu_3 s_2$, where  $s_{1,2,3}$ are pauli matrices representing the spin degree and $\mu_{1,2,3}$ are pauli matrices representing the ``orbital'' degree. (The meaning of ``orbital'' is twisted in BdG Hamiltonian.)
The complex structure $Y$ of $p2mm$ is illustrated in Fig.~\ref{fig:p2mm}a, where the only 2-cell in the $G$-orbits $Y_2/G$ is $\sigma_1$, the four 1-cells in $Y_1/G$ are $\tau_{i=1,2,3,4}$, and the four 0-cells in $Y_0/G$ are $\lambda_{i=1,2,3,4}$.

Now we decorate the 2-cell $\sigma_1$ with the following BdG Hamiltonian
\begin{equation}
\hat{H}^{(L)}(\mathbf{k}) = k_x \mu_1 s_3 + k_y \mu_2 s_0 + \left(M- k_x^2- k_y^2\right)\mu_3 s_0, \label{eq:TCS-H1}
\end{equation}
where $M>0$.
Due to the $\hat{M}_y$ symmetry, the Hamiltonian decorated on the 2-cell $\sigma_3$ should be $\hat{M}_y \hat{H}^{(L)}(k_x,-k_y) \hat{M}_y^{-1}$, which is identical with Eq. (\ref{eq:TCS-H1}).
Therefore, we extend the Hamiltonian on $\sigma_1$ (Eq. (\ref{eq:TCS-H1})) to $\sigma_3$.
Apparently, there is no boundary state on $\tau_2$.
Due the symmetry $\hat{M}_x$, the Hamiltonian on $M_x\sigma_1$ and $C_2\sigma_1$ can be derived as
\begin{align}
&\hat{H}^{(R)}(\mathbf{k}) = \hat{M}_x \hat{H}^{(L)}(-k_x,k_y) \hat{M}_x^{-1} \nonumber\\
=& k_x \mu_1 s_3 - k_y \mu_2 s_0 + \left(M-k_x^2-k_y^2\right)\mu_3 s_0, \label{eq:TSC-H2}
\end{align}
which is different from from Eq. (\ref{eq:TCS-H1}).
In order to determine the boundary state between $H^{(L)}$ and $H^{(R)}$, we consider to insert an infinite barrier potential on the edges $\tau_1$ and $M_y \tau_1$ and the vertex $\lambda_1$.
We assume the boundary Hamiltonian on $\tau_1$ is
$\hat{H}^{(\tau_1)} = \mu_0 s_3 k_y$, where the upper block ($\mu_3=1$) is the boundary state of $H^{(L)}$ and the lower block ($\mu_3=-1$) is the boundary state of $H^{(R)}$.
The time-reversal symmetry acts locally on the two boundaries and hence can be represented by $\hat{T}^{(b)}=is_2 K$.
The particle-hole symmetry also acts locally on the two boundaries, and hence can be represented by $\hat{P}^{(b)}=K$.
The mirror symmetry $M_x$ interchanges the two blocks and hence must be proportional to $i\mu_1$ or $i\mu_2$.
$M_x$ need to commute with $H^{(\tau_1)}, \hat{T}^{(b)}, \hat{P}^{(b)}$ and square to $-1$.
We find that the only choice is $\hat{M}_x = i\mu_2$.
We then soften the barrier to introduce coupling between the two blocks.
The only symmetry-allowed mass term that gaps $\hat{H}^{(\tau_1)}$ is $\mu_2 s_1$.
Hence we model the gapped state on $\tau_1$ as
\begin{equation}
\hat{H}^{(\tau_1)}(k_y) = \mu_0 s_3 k_y + m \mu_2 s_1.\label{eq:conn-mass}
\end{equation}
Now we study how $\hat{H}^{(\tau_1)}$ transforms under the $M_y$ operation.
Using the constraints $\{\hat{T}^{(b)},\hat{M}_y^{(b)}\}=0$, $[\hat{P}^{(b)},\hat{M}_y^{(b)}]=0$, $\{\hat{M}_x^{(b)},\hat{M}_y^{(b)}\}=0$, $\hat{M}_y^{(b)2}=1$, $\hat{M}_y^{(b)}$ can be chosen as $\mu_{1,3}s_{1,3}$.
Since $M_y$ does not interchange the two blocks, we only consider the two options $\hat{M}_y^{(b)\prime}=\mu_3 s_1$, $\hat{M}_y^{(b)\prime\prime}=\mu_3 s_3$.
Correspondingly, the boundary Hamiltonian on $M_y\tau_1$ under the two $M_y$ representations are given by
\begin{align}
& \hat{H}^{(M_y\tau_1)\prime}(k_y) = \hat{M}_y^{(b)\prime}\hat{H}^{(\tau_1)}(-k_y)\hat{M}_y^{(b)\prime-1} \nonumber\\
=& \mu_0 s_3 k_y - m\mu_2 s_1,
\end{align}
\begin{align}
& \hat{H}^{(M_y\tau_1)\prime\prime}(k_y) = \hat{M}_y^{(b)\prime\prime}\hat{H}^{(\tau_1)}(-k_y)\hat{M}_y^{(b)\prime\prime-1} \nonumber\\
=& -\mu_0 s_3 k_y + m \mu_2 s_1,
\end{align}
respectively.
Therefore, either the kinetic term or the mass term will be flipped under $M_y$, leading to a gapless domain wall at $\lambda_1$.
This serves as a nontrivial connector on the 1-cells in the vertical direction in Fig. \ref{fig:p2mm}a.

In the end, we consider an example of $p+ip$ topological superconductor with nontrivial connector on the 1-cells.
We consider a $p+ip$ superconductor in the wallpaper group $p2$.
The cell decomposition of $p2$ is shown in Fig. \ref{fig:p2mm}b, where the only 2-cell in the $G$-orbits $Y_2/G$ is $\sigma_1$, the three 1-cells in $Y_1/G$ are $\tau_{i=1,2,3}$, and the four 0-cells in $Y_0/G$ are $\lambda_{i=1,2,3,4}$.
We assume the Hamiltonian on $\sigma_1$ as
\begin{equation}
\hat{H}^{(\sigma_1)}(\mathbf{k}) = k_x s_1 + k_y s_2 + (M- k_x^2 -  k_y^2)s_3,
\end{equation}
where $M>0$ and the particle-hole symmetry is represented as $\hat{P} = s_x K$.
The Hamiltonian on $C_2 \sigma_1$ can be generated by acting the $C_2$ operation on the above Hamiltonian.
We consider the $C_2$ operator $\hat{C}_2=s_0$, it commutes with $\hat{P}$ and squares to 1.
Thus the Hamiltonian on $C_2\sigma_1$ is
\begin{align}
& \hat{H}^{(C_2\sigma_1)}(\mathbf{k}) = \hat{C}_2 \hat{H}^{(\sigma_1)}(-\mathbf{k})\hat{C}_2^{-1}\nonumber\\
=& -k_x s_1 - k_y s_2 + (M- k_x^2 -  k_y^2)s_3.
\end{align}
In order to study the boundary state on $\tau_1$ and $C_2\tau_1$, we consider to instert an infinite barrier potential on $\tau_1$, $C_2\tau_1$, and $\lambda_1$.
Since the $H^{(\sigma_1)}(\mathbf{k}) $ and $H^{(C_2\sigma_1)}(\mathbf{k}) $ have the same chirality (rotation does not change chirality), the majorana chiral modes from the two sides must move in opposite directions.
We assume the boundary Hamiltonian on $\tau_1$ as $\hat{H}^{(\tau_1)} = k_y s_3$, where the $s_3=1$ state comes from the boundary of $\sigma_1$ and the $s_3=-1$ state comes from boundary of $C_2\sigma_1$.
Since the particle-hole symmetry acts locally on the two boundaries, we choose its representation as $\hat{P}^{(b)}=K$.
Now we soften the barrier to introduce coupling between the two majorana modes.
The only symmetry-allowed mass term is $s_2$.
Hence we model the gapped state on $\tau_1$ as
\begin{equation}
\hat{H}^{(\tau_1)}(k_y) = s_3 k_y + m s_2.
\end{equation}
Now we study how $\hat{H}^{(\tau_1)}$ transforms under the $C_2$ rotation.
Since $C_2$ interchanges the two boundaries and commutes with $\hat{P}^{(b)}$, its representation must be off-diagonal and real, i.e., $\hat{C}_2^{(b)} = s_1$.
The boundary Hamiltonian on $C_2\tau_1$ is hence obtained as
\begin{align}
& \hat{H}^{(C_2\tau_1)}(k_y) = \hat{C}_2^{(b)} \hat{H}^{(\tau_1)}(-k_y)\hat{C}_2^{(b)-1}\nonumber\\
=& s_3 k_y - m s_2.
\end{align}
The mass is flipped under the $C_2$ rotation, leading to a gapless domain wall at $\lambda_1$.
This serves as a nontrivial connector on the 1-cells.

\subsection{Second-page no-open-edge conditions}
\label{sec:noe2}

For a second-page TCS $[\psi]\in E^p_{p,2}$, the second-page no-open-edge condition demands that all $(p-2)$-cells can be filled with gapped symmetric connectors.
The connectors decorated on $(p-2)$-cells, collectively denoted by $\psi_{p-2}$, must satisfy the bulk-boundary relation similar to \eqref{eq:psi-1-0}:
\begin{equation}
\label{eq:psi-2-1}
d\psi_{p-2}=\partial\psi_{p-1}.
\end{equation}
Hence, the existence of such connectors is determined by the condition that the r.h.s. of Eq.~\eqref{eq:psi-2-1} belongs to the trivial SPT phase on each cell, $\partial\psi_{p-1}\sim0$.

We introduce a linear map $d^p_{p,2}:E^p_{p,2}\rightarrow E^{p-1}_{p-2,2}$ to represent this no-open-edge condition.
As in Sec.~\ref{sec:simplified}, $d^p_{p,2}$ will be abbrivated to $d_2$ if the domain of the map is clear from the context.
For each element $[\psi]$ in $E^q_{p,2}$, we choose a particular wave function $\psi_p$ for the building blocks.
Then, we choose an arbitrary solution $\psi_{p-1}$ of Eq.~\eqref{eq:psi-1-0}.
The image of $d_2$ map is then defined as
\begin{equation}
  \label{eq:d2}
  d_2[\psi] = [\partial\psi_{p-1}],\quad
  d\psi_{p-1}=\partial\psi_p.
\end{equation}

Several remarks are in order:
First, the domain of the $d_2$ maps are the $E_2$ modules, because $\psi_{p-1}$ only exists if $[\psi]$ belongs to the $E_2$ module, where the cocycle condition $d_1[\psi]=0$ guarantees the existence of $\psi_{p-1}$.
Second, we explain why the images of $d_2$ maps are the $E_2$ modules. The meaning of this assertion is twofold:
On one hand, $d_2[\psi]=[\partial\psi_{p-1}]$ satisfies the cocycle condition $d_1d_2[\psi]$ because $\partial^2=0$.
Therefore, it indeed belongs to $E_2$.
On the other hand, the equivalence relation of $d\psi_{p-2}=\partial\psi_{p-1}\sim0$ should be understood as the one in $E^{p-1}_{p-2,2}$:
Instead of requiring the obstruction $[\left(\partial\psi_{p-1}\right)|_\sigma]$ to vanish on every $(p-2)$-cell $\sigma$, we only require that $\partial\psi_{p-1}$ can be trivialized by a bubbling process $\tilde\psi\in E^{q+1}_{p-1,1}$: $\partial\psi_{p-1}\sim d_1[\tilde\psi]=\partial\tilde\psi_p$.
This is because when extending $\psi$ to $(p-1)$-cells, we can choose $\psi_{p-1}^\prime=\psi_{p-1}-\tilde\psi_p$ instead of $\psi_{p-1}$ as the connectors.
This choice of connectors then satisfies $\left(\partial\psi_{p-1}^\prime\right)|_\mu\sim0$ on every $(p-2)$-cell.

In summary, the second-page no-open-edge condition, which tests whether $[\psi]$ can be filled with gapped symmetric connectors on $(p-2)$-cells, is expressed as
\begin{equation}
  \label{eq:d2=0}
  d_2[\psi]=0.
\end{equation}

\subsubsection{Bosonic example}
\label{sec:noe2-b}
We now use this no-open-edge condition to examine the example introduced in Sec.~\ref{sec:conn-b}.
We will see that the $d_2$ map is nontrivial in this example.
We consider the result of $d_2[\psi]$ on the 0-cell $\mu$ shown in Fig.~\ref{fig:mag}.
Recall that using the wave-function realization $\psi$ we chose in Eq.~\eqref{eq:ta-1}, the connectors $\psi_1$ are given as follows:
$\psi_1|_{\tau_{12}}=\psi_1|_{\tau_{34}}=0$; $\psi_1|_{\tau_{13}}$ is given by Eq.~\eqref{eq:ta-13}.
Hence, $\psi_1|_{\tau_{24}}$ is constrainted to be $t_x^{-1}\cdot\psi_1|_{\tau_{13}}$ and has the following form,
\begin{equation}
  \label{eq:ta-24}
  \psi_1|_{\tau_{24}}(g_0,g_1,g_2)=
[n_{tx}(g_0)+1]\beta(g_0,g_1,g_2).
\end{equation}

Using these results of $\psi_1$, we can compute $d_2[\psi]$ using the definition in Eq.~\eqref{eq:d2},
\begin{equation}
  \label{eq:d2psi=beta}
  d_2[\psi]|_\tau
  \sim\psi_1|_{\tau_{13}}+\psi_1|_{\tau_{24}}+\psi_1|_{\tau_{12}}
  +\psi_1|_{\tau_{34}}\sim[\beta].
\end{equation}

Here, $\beta$ is exactly the 2-cocycle representing the projective representation of $T^2=-1$, or a Kramers doublet.
Hence, in this case, $d_2: E^3_{2,2}\rightarrow E^2_{0,2}$ is a nontrivial map that sends $\hat\alpha$ to the nontrivial element $[\tilde\psi]$ in $E^2_{0,2}$:
\begin{equation}
  \label{eq:d2psi=tpsi}
  d_2[\psi] = [\tilde\psi].
\end{equation}
Here, $[\tilde\psi]$ is an anomalous pattern represented by the following decomposition on 0-cells: $[\tilde\psi]_\mu=[\beta]$ on all 0-cells in $Y$.
This implies that $[\psi]$ is actually not a 2D TCS because it does not satisfy the second-page no-open-edge condition:
it has open edges, actually Kramers doublets, on the 0-cells.

\subsubsection{Free-fermion example}
\label{sec:noe2-ff}

Next, we visit the free-fermion example in Sec.~\ref{sec:conn-ff}.
In this example, we use the connector in Eq.~\eqref{eq:conn-mass} to gap out the 1-cells in the $y$-direction in Fig.~\ref{fig:p2mm}.
Furthermore, the symmetry condition in Eq.~\eqref{eq:a'=ga2} implies that mass terms on different 1-cells are related by symmetries.
In particular, using the projected symmetry operation $\hat M_y^{(b)}=s_2$, we see that the mass term must change sign under $\hat M_y^{(b)}$.
Therefore the mass terms on the two cells $\tau_1$ and $\tau_3$, related by $M_y$, must have opposite signs.
Such two opposite mass terms would left a symmetry protected gapless point at $y=0$, {\it i.e.}, the 0-cell $\lambda_1$.
Replacing the mirror symmetries above with mirror symmetries on $x=-1/2$ and $y=1/2$, and following the same analysis, one can easily show that all the 0-cells in $Y_0/G$, $\lambda_{1,2,3,4}$, are gapless.

\subsection{Second-page bubbling equivalences}
\label{sec:bub2}

The second-page bubbling equivalence can be computed using a similar $d_2$ map.%
Each element $[\psi]\in E^{p+1}_{p+2,2}$ represents such a bubbling process, where SPT bubbles are generated on $(p+2)$-cells.
Unlike the bubbling processes studied in Sec.~\ref{sec:cobdry}, it leaves the $(p+1)$-blocks intact, and changes the SPT phases on the $p$-cells.

In order to compute the changes to the $p$-cells, we also need to consider additional bubbles generated on lower-dimensional cells.
On a $p'$-cell $\tau\in Y_{p'}$ ($p'<p+2$), we can also generate a bubble $\psi|_\tau$, which we will refer to as a $p'$-bubble.
Different from the ($p+2$)-bubbles, the lower-dimensional bubble can have a nontrivial filling: $d(\psi|_\tau)\neq0$, meaning that the process not only changes the wave functions on $\partial\tau$ by $\psi|_\tau$, but also changes the wave function on $\tau$ by $d(\psi|_\tau)$.
Such a process is allowed because the bubble and its filling satisfy the bulk-boundary relation reviewed in Appendix~\ref{app:spt-review}, and together form a gapped symmetric state on $\tau$.
In this way, a general bubbling process contains not only $p$-bubbles, denoted by $\psi_p$, but also $p'$-bubbles $\psi_{p'}$ for all $p'<p$.
On $p'$-cells, the total changes made by the bubbling include the $(p'+1)$-bubbles and the filling of the $p'$-bubbles:
\begin{equation}
  \label{eq:D_psi}
  \Delta\psi_{p'} = \partial\psi_{p'+1}\boxplus d\psi_{p'}.
\end{equation}

We now compute the changes to $p$-cells for a bubbling process $[\psi]\in E^{p+1}_{p+2,2}$.
First, we choose a wave-function realization $\psi_{p+2}$ of the $(p+2)$-bubbles.
Since the bubbling process leaves $(p+1)$-blocks intact, the $(p+1)$-bubble $\psi_{p+1}$ must satisfy
\begin{equation}
  \label{eq:psi-1-0b}
  \partial\psi_{p+2}\boxplus d\psi_{p+1} = 0.
\end{equation}
The existence of solutions of this equation is provided by the fact that $[\psi]\in E^{p+1}_{p+2,2}$ satisfies $d_1[\psi] = 0$.
We then choose an arbitrary solution of $\psi_{p+1}$, and the SPT phases the process generates is $d\psi_p\boxplus\partial\psi_{p+1}\sim[\partial\psi_{p+1}]$.
Therefore, this process can be represented by the following $d_2$ map,
\begin{equation}
  \label{eq:d2b}
  d_2[\psi] = [\partial\psi_{p+1}],\quad d\psi_{p+1} = -\partial\psi_{p+2}.
\end{equation}
Note, that the $d_2$ map defined here is different from the one defined in Sec.~\eqref{eq:d2} by a minus sign in the constraint equation.
Hence, we can use a generic definition to unify the two $d_2$ maps:
\begin{equation}
  \label{eq:d2c}
  \begin{split}
  d^q_{p,2}:&E^q_{p,2}\rightarrow E^{q-1}_{p-2,2}:\\
  &[\psi]\mapsto[\partial\psi_{p-1}],\quad
  d\psi_{p-1}=(-1)^{q-p}\partial\psi_p.
  \end{split}
\end{equation}
We notice that a nontrivial example of $d_2$ bubbling process is provided in Sec.~\ref{sec:lsm}.

Taking into account the no-open-edge conditions and bubbling equivalences given by the $d_2$ map, the third-page approximation of the TCS classification is given by the cohomology group of $d_2$:
\begin{equation}
\label{eq:E3}
E^q_{p,3} = \frac{\ker d^q_{p,2}}{\img d^{q+1}_{p+2,2}}.
\end{equation}

\subsection{Higher-page results}

This process can be generalized to higher pages.

To consider the $r$-th page no-open-edge conditions, we start with an $r$-th page assembly $\psi\in E^p_{p,r}$.
On previous pages, we have constructed the connectors $\psi_{p-1}$, ..., $\psi_{p-r+1}$.
The no-open-edge condition on the previous page, $d_{r-1}[\psi]=0$, guarantees that the equation $d\psi_{p-r}=\partial\psi_{p-r+1}$ has solutions.
We then pick a solution of $\psi_{p-r}$, and define $d_r[\psi]=[\partial\psi_{p-r}]\in E^{p-r+1}_{p-r,r}$.
The no-open-edge condition is given by $d_r[\psi]=0$.

Similarly, consider an $r$-th page bubbling process $[\psi]\in E^{p+r-1}_{p+r,r}$.
On previous pages, we have chosen lower-dimensional bubbles $\psi_{p+r-2}$, ..., $\psi_{p+2}$, such that the process generates nothing on cells with dimensions higher than $p+2$.
In order to get a process that also generates nothing on $(p+1)$-cells, we choose $\psi_{p+1}$ satisfying $d\psi_{p+1}\boxplus\partial\psi_{p+2}=0$.
The existence of solutions is provided by $d_{r-1}[\psi]=0$.
The resulting trivial TCS is then given by $d_r[\psi]$ defined as $[\partial \psi_{p+1}]$.

Similar to Eq.~\eqref{eq:d2c}, we write a unified definition for $d_r$:
\begin{equation}
  \label{eq:drc}
  \begin{split}
  &d^q_{p,r}:E^q_{p,r}\rightarrow E^{q-r+1}_{p-r,r}:\\
  &[\psi]\mapsto[\partial\psi_{p-r+1}],\quad
  d\psi_{p-r+1}=(-1)^{q-p}\partial\psi_{p-r+2}.
  \end{split}
\end{equation}

Therefore, the classification on the next page is given by the cohomology group of $d_r$:
\begin{equation}
  \label{eq:Er+1}
  E^q_{p,r+1}=\frac{\ker d^q_{p,r}}{\img d^{q+r-1}_{p+r,r}}.
\end{equation}

Iteratively, this process computes a series of pages $E^q_{p,1},E^q_{p,2},\ldots$, where $E^p_{p,r}$ provides a series of finer and finer approximations to the classification of $p$-block TCSs.
Since we are eliminating false and redundant entries on each page,  the list of candidate assemblies are getting smaller and smaller, $E^q_{p,1}\supseteq E^q_{p,2}\supseteq\cdots$.
In the limit of $r\rightarrow\infty$, this series of approximations reveals the true answer of the classification problem, which we denote by $E^q_{p,\infty}$.
In particular, $E^p_{p,\infty}$ classify all $p$-block TCSs.
In fact, this process only takes a finite number of steps to converge to $E^q_{p,\infty}$, because the $d_r$ map reduces the spatial dimension of the cells by $r$ and necessarily becomes trivial once $r$ exceeds the dimension of $Y$.

\subsection{Recombing states with different $d_b$ through group extension}
\label{sec:gext}

In our topological-crystal constructions, we first divide $d$-dimensional SPTs into TCSs with different building-block dimensions $d_b=0,1,\ldots d$.
We then compute the classification for each $p$-blocks separately.
The classification for $p$-block TCSs is then given by $E^p_{p,\infty}$, which is calculated by a series of cohomology-group computations.
Next, to obtain the full classification of crystaline SPT states, we need to recombine results of $d_b=1,2,\ldots d$, in a way similar to Eq.~\eqref{eq:sum1}. We also need to include results of $d_b=0$ if we want to recover all bosonic SPTs in $H^{d+1}[G,\uone_T]$.
However, as briefly mentioned in Sec.~\ref{sec:further},
such recombination may not be a simple direct sum but a nontrivial group extension.
In this section, we explain how this group extension is computed in general.
Appendix~\ref{app:trivial} will use this method to prove that, for the simple cases $G=SG\times G_0$, the group extension is always trivial and one can just take a direct sum.

We begin by recalling that a TCS $E^p_{p,\infty}$ are labeled by different building blocks on $p$-cells, but each state $[\psi]$ also contains connectors on all lower-dimensional cells.
The lower-dimensional decorations will affect the results of adding (stacking) two SPTs states if the decoration cancels on higher-dimensional cells.

To be more specific, consider an order-$n$ TCS $[\psi]\in E^p_{p,\infty}$, such that $n[\psi]\sim0$ as in $E^p_{p,\infty}$.
This implies that stacking $n$ coplies of $[\psi]$ results in a state $[\tilde\psi]=n[\psi]$ which is trivial if viewed as an element in $E^p_{p,\infty}$.
In other words, $[\tilde\psi]$ has trivial decorations on all $p$-cells.
However, $[\tilde\psi]$ may have nontrivial decorations on lower-dimensional cells, and thus should be viewed as a nontrivial topological crystal with a lower building-block dimension.
To compute this, recall that in the topological crystal $[\psi]$, the subleading terms $\psi_{p'}$, representing the decoration on $p'$-cells, are obtained in the spectral-sequence computation.
Using these subleading terms, the decorations on lower-dimensional cells in $\tilde\psi=n\psi$ is computed as
\begin{equation}
\label{eq:mu-r}
\tilde\psi_{p-r}=n\psi_{p-r}.
\end{equation}
One can then look $\tilde\psi_{p-r}$ up in $E^{p-r}_{p-r,\infty}$ to see whether it is nontrivial.
The smallest $r$ such that $\tilde\psi_{p-r}$ is nontrivial then indicates $\tilde\psi=n\psi$ is a nontrivial TCS with $(p-r)$-blocks.
When this happens, combining $E^p_{p,\infty}$ and $E^{p-r}_{p-r,\infty}$ then becomes a nontrivial group-extension problem instead of a direct sum.

We notice that the answer of whether $n[\psi]$ is a nontrivial SPT state may depend on the choice of the subleading terms of the generator $\psi$ (on the contrary, the computation of $E^p_{p,r}$ does not depend on this choice).
However, the final result of the group-extension problem is independent of the choice of the generators.
In Appendix~\ref{app:trivial}, we will show that for the simple cases of bosonic TCSs $G=SG\times G_0$, if we choose to ignore the 0-block TCS, there exists a simple choice of $\psi_{p-1}$ such that the group-extension problem becomes trivial, and a naive direct sum in Eq.~\eqref{eq:sum1} gives the correct classification.

\section{Examples}
\label{sec:examples}
The algorithm to classify TCSs outlined in above sections can be automated for the bosonic case, using the formulation given in Appendices~\ref{app:gaction} and \ref{app:transfer}.
For simplicity, here we only consider the case when the total group is a direct product of space (wallpaper) group and a local symmetry group, {\it i.e.}, $G=SG\times G_0$.
In this case, since $d_2$ map is trivial, we only need to take care of the first-page no-open-edge condition and the bubble equivalence.
By an automated script, we have enumerate the bosonic TCSs with seveal local symmetry groups in {\it all} the wallpaper groups and {\it all} the space groups.
The main results can be found in Tables \ref{tab:2D} and \ref{tab:3D} in Appendix~\ref{app:wgsg}, respectively.

\begin{figure}
\begin{centering}
\includegraphics[width=1.0\linewidth]{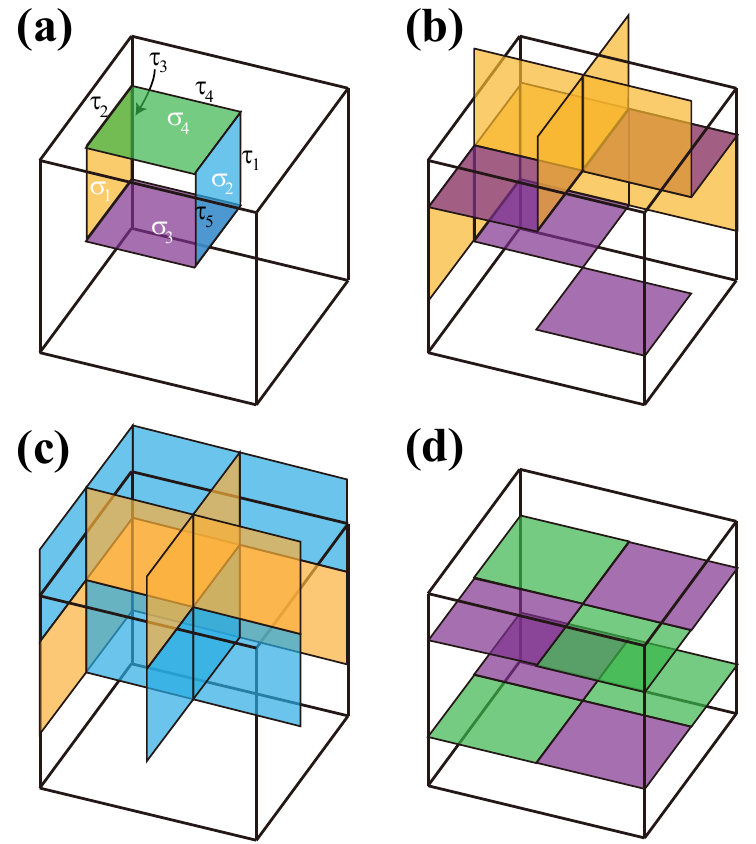}
\par\end{centering}
\caption{
A 2D construction beyond layer construction.
(a) The inequivalent 2-cells and 1-cells of the space group $P\bar4n2$.
(b-d) The three $\mathbb{Z}_2$ generators for the crystalline states.
(b) is beyond layer construction, and (c-d) are equivalent with layer constructions.
The equivalent 2-cells are colored in the same colors.
See Appendix \ref{app:P-4n2} for the definitions of the 2-cells and 1-cells.
}\label{fig:nonLC}
\end{figure}

In our results, we find that, although some of the 2D constructions of TCSs are equivalent to decoupled straight lines or straight planes, some of them, however, are beyond such simple layer constructions.
Here we give a bosonic example of a TCS beyond layer constructions, which has a geometric structure similar to an example studied in Ref.~\cite{Song2018}.
We consider the space group $P\bar4n2$ with the local symmetry group $Z_2$.
The 2-cells and 1-cells are shown in Fig. \ref{fig:nonLC}.
The details of the 2-cells and 1-cells are given in Appendix \ref{app:P-4n2}.
According to Ref. \cite{ChenSPTScience,ChenSPTPRB,levin12}, $Z_2$ symmetry protects a $\mathbb{Z}_2$ 2D SPT.
Thus we can decorate each 2-cell with a such a 2D SPT.
As discussed in Appendix~\ref{app:trivial}, the anomalies of these 2D SPTs can cancel each other on the 1-cells where they meet.
Thus, the no-open-edge  condition reduces to the constraint that there should be even number of 2D SPT ending at each 1-cell \cite{Song2018}.
On the other hand, the bubble equivalence is  trivial because on the 2-cells the $\mathbb{Z}_2$ bubble is always canceled by its symmetry partner \cite{Song2018}.
Therefore the classification of bosonic TCS in this case is just given by the allowed decorating configurations.
The boundaries of the four inequivalent 2-cells are
\begin{equation}
\sum_{g\in SG} \partial g \sigma_1 = \sum_{t\in T} t\tau_2 + t\tau_4,\label{eq:sigma1_boundary}
\end{equation}
\begin{equation}
\sum_{g\in SG} \partial g \sigma_2 = \sum_{t\in T} t\tau_2 + t\tau_4,
\end{equation}
\begin{equation}
\sum_{g\in SG} \partial g \sigma_3 = \sum_{t\in T} t\tau_2 + t\tau_4,
\end{equation}
\begin{equation}
\sum_{g\in SG} \partial g \sigma_4 = \sum_{t\in T} t\tau_2 + t\tau_4,\label{eq:sigma4_boundary}
\end{equation}
where $T$ represents the translation group.
Since all the 2-cells have the same boundary, in order to cancel the anomalies on the 1-cell we only need to decorate two of the four 2-cells.
There are three independent decorations: $\sigma_1+\sigma_3$, $\sigma_1+\sigma_2$, $\sigma_3+\sigma_4$.
As shown in Fig. \ref{fig:nonLC}b-d, $\sigma_1+\sigma_2$ and $\sigma_3+\sigma_4$ decompose into horizontal and vertical planar layers, whereas $\sigma_1+\sigma_3$ does not.
In fact, using similar argument in Ref. \cite{Song2018}, one can show that $\sigma_1+\sigma_3$ is not equivalent with any layer construction.


\begin{figure}[htbp]
  \includegraphics{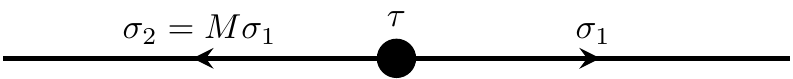}
  \caption{Cell decomposition for a 1D system with a mirror-reflection symmetry.}
  \label{fig:cell1d}
\end{figure}

Next, we demonstrate the application of our real-space recipe to classify interacting-fermion SPTs by revisiting an example discussed in Ref.~\cite{Lapa2016}.
In this example, we consider a 1D superconducting system with a time-reversal symmetry $T$ and a mirror-reflection symmetry $M$ that reverses the whole 1D system, and we assume that the fermions transform as spinless fermions under both $T$ and $M$ (i. e. fermions carry linear representations $T^2=M^2=+1$).
When considering interacting-fermion classification, being a superconducting system means there is no U(1) charge-conservation symmetry.
To classify topological superconductors in such a system, we divide the 1D system into two 1-cells $\sigma_{1,2}$ and a 0-cell $\tau$, as shown in Fig.~\ref{fig:cell1d}.
The 0-cell is located at the reflection center, and the two 1-cells are mapped to each other by the reflection symmetry $M$.
In 1D, the TCSs we are interested in are given by $E^1_{1,\infty}$, denoting decorating 1D blocks on 1-cells (we are ignoring 0D decorations in $E^0_{0,\infty}$, as we always do in this paper.)

On the first page, $E^1_{1,1}$ is given by Eq.~\eqref{eq:E1}.
Here, $Y_1/G$ contains only one $G$-orbit containing both $\sigma_{1,2}$.
Choosing $\sigma_1$ to represent this orbit, $E^1_{1,1}$ is given by $\Phi^1(G_{\sigma_1})$, which contain 1D SPT states protected by the onsite symmetry group $G_{\sigma_1}=\mathbb Z_2^T$.
It is well-known that $\Phi^1(\mathbb Z_2^T)$, representing the phases 1D topological superconductors with $T^2=+1$, has a $\mathbb Z_8$ classification, where the root state is the 1D Kitaev chain~\cite{Fidkowski2010,Fidkowski2011}.
Hence, we have $E^1_{1,1}=\mathbb Z_2$.

Next, we compute the first-page no-open-edge condition.
(There is no bubbling equivalence in this case, because there is no 2-cell.)
This condition is described by the first-page derivative $d^1_{1,1}: E^1_{1,1}\rightarrow E^1_{0,1}$.
Here, the codomain of this derivative is the space of anomaly patterns on the 0-cell $\tau$: $E^1_{0,1}=\Phi^1(G_\tau)$.
As the reflection center, $G_\tau=\mathbb Z_2^T\times\mathbb Z_2^M$, where $M$ is a unitary onsite symmetry.
Correspondingly, the classification of anomaly patterns is given by $\Phi^1(G_\tau)=\mathbb Z_8\oplus\mathbb Z_4$, where the $\mathbb Z_8$ and the $\mathbb Z_4$ subgroups are generated by the same Kitaev chain and an additional fermionic SPT state with complex-fermion decorations, respectively~\cite{Turzillo2019}.
The second root state does not play a role in our calculation, so its details will not be discussed here.
Representing elements of $E^1_{1,1}$ and $E^1_{0,1}$ by a mod-8 integer $n$ and a pair of a mod-8 integer and a mod-4 integer $(n_1,n_2)$, respectively,
the derivative $d^1_{1,1}$ has the form $d^1_{1,1}(n) = (2n, 0)$.
The first component of $d^1_{1,1}(n)$ is $2n$, because $\sigma_1$ and $\sigma_2$ should be decorated by the same state denoted by $n$ as the result of the symmetry condition, and their total contribution to the anomaly on $\tau$ is thus $2n$.
The second component of $d^1_{1,1}(n)$ can be computed explicitly using the approach in Appendix~\ref{app:spt:if}, but this result does not affect the classification of $E^1_{1,\infty}$, and hence will not be discussed here.
Using the explicit form of $d^1_{1,1}$, we see that the no-open-edge condition $d^1_{1,1}(n)=0$ demands that $2n=0 \mod8$, or $n=0,4$.
Therefore, the second-page result is $E^1_{1,2}=\ker d^1_{1,1}=\mathbb Z_2$, generated by $n=4$, representing decorating four copies of Kitaev chains on $\sigma_{1,2}$.
In 1D, $E^1_{1,2}=E^1_{1,\infty}$ is the final answer.
Hence, the topological superconductors in such a 1D system has a $\mathbb Z_2$ classificationm, and the root state is the interaction-enabled topological superconductor state studied in Ref.~\cite{Lapa2016}.

The steps described above can be generalized to study interacting-fermion SPT states with more complex space-group symmetries.
Some examples of using similar ideas to study real-space construction of fermionic topological crystalline states can be found in Refs.~\cite{Lapa2016,Song2017,MChengRotSPT2018X}
We shall leave a fully automated implimentation of our real-space recipes to future works.


\section{Application to HOLSM Theorems}
\label{sec:lsm}

In this section, we apply our TCS constructions to obtain generalized HOLSM-type theorems.
We first review the original HOLSM Theorem and its generalizations to different onsite symmetry groups, and discuss how to understand them using our TCS constructions.
We then revisit the first example given in Sec.~\ref{sec:noe2}, and discuss how to reinterpret it as an SPT-enforcing HOLSM Theorem~\cite{DyonLSM, ElseLSM}.
Last, we will give the general relation between TCS constructions and generalized HOLSM Theorems.
For simplicity, we first discuss the 2D examples, then generalize our results to 3D.

The original HOLSM Theorem asserts that in a 2D lattice with translation symmetries and spin-rotation symmetries, if there is an odd number of spin-$\frac12$ per unit cell, the system cannot have a symmetric gapped unique ground state.
This theorem is later generalized to the cases where the spin-rotation symmetry and the spin-$\frac12$ objects are replaced by an arbitraty onsite symmetry group $G_0$ and a nontrivial projective representation of $G_0$, respectively.
In this section, we will refer to the original and these generalizations as the ``generalized HOLSM Theorems''.

Using our TCS framework, we can view the distribution of projective representations as a nontrivial anomaly pattern $[\tilde\psi]$ in the module $E^1_{0,\infty}$.
In our language, the total symmetry group is $G=SG\times G_0$, where the space group $SG=\mathbb Z^2$.
Hence, the $G$-complex $Y$ we construct is the same as the one shown in Fig.~\ref{fig:mag}, with one 0-cell per unit cell.
The local symmetry group on the 0-cell is simply $G_0$.
The nontrivial projective representation can be translated to a nontrivial element $[\beta]=H^2[G_0,\uone_T]=\Phi^1(G_0)$, which is decorated to the 0-cells, as $[\tilde\psi]_\mu = [\beta]$.
In this way, the distribution of a nontrivial projective representation is translated to an anomaly pattern $[\tilde\psi]$.

We now argue that, including this $[\tilde\psi]$, every nontrivial element in $E^1_{0,\infty}$ represents an anomaly pattern that cannot be gapped out by a symmetric unique ground state.
This is done by reinterpreting the no-open-edge conditions we introduced in Sec.~\ref{sec:full}.
Consider an element $[\tilde\psi]$ in $E^1_{0,r}$ that is trivialized by the $d_r$ map,
\begin{equation}
  \label{eq:drpsi=tpsi}
  d_r[\psi] = [\tilde\psi] .
\end{equation}
In Sec.~\ref{sec:full}, we interpreted this relation as the fact that, the assembly $[\psi]$ does not satisfy the no-open-edge condition and cannot be realized as a TCS, because assembling it will result in anomaly patterns specified by $[\tilde\psi]$ on the 0-cells.
However, if the physical Hilbert space already contains an anomaly pattern $[\tilde\psi]$ on the 0-cells,
the assembly $-[\psi]$ can be realized in such physical systems, because the obstruction $d_r(-[\psi])=-[\psi]$ is now canceled by the background anomaly pattern in the Hilbert space.
Therefore, Eq.~\eqref{eq:drpsi=tpsi} also implies that the anomaly pattern $-[\psi]$ can be gapped out by the TCS assembly $[\psi]$.
In other words, it reveals a UV/IR anomaly matching between the TCS assembly (which can be viewed as the IR limit) and the anomaly pattern (which can be viewed as the UV limit).

A corollary of this reinterpretation is that a nontrivial element in $E^1_{0,\infty}$ cannot be gapped out by any such TCS assembly, and therefore cannot realize a symmetric gapped unique ground state.

We can also revisit the first example in Sec.~\ref{sec:noe2} using this alternative interpretation.
Recall that Eq.~\eqref{eq:d2psi=tpsi} indicates that $[\psi]$ does not represent a valid 2D $G$-SPT state:
tiling the 2D plane with the SPT phase $[\alpha]$ will leave one gapless Kramers doublet in each unit cell.
However, if we start with a model that has one Kramers doublet per unit cell in the original Hilbert space,
this will cancel the anomaly of $d_2[\psi]$ and allows the construction of the 2D SPT $[\alpha]$.
The construction of the trivial SPT state in such system, however, becomes impossible, because the anomaly would be left uncanceled.
In other words, the existence of a nontrivial anomaly pattern in the Hilbert space requires that a gapped unique ground state must be a nontrivial SPT state.
This is precisely the theorem proved in Ref.~\cite{DyonLSM}, which we will refer to as an SPT-enforcing theorem.

In general, we can express these HOLSM-like constraints as the following:
In our language, a spatial distribution of projective representations is represented by an anomaly pattern $[\tilde\psi]$ in $E^1_{0, 1}$.
If $[\tilde\psi]$ is trivialized through $d_1$ by a $[\psi]\in E^1_{1,1}$ as $[\tilde\psi]=d_1[\psi]$, then $[\tilde\psi]$ can be gapped out by the 1-block TCS assembly $[\psi]$.
If $[\tilde\psi]$ is a nontrivial element in $E^1_{0,2}$ but is trivialized through $d_2$ by $\hat\alpha\in E^2_{2,2}$ as $[\tilde\psi]=d_2[\psi]$, then $[\tilde\psi]$ can be gapped out by the 2-block TCS assembly $[\psi]$, which must be a strong SPT state (i.e. it is protected solely by $G_0$) because 2-cells in a 2D space must have $G_\sigma=G_0$.
Furthermore, if $[\tilde\psi]$ is a nontrivial element in $E^1_{0,3}=E^1_{0,\infty}$, then it cannot be gapped out by an SPT state, and will give the consequences of the original LSM Theorem.
Therefore, the spectral sequence introduced in Sec.~\ref{sec:full} provides a way to compute these constraints.

These constraints can be further generalized to 3D.
There, an anomaly pattern, which is an element in $E^1_{0, 1}$, can be trivialized through $d_1$ by a TCS assembly in $E^2_{1, 1}$, through $d_2$ by a TCS assembly in $E^2_{2, 2}$, through $d_3$ by a TCS assembly in $E^3_{3, 3}$, or cannot be trivialized at all.
Among these possiblities, $E^3_{3, 3}$ represents strong 3D SPT states.

\begin{figure}
  \includegraphics[width=\columnwidth]{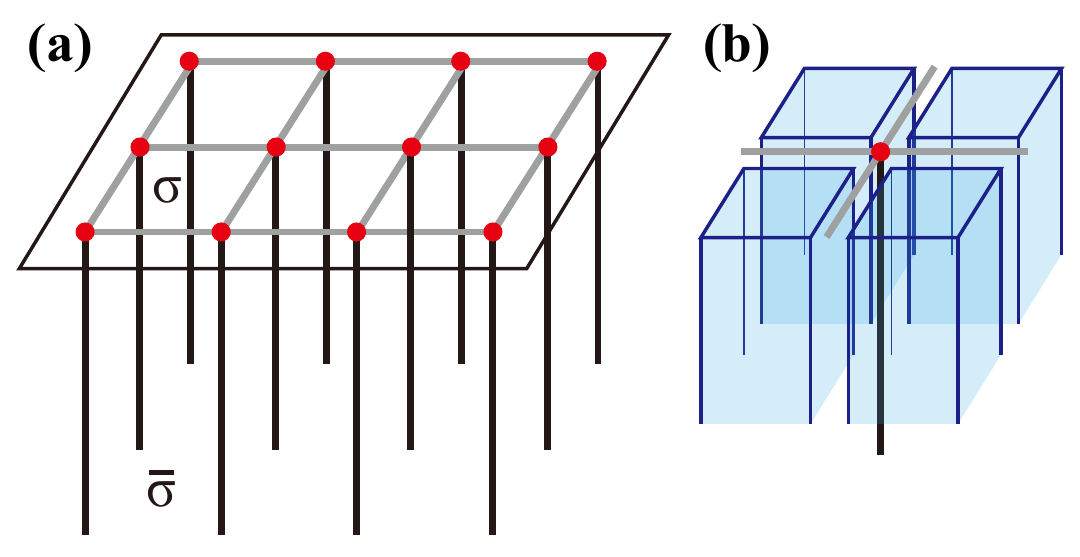}
  \caption{Mapping between HOLSM in 2D and TCS in 3D.
  (a) Mapping between a 2D $G$-complex $Y$ and a 3D $G$-complex $\bar Y$.
  $Y$ can be viewed as the boundary of $\bar Y$.
  (b) Illustration of a 3-block bubbling process in $\bar Y$.
  On the second page, such a process trivializes a 1-block assembly in $\bar Y$ and generates a 2-block SPT in $Y$.}
  \label{fig:2d3d}
\end{figure}

The UV/IR anomaly matching condition in Eq.~\eqref{eq:drpsi=tpsi}, and the resulting SPT-enforcing HOLSM Theorems, can be understood by viewing the anomaly pattern as the surface anomaly of a 3D bulk state.
We first explain this for the simple cases of the generalized HOLSM Theorem with $G=G_0\times \mathbb Z^2$.
Since the projective representation $[\beta]$ is the edge state of a 1D SPT state, the 3D bulk state can be constructed by decorating one such SPT chain in each 3D unit cell, as shown in Fig.~\ref{fig:2d3d}(a).
Here, we argue that the 3D bulk state is closely related to $[\tilde\psi]$, and can be constructed mathematical from it.
The 3D bulk has the same symmetry $G$ as its surface.
Hence, we construct a 3D topological space $\bar Y$ compatible with $SG$.
$\bar Y$ has one 3-cell, two 2-cells and one 1-cell in each unit cell (it has no 0-cells).
The aforementioned 3D bulk state is represented by an SPT pattern $[\bar\psi]\in \bar E^1_{1,\infty}$, constructed by decorating one Haldane chain on each 1-cell.
Here, the bar on $\tilde E^q_{p,r}$ indicates that it is the spectral sequence constructed for $\bar Y$ instead of $Y$.
It is easy to realize that there is a one-to-one correspondence between $p$-cells in $Y$ and $(p+1)$-cells in $\bar Y$, which can be expressed as an isomorphism $Y_p\simeq\bar Y_{p+1}$.
We denote the corresponding cells in $Y_p$ and $\bar Y_{p+1}$ as $\sigma$ and $\bar\sigma$, respectively.
Using $\bar Y$, the SPT pattern $[\bar\psi]$ can be expressed as decorating one Haldane chain to each 1-cell in $\bar Y$.
Next, we notice that, mathematically, the anomaly decorated to each $\sigma\in Y_0$ and the 1D SPT decorated to each $\bar\sigma\in\bar Y_1$ are represented by the same 2-cocycle $[\beta]$ in $H^2[G_0,\uone_T]$.
Hence, $[\tilde\psi]$ and $[\bar\psi]$ can be converted to each other by copying the decoration between corresponding cells.
Mathematically, this is described by the fact that the isomorphism $Y_p\simeq\bar Y_{p+1}$ naturally induces an isomorphism $E^1_{0,\infty}\simeq\bar E1_{1,\infty}$: the isomorphsm map, denoted by $L:E^1_{0,\infty}\rightarrow\bar E^1_{1,\infty}$, is given by
\begin{equation}
\label{eq:simeqpp}
L([\tilde\psi])|_{\bar\sigma}=[\tilde\psi]|_\sigma.
\end{equation}
Using this isomorphism, the relation between the anomaly pattern $[\tilde\psi]$ and the corresponding bulk state $[\bar\psi]$ is then given by $\bar\psi=L([\tilde\psi])$.

The surface-bulk correspondence illustrated above can be generalized to arbitrary symmetry groups and dimensions.
We consider a 2D surface and a 3D bulk with the same symmetry group $G$.
Here, the space group $SG=G/G_0$ is a 2D wallpaper group instead of a 3D space group.
Similar to the previous example, the cellular decomposition for the 3D bulk can be constructed from the one of the 2D surface using the isomorphism $\bar Y_{p+1}\simeq Y_p$.
Again, this induces an isomorphism between $E^q_{p,r}$ and $\bar E^q_{p+1,r}$ for arbitrary $p$, $q$ and $r$ through the definition in Eq.~\eqref{eq:simeqpp}.
To understand the physical meaning of this correspondence, we notice that the $p$-cell $\sigma\in Y_p$ is an edge of the $(p+1)$-cell $\bar\sigma\in\bar Y_{p+1}$.
In this way, for $[\tilde\psi]\in E^{p+1}_{p,r}$, representing an $r$-th-page anomaly pattern on the surface, $L([\tilde\psi])$ represents decorating on $\bar\sigma$ the bulk SPT state $[\tilde\psi]|_{\sigma}$ that corresponds to the anomaly $[\tilde\psi]|_{\sigma}$ on its boundary $\sigma$.
Hence, $L([\tilde\psi])\in\bar E^{p+1}_{p+1,r}$ is the bulk SPT state corresponding to the surface anomaly pattern $\hat\omega$.
Similarly, consider a surface TCS $[\psi]\in E^{p+1}_{p,r}$.
$L([\psi])\in\bar E^{p+1}_{p+1,r}$ is a bubbling pattern that generates SPT states $[\psi]$ on the edges of cell $\bar\sigma$, including $\sigma$ itself.
Hence, the 3D bubbling pattern $L([\psi])$ generates the 2D TCS $[\psi]$ on the surface.

In summary, the isomorphism $L$ defined above allows us to express the correspondence between 2D anomaly patterns and 3D TCSs, and between 2D TCSs and 3D bubbling patterns.
It also allows us to convert the 2D no-open-edge conditions to 3D bubbling equivalences.
Consider a $d_r$ map between a 2D assembly and a 2D anomaly pattern, as in Eq.~\eqref{eq:drpsi=tpsi}.
The $L$-isomorphism maps the r.h.s. to a 3D TCS that can host the anomaly pattern on its 2D surface, and the r.h.s. to a 3D bubbling pattern that generates the 2D assembly on its surface [see Fig.~\ref{fig:2d3d}(b)].
In this way, the relation \eqref{eq:drpsi=tpsi} also describes a 3D bubbling equivalence,
\begin{equation}
  \label{eq:dr-L}
    d_rL([\psi]) = L([\tilde\psi]) .
\end{equation}
This dimensional-shifting correspondence is consistent with our understanding that the TCS $L([\tilde\psi])$ is actually trivial because its surface anomaly $[\tilde\psi]$ can be realized as a gapped symmetric state.

As an example, we apply this bulk-boundary correspondence again to the first example in Sec.~\ref{sec:noe2}.
The $L$-isomorphism maps the relation in Eq.~\eqref{eq:d2psi=tpsi} to the following bubbling relation,
\begin{equation}
  \label{eq:d2-L}
  d_rL([\psi]) = L([\tilde\psi]) .
\end{equation}
This indicates that $L([\tilde\psi])$ is a 3D TCS trivialized by the second-page bubbling process $L([\psi])$.
This provides an explicit example illustrating that it is necessary to consider higher-dimensional trivialization processes as discussed in Sec.~\ref{sec:further}, which can be computed using the higher-page $d_r$ maps introduced in Sec.~\ref{sec:bub2}.

\section{Conclusion}
\label{sec:conclusion}

In this work, we systematically study real-space construction of TCSs.
Starting from building blocks made of SPT states protected by the little symmetry group, we construct a TCS by examining the no-open-edge conditions, the bubbling equivalences and solving the group-extension problems.
These steps form a thoughout framework to compute TCS classification.
In particular, for bosonic TCS, we prove that, for any symmetry group, this framework gives exactly the same results as the group-cohomology formula in Ref.~\cite{ThorngrenElse2018}.
For the simple cases of $G=SG\times G_0$, the computation is greatly simplified, as discussed in Sec.~\ref{sec:simplified}.
To demonstrate our framework, we develop an automated code to compute bosonic TCS protected by the direct produce of typical onsite symmetries and any of the 2D and 3D crystalline symmetry groups.

Our framework also applies to interacting fermions.
In deed, we give several examples to demonstrate this application.
The general computation, however, is much more elusive than its bosonic counterpart, due to the complex structure of the space of fixed-point wave functions of onsite symmetries, denoted by $\Psi^d(G_\sigma)$.
This is further complicated by the fact that the stacking operation of fermionic wave functions do not necessarily commute, as pointed out in Sec.~\ref{sec:full}.
However, the recent progresses in the classification of fermionic SPTs protected by onsite symmetries~\cite{GuSuH,Kapustin2015,Gaiotto2016,Kapustin2017,QRWangBSuH,Cheng2018,LanFSPT2018X} should allow such computation to be carried out.
We will leave this to future works.

In Sec.~\ref{sec:lsm}, we point out that our framework can also be used to study generalized HOLSM Theorems, especially those enforcing nontrivial SPT states.
It will be interesting to apply it to look for new SPT-enforcing HOLSM Theorems in more general symmetry groups that mix crystalline and onsite symmetries.
We will also leave this to future works.

\emph{Data availability.} Data of bosonic-SPT classification protected by 2D wallpaper groups and 3D space groups together with various onsite symmetry groups can be found in Appendix~\ref{app:wgsg}. Other data that support the findings of this study are available from the corresponding author upon request.

\emph{Code availability}
All numerical codes in this paper are available upon request to the authors.

\emph{Note added:} As this work was being finalized for posting on the arXiv, Refs.~\cite{LuX,KenX} appeared, which contains some related results, and we also became aware of another indenpendent work by~\citet{ElseX}.\
\begin{acknowledgements}
  YQ is grateful to Zheng-Cheng Gu, Zheng-Xin Liu, Shenghai Jiang and Ying Ran for invaluable discussions. YQ also thanks Aspen Center for Physics for hospitality, where part of this work was performed. SZD and CF acknowledge support from Minstry of Science and Technology of China under grant numbers 2016YFA0302400, 2016YFA0300600, from National Science Foundation of China under grant number 11674370, and from Chinese Academy of Sciences under grant number XXH13506-202.
  YQ acknowledges support from Minstry of Science and Technology of China under grant numbers 2015CB921700, and from National Science Foundation of China under grant number 11874115.
\end{acknowledgements}

\emph{Author contributions.} C.F. and Y.Q. conceived the project and developed the theoretical ideas. Y.Q. derived the spectral-sequence formulas. Z.S. implemented the algorithm and obtained all classification. All authors contributed to the writing of the manuscript.

\emph{Competing interests} The authors declare no competing interests.

\appendix

\section{Review of onsite-symmetry SPT classification}
\label{app:spt-review}

In this section, we briefly review the classification of SPT states protected solely by onsite symmetries.
We first give some general and abstract description of such classification in various systems, including bosonic systems (which are necessarily interacting), free-fermion systems and interacting-fermion systems.
We then focus on the classification of bosonic SPTs, which can be described by the group-cohomology theory (plus some exceptions of beyond-group-cohomology states).
This will be the main focus of this appendix.
At the end of this appendix, we will briefly review the structure of fermionic SPT classification.

\subsection{Abstract description}
\label{app:spt:abs}

We first introduce some abstract notations that apply to both bosonic, free-fermion and interacting fermion systems, and list some common properties shared by all these systems.

We use the notation $\Phi^d(G)$ to denote the set of $d$-dimensional SPT phases protected by the onsite symmetry group $G$.
Since SPT phases are equivalence classes of states, we use $[\alpha]$ to denote them, and use $\sim$ to denote two identical phases.
The phases in $\Phi^d(G)$ actually form an Abelian group:
First, two phases $[\alpha]$ and $[\beta]$ can be added together.
Denoted by $[\alpha]+[\beta]$, the sum represents the resulting phase obtained by stacking the two phases.
The addition is commutative:
\begin{equation}
\label{eq:psi-add-comm}
[\alpha]+[\beta]\sim[\beta]+[\alpha].
\end{equation}
Second, there is a trivial phase, which we denote by $0$.
Third, each phase $[\psi]\in\Phi^d(G)$ has an inverse, denoted by $-[\psi]$, such that stacking them together gives the trivial phase.
In other words, phases in $\Phi^d(G)$ are ``invertible'' and have no topological ground-state degeneracy.
This Abelian group can also be viewed as a $\mathbb Z$-module, or a linear space with integral coefficients.

In general, the classification $\Phi^d(G)$ at different dimensions can be computed using a generalized cohomology theory.
Intuitively, we may understand this mathematical structure using fixed-point wave functions (or partition functions) of the SPT phases.
Here, fixed-point means that the wave functions or partition functions have zero correlation lengths, which make them suitable to describe gapped phases.
In particular, we need to consider fixed-point wave functions in the geometry where a $d$-dimensional bulk terminates at a $(d-1)$-dimensional boundary.
We need to consider not only the bulk wave functions, but also wave functions on the boundary of another bulk wave function.
All such wave functions form a set, which we denote by $\Psi^d(G)$.
Similar to the notation $\Phi^d(G)$, the superscript $d$ denotes the dimensionality of the wave function and $G$ denotes the onsite symmetry group.

The set $\Psi^d(G)$ has a similar structure as $\Phi^d(G)$, with a crucial difference:
For fermionic systems, the stacking operation is not commutative, due to the statistical phases generated by exchanging fermionic operators.
For this reason, we denote this stacking operation by $\boxplus$ instead.

Another important structure of $\Psi^d(G)$ is the coboundary operator $d$, which represents the relation between two wave functions in the bulk-boundary geometry.
Given a $d$-dimensional bulk wave function $\alpha\in\Psi^d(G)$, not all $(d-1)$-dimensional wave functions can be realized on the boundary.
If a wave function $\beta$ can be put on the boundary of $\alpha$, we denote this relation by
\begin{equation}
  \label{eq:db=a}
  d\beta = \alpha.
\end{equation}
Here, $d^{d-1}:\Psi^{d-1}(G)\rightarrow\Psi^d(G)$ is the coboundary map.
The superscript $d-1$ indicates the domain of the map, and it is often omitted when it can be inferred from the context.

The two operations $\boxplus$ and $d$ should satisfy some consistency conditions.
First, we require that the coboundary operation commutes with stacking: $d\psi_1\boxplus d\psi_2 = d(\psi_1\boxplus\psi_2)$.
In other words, $d:\Psi^p(G_\sigma)\rightarrow \Psi^{p+1}(G_\sigma)$ is a group homomorphism.
This is because we can stack the bulk and boundary separately, and the resulting states still satisfies the bulk-boundary correspondence.
Second, we require that the commutator between two states, $[\psi_1,\psi_2]=\psi_1\boxplus\psi_2 - \psi_2\boxplus\psi_1$, is a coboundary of another state in $(p-1)$ dimension.
This is because although $\psi_1\boxplus\psi_2$ and $\psi_2\boxplus\psi_1$ may be different states, they should still belong to the same topological phase.

With these conditions satisfied, the SPT classification $\Phi^d(G)$ can be computed from $\Psi^d(G)$ using the coboundary maps.
First, a standalone $d$-dimensional wave function must satisfy the cocycle equation $d\alpha=0$, because it can be viewed as the boundary wave function of a $(d+1)$-dimensional vacuum.
Second, if a $(d-1)$-dimensional wave function $\beta$ can be put on the boundary of a $d$-dimensional wave function $\alpha$: $\alpha=d\beta$, then $\alpha$ belongs to the trivial phase.
This is because as a fixed-point wave function, $\beta$ represents a gapped symmetric boundary state, and the existence of such a gapped symmetric boundary indicates that the bulk $\alpha$ belongs to the trivial SPT phase.
Furthermore, two wave functions differ only by a coboundary $d\beta$ belong to the same SPT phase.
One important property of $d$ is that it satisfies the relation $d^2=0$, meaning that if $\alpha=d\beta$ is a bulk wave function, $\alpha$ is naturally a standalone $d$-dimensional wave function and satisfies $d\alpha=0$.
Using these results, we can compute the SPT classification using the cohomology group of the $d$ map,
\begin{equation}
  \label{eq:Phi=hom-d}
  \Phi^d(G)=\frac{\ker d^d}{\img d^{d-1}}.
\end{equation}
Here, the $\ker d$ in the numerator means that a wave function representing an SPT phase must satisfy the cocycle equation $d\alpha=0$;
the $\img d$ in the denominator means that two wave functions differ by a coboundary $d\beta$ represents the same phase.

\subsection{Bosonic SPT and group cohomology}
\label{app:spt:boson}

We first briefly review the definition of group cohomology for a discrete group G.
The group cohomology can be computed using the homogeneous cochains, which are functions that map $(p+1)$ numbers of group elements to the coefficient ring $\uone_X$ (here, $X$ denotes a generic group action, which will be explained below),
\begin{equation}
\label{eq:pcochain}
\alpha(g_0,\ldots,g_p)\in\uone.
\end{equation}
The word ``homogeneous'' means $\alpha$ satisfies the homogeneous condition,
\begin{equation}
\label{eq:homogeneous}
\alpha(gg_0,\ldots,gg_p)
=\rho_X(g)\alpha(g_0,\ldots,g_p),
\end{equation}
where $\rho_X$ denotes an arbitrary group action.
In this work, we consider the following symmetry actions given by three $\mathbb Z_2$ gradings of the symmetry group $G$:
First, we use $\rho_T(g)=\pm1$ to denote whether $g\in G$ is a time-reversal operation: $\rho_T(g)=\pm1$ if $g$ is time-reversal even (odd), respectively.
Second, we use $\rho_P(g)=\pm1$ to denote whether $g$ reverses the spatial orientation: a proper transformation, including a translation, a rotation and a skew rotation, has $\rho_P(g)=+1$; an improper transformation, including a mirror-reflection, a 3D inversion and a glide reflection, has $\rho_P(g)=-1$.
Finally, we use $\rho_{PT}$ to denote $\rho_{PT}(g)=\rho_P(g)\rho_T(g)$.
Furthermore, we use $\uone_T$, $\uone_P$ and $\uone_{PT}$ to denote $\uone$ coefficient modules with the corresponding symmetry actions: $g\in G$ acts as a unitary (antiunitary) operator on coefficients in $\uone_X$ if $\rho_X(g)=\pm1$, respectively.

An $\alpha$ satisfying Eqs.~\eqref{eq:pcochain} and \eqref{eq:homogeneous} is called a $p$-cochain.
Notice that, in this paper, we treat \uone coefficients as phase angles modulo $2\pi$, instead of a phase factor.
In this notation, the cochains form an additive group instead of a multiplicative group.
All $p$ cochains form an Abelian group (actually a $\mathbb Z$-linear space), denoted by $C^p[G, \uone_X]$.

We now define a coboundary map $d^p:C^p[G,\uone_X]\rightarrow C^{p+1}[G,\uone_X]$, with the following explicit formula,
\begin{equation}
\label{eq:pcobdry}
d^p\alpha(g_0,\ldots,g_{p+1})
=\sum_{k=0}^{p+1}(-1)^k
\alpha(g_0,\ldots,\hat g_k,\ldots,g_{p+1}),
\end{equation}
where $\hat g_k$ means the element $g_k$ is skipped.
The superscript $p$ in $d^p$ denotes the cochain space it acts upon, and as in the main text, we often omit it when it can be determined from the context.
The cochain map defined in Eq.~\eqref{eq:pcobdry} satisfies the condition
\begin{equation}
\label{eq:dsq=0}
d^pd^{p-1}=0.
\end{equation}
Hence, linked by $d^p$, the cochain spaces $C^p[G,\uone_X]$ form a cochain complex,
\begin{widetext}
\begin{equation}
\label{eq:cochain-cplx}
\cdots\rightarrow C^{p-1}[G,\uone_X]
\xrightarrow{d^{p-1}}
C^p[G,\uone_X]
\xrightarrow{d^p}C^{p+1}[G,\uone_X]
\rightarrow\cdots.
\end{equation}
\end{widetext}
The group cohomology of $G$ is defined as the cohomology group of this cochain complex,
\begin{equation}
\label{eq:def-hp}
H^p[G,\uone_X]=\frac{\ker d^p}{\img d^{p-1}}.
\end{equation}
The numerator $\ker d^p$ contains cochains satisfying the cocycle condition $d\omega=0$, which are then called $p$-cocycles.
The denominator implies that two cocycles differ by a coboundary $d\mu$ are considered as the same cohomology class.
The cohomology group is then the quotient group of the cocycles over the coboundary equivalence.

For an onsite symmetry group $G_0$, the cohomology classes in $H^{d+1}[G_0,\uone_T]$ are in one-to-one correspondence to $d$-dimensional SPT states (there are a few beyond-group-cohomology SPT states that are not described by any cohomology classes.) This correspondence can be explicitly demonstrated using the group-cohomology models, which we briefly review below.
These onsite SPT states are used in the main text as the building block of the topological crystals.

A group-cohomology model lives on a $d$-dimensional lattice, with a triangularization and a branching structure.
The vertices of the lattice is organized into $d$-dimensional simplexes (triangles in 2D and tetrahedra in 3D).
The branching structure is a set of orientations on all links between vertices, satisfying the condition that the links do not form any oriented loop.
We also require that the orientations are invariant under the action of $SG$.
It can be shown that one can always choose such a branching structure~\cite{ThorngrenElse2018}.
The Hilbert space of the model consists of a local Hilbert space on each vertex, spanned by basis vectors $|g_i\rangle$ corresponding to group elements $g_i\in G$.
Using a cocycle $\alpha\in H^{d+1}[G, \uone_{PT}]$, one can construct the following fixed-point wave function on such a lattice,
\begin{widetext}
\begin{equation}
\label{eq:Psi-alpha}
|\Psi[\alpha]\rangle
=\sum_{\{g_i\}}\prod_{\Delta i_1\cdots i_d}
\exp\left\{is(i_1,\ldots,i_d)\alpha(g_0, g_{i_1},\ldots g_{i_d})\right\}
|g_1g_2\cdots g_N\rangle,
\end{equation}
\end{widetext}
where the product runs over all $d$-dimensional simplexes in the lattice, and $s(i_1,\ldots i_{d+1})=\pm1$ denotes whether the orientation of the simplex determined from the branching structure is the same or opposite to an overall orientation of the manifold.
Recall that, in this paper, we treat \uone coefficients as phase angles modulo $2\pi$, instead of a phase factor.
Hence, the cocycle $\alpha$ appears in the exponent in the above equation.
In Eq.~\eqref{eq:Psi-alpha}, $g_0$ is an arbitrary but fixed group element in $G$.
One can shown that the wave function is independent of the choice of $g_0$, and it is invariant under the action of $G$ on a $d$-dimensional manifold without a boundary, if we assume $G$ acts in the following way: $g|g_i\rangle=|gg_{i^\prime}\rangle$, where $i^\prime$ is the image of $i$ under the action of $g$.

We now consider a more complicated geometry, where we simutaneously decorate cochains both inside a $d$-dimensional bulk and on its $(d-1)$-dimensional boundary.
We start with the simplest case, where the bulk is one $d$-simplex, and the boundary has $(d+1)$ pieces of $(d-1)$-simplices.
We decorate the bulk and the surface with wave functions constructed from a $(d+1)$-cochain $\alpha$ and a $d$-cochain $\beta$, respectively.
Choosing the orientation of the bulk to be $+1$, the phase factor contributed from the bulk wave function is $\exp\{i\alpha(g_0,g_1,\ldots g_{d+1})\}$.
It is straigntforward to check that the phase factors from all surface $d$-simplices is given by $d\beta$, as
$\exp\{-id\beta(g_0,g_1,\ldots g_{d+1})-i\beta(g_1,\ldots g_{d+1})\}$.
Hence, if the two cochains satisfy
\begin{equation}
\label{eq:a=db}
\alpha = d\beta,
\end{equation}
the total phase factor is then $\exp\{-i\beta(g_1,\ldots g_{d+1})\}$, which is invariant under $G_0$-action, because $\beta$ satisfies the homogeneous condition \eqref{eq:homogeneous}.

This can be generalized to an arbitrary $d$-dimensional bulk, which are divided into many $d$-simplices.
For each simplex, the phase factors attached to the bulk and its surface is $G_0$-invariant.
When multiplying the phase factors from different simplices together, the phase factors on the interior $(d-1)$-simplices cancel with each other, and we obtain a wave function that has phase factors from each $d$-simplex in the bulk, and each $(d-1)$-simplex on the boundary.
Therefore, the wave function on the entire bulk and surface is also $G_0$-invariant, if the condition \eqref{eq:a=db} is satisfied.

This setup can be used to demonstrate the bulk-boundary correspondence of SPT states.
To realize gapped and symmetric states on both the bulk and the boundary, the cochains decorated to the bulk and the boundary must satisfy the condition in Eq.~\eqref{eq:a=db}.
If the bulk is empty, i. e. $\alpha=0$, the decoration on the boundary must be a cocycle, since Eq.~\eqref{eq:a=db} then becomes the cocycle condition $d\beta=0$.
This is consistent with our understanding that a cocycle in $H^d[G,\uone_T]$ represents a $(d-1)$-dimensional symmetric gapped SPT state.
If the bulk is not empty: $\alpha\neq0$, the cochain $\beta$ is then not a cocycle, as $d\beta=\alpha\neq0$.
This further implies that $\alpha$ is a trivial cocycle in $H^{d+1}[G,\uone_T]$, since $d\beta$ is a coboundary.
On the other hand, if $\alpha$ is a nontrivial cocycle, then there is no solution of $\beta$ that would satisfy Eq.~\eqref{eq:a=db}.
This means that the boundary of a nontrivial SPT state cannot be both gapped and symmetric.
In this case, the nontrivial SPT state in the bulk induces a symmetry anomaly on the boundary, which obstructs the construction of a gapped symmetric wave function.

In summary, the group-cohomology models demonstrate that the cochains satisfied the general property we need in Sec.~\ref{app:spt:abs}, and can be used to represent fixed-point wave functions of onsite-symmetry SPT states.
In particular, we can take $\Psi^d(G)=C^{d+1}[G, \uone_T]$. Correspondingly, the SPT classification is given by $\Phi^d(G)=H^{d+1}[G,\uone_T]$.

In our paper, we need a more general type of cochain functions, which can be used to derive the same group-cohomology classes $H^p[G, \uone_X]$.
Here, we consider two groups, $H$ and $G$, where $H$ is a subgroup of $G$: $H\subset G$.
To compute the group cohomology of $H$, we consider cochains similar to Eq.~\eqref{eq:pcochain}, but the variables $g_i$ take values in $G$ instead of $H$.
Different from regular cochains, we only require the $G$-valued cochains to be invariant under the action of $H$,
\begin{equation}
\label{eq:hom-H}
\alpha(hg_0,\ldots,hg_p)
=\rho_X(h)\alpha(g_0,\ldots,g_p),\quad
\forall h\in H.
\end{equation}
Hence, we will refer to these cochains as the $G$-valued-$H$-invariant cochains.
We denote the set of these cochains by $C^p_H[G,\uone_X]$, which is also a $\mathbb Z$-module.
In general, $\alpha$ is not invariant under the action of $g\in G$, and the $G$-action on the $G$-valued-$H$-invariant cochains will induce a useful symmetry action that we will discuss in more details in Appendix~\ref{app:gaction}.

The coboundary maps $d^p$ can be defined in a similary way as in Eq.~\eqref{eq:pcobdry}.
With these coboundary maps, the $G$-valued-$H$-invariant cochains also form a cochain complex.
It can be shown using the fundamental lemma of homological algebra (see Lemma 7.4 of Ref.~\cite{Brown}] that the cohomology group of the $G$-valued cochain complex gives the same group cohomology of $H$.

In summary, one can choose to compute $H^p[H,\uone_X]$ using $G$-valued-$H$-invariant cochains where $H\subset G$.
In practise, this seems to be a redundant way to compute $H^p[H,\uone_X]$, but as we see in the main text, this is a useful tool in the spectral-sequence computation, because it allows us to compute the symmetry actions on the cochains and the transfer map, which we will introduce in Appendices~\ref{app:gaction} and \ref{app:transfer}, respectively.
In the main text, we will use $G$-valued-$G_\sigma$-invariant cochains to represent states decorated on a cell $\sigma$, where $G_\sigma$ is the local symmetry group of $\sigma$.

We can also use $G$-valued-$H$-invariant cocycles to construct group-cohomology models, which will be useful for Appendices~\ref{app:gaction} and \ref{app:transfer}.
Since the cochains are valued in $G$, we construct local Hilbert space as $|g\rangle$, where $g\in G$.
We can then use a $G$-valued-$H$-invariant cocycle $\alpha$ to construct the wave function in Eq.~\eqref{eq:Psi-alpha}.
Since $\alpha$ is only invariant under $H$, the constructed wave function is also only invariant under $H$ but not $G$.
Hence, although we chose an enlarged local Hilbert space, the constructed wave functions still represent $H$-SPT states.

\subsection{Interacting-fermion SPTs}
\label{app:spt:if}

In this section, we briefly review the classification of SPT phases in interacting-fermion systems, using the notations outlined in Sec.~\ref{app:spt:abs}

First, the space of fixed-point wave functions $\Psi^d(G)$, as a set, is constructed by combining the following cochain spaces:
\begin{equation}
  \label{eq:psid-def}
  \Psi^d(G) = C^0(G_b, \Omega^{d+1})\times C^1(G_b, \Omega^d)\times\cdots\times
  C^{d+1}(G_b, \Omega^0).
\end{equation}
Here, $G_b$ is the bosonic subgroup of the total symmetry group $G$:
$G$ always contains the fermion-parity symmetry group $\mathbb Z_2^f$ as a normal subgroup, and $G_b=G/\mathbb Z_2^f$ is the quotient group.
$\Omega^d$ denotes the Abelian group of $(d+1)$-dimensional invertible topological phases.
For fermionic systems, nontrivial entries of $\Omega^d$ are $\Omega^0=\uone_T$, $\Omega^1=\mathbb Z_2$, $\Omega^2=\mathbb Z_2$ and $\Omega^3=\mathbb Z_T$.
Therefore, an element $\psi\in\Psi^d(G)$ is expressed as a tuple $(\psi^0,\psi^1,\ldots,\psi^{d+1})$, where $\psi^p\in C^p(G, \Omega^{d-p+1})$.

The stacking operation, however, is not a simple componentwise addition of two tuples:
$\psi_1\boxplus\psi_2\neq (\psi_1^{d+1}+\psi_2^{d+1},\ldots,\psi_1^0+\psi_2^0)$.
Instead, the result on the $p^{\text{th}}$ layer will be twisted by layers with smaller $p$.
Such twisting has the following general form,
\begin{equation}
  \label{eq:add-twist}
  \begin{split}
    (\psi_1\boxplus&\psi_2)^p = \psi_1^p + \psi_2^p\\
    &+ \mathcal A^{p,d-p+1}(\psi_1^0,\psi_2^0,\ldots,\psi_1^{p-1}, \psi_2^{p-1}),
  \end{split}
\end{equation}
where $\mathcal A^{p,d-p+1}$ is a generic function describing the twisting.

Next, we consider the coboundary map $d: \Psi^d(G)\rightarrow\Psi^{d+1}(G)$.
Because the coboundary operation commutes with the stacking, it is suffice to consider the coboundary of a state that has only a single nonvanishing layer $\psi=(0, \ldots, 0, \psi^p, 0,\ldots,0)$.
Generally, the coboundary of such a single-layer state has the following form,
\begin{equation}
  \label{eq:cobdry-D}
  (d\psi)^{p'}=\mathcal D^{p, d-p+1}_{p'-p}(\psi^p).
\end{equation}
Here, $\mathcal D^{p,q}_r$ is a function from $C^p(G_b, \Omega^q)$ to $C^{p+r}(G_b, \Omega^{q-r+1})$, and it encodes the form of coboundaries.
It is only nonvanishing for $r > 0$, and for $r=1$, it is simply $\mathcal D^{p,q}_0(\psi^p)=d\psi^p$.

As an example, we list explicit forms of $\mathcal A^{pq}$ and $\mathcal D^{pq}_r$ for the case of $p+q=3$ (corresponding to $d=2$), and $G=G_b\times\mathbb Z_2^f$.
Derivation of these results can be found in Refs.~\cite{QRWangBSuH,QRWang2018X,JohnMorgan3D}.
First, we list the nontrivial entries of $\mathcal D^{pq}_r$.
\begin{align}
  \mathcal D_2^{21} &= \frac12\psi^2\cup\psi^2
  +\frac12\psi^2\cup_1d\psi^2
  +\mathcal O_4'(d\psi^2);\\
  \mathcal D_2^{12} &= s\cup\psi^1\cup\psi^1,
\end{align}
where $O_4'(\psi^2)$ is a 4-cochain with the following form:
\begin{equation}
  \label{eq:O4p}
  \begin{split}
  O_4'(\psi^2)(01234)
  =\frac12d\psi^2(0124)d\psi^2(0234)\\
  -\frac14\{d\psi^2(0123)[1-d\psi^2(0124)] \mod 2\}.
\end{split}
\end{equation}
Here, we follow Ref.~\cite{QRWang2018X} and use the simplified notation $\alpha(i_1\cdots i_m)=\alpha(g_{i_1},\ldots,g_{i_m})$.
$s$ denotes the one-cocycle $s(g)=\pm1$ indicating whether $g$ is a unitary or an antiunitary operation.

Next, we list the nontrivial entries of $\mathcal A^{pq}$.
For simplicity, the parameters of the function are omitted, as they can be inferred from Eq.~\eqref{eq:add-twist}.
\begin{align}
  \mathcal A^{21} =& \psi^1_1\cup\psi^1_2;\\
  \nonumber
  \mathcal A^{30} =& \frac12\psi^2_1\cup_1\psi^2_2 + \frac12 (\psi^1_1\cup\psi^1_2)\cup_1(\psi^2_1+\psi^2_2)\\
  &+\frac12 \psi^1_1\cup\left(\psi^1_1\cup_1\psi^1_2\right)\cup\psi^1_2.
\end{align}
The above results of $A^{pq}$ only apply to the limited cases when all symmetry operations in $G_b$ are unitary.








\section{Symmetry actions on cochains and cocycles}
\label{app:gaction}

In this appendix, we derive the mathematical form of symmetry actions relating cochains representing fixed-point wave functions of bosonic SPT states decorated on symmetry-related cells.

We consider two $p$-cells $\sigma$ and $\sigma^\prime$, which are related by a symmetry operation $g_{\sigma\sigma'}\in G$ as $\sigma^\prime = g_{\sigma\sigma'}\sigma$.
The SPT states decorated on $\sigma$ are classified by the cohomology group $H^{p+1}[G_\sigma, \uone_T]$.
To study the $G$ action on these cohomology classes, we represent these states using the group-cohomology models with $G$-valued-$G_\sigma$-invariant cocycles $C^{p+1}_{G_\sigma}[G,\uone_T]$ introduced in Appendix~\ref{app:spt-review}.
We consider a triangularization of $\sigma$ with vertices $i_1,i_2,\ldots,i_N$.
On $\sigma^\prime$, we construct a symmetry-related triangularization, where each vertex $i_a^\prime$ is given by the image of $i_a$: $i_a^\prime = gi_a$.
On each vertex, the local Hilbert space is spanned by basis states $|g\rangle$, where $g\in G$.
The symmetry group $G$ acts on this Hilbert space in the following way:
A local-symmetry operation $h\in G_\sigma$ acts locally on the Hilbert space on each vertex $i$ on $\sigma$:
$h|g_i\rangle_i = |hg_i\rangle_i$.
A symmetry operation $g$ in the left coset $g_{\sigma\sigma'}G_0$ maps the local Hilbert space on $i$ to $i^\prime$, in addition to multiplying the onsite group element:
$g|g_i\rangle_i=|gg_i\rangle_{i'}$.
Symmetry elements not in the two cosets $G_0$ and $g_{\sigma\sigma'}G_0$ maps $\sigma$ to other cells, and therefore are not considered here.
As discussed before in Appendix~\ref{app:spt-review},
the symmetries in $G_\sigma$ requires that the cocycles decorated on $\sigma$ must be invariant under $G_\sigma$, which then implies that they are classified by $H^{p+1}[G_\sigma,\uone_T]$.

The symmetries in $g_{\sigma\sigma'}G_0$, on the other hand, gives constraints between the decorations on $\sigma$ and $\sigma^\prime$, which we now discuss.
Assume that $\sigma$ is decorated with the wave function $|\Psi[\alpha]\rangle$ as in Eq.~\eqref{eq:Psi-alpha}, constructed with a $G$-valued-$G_\sigma$-invariant cochain $\alpha$.
The action $g_{\sigma\sigma^\prime}$ maps the state $|g_{i_1}\rangle_{i_1}\otimes\cdots\otimes|g_{i_N}\rangle_{i_N}$ on $\sigma$ to the state $|g_{\sigma\sigma^\prime}g_{i_1}\rangle_{i_1'}\otimes\cdots\otimes|g_{\sigma\sigma^\prime}g_{i_N}\rangle_{i_N'}$ on $\sigma^\prime$.
Therefore, it maps $|\Psi[\alpha]\rangle$ to the following wave function on $\sigma^\prime$,
\begin{widetext}
  \begin{align}
  \nonumber
  |\Psi^\prime\rangle
  =&\sum_{\{g_{i_a}\}}\prod_{\Delta i_1\cdots i_p}
  \exp\left\{i\rho_T(g_{\sigma\sigma^\prime})s(i_1,\ldots,i_p)\alpha(g_0, g_{i_1},\ldots, g_{i_{p}})\right\}|g_{\sigma\sigma^\prime}g_{i_1}\rangle_{i_1'}\otimes\cdots\otimes|g_{\sigma\sigma^\prime}g_{i_N}\rangle_{i_N'}\\
  =&\sum_{\{g_{i_a^\prime}\}}\prod_{\Delta i_1^\prime\cdots i_{p}^\prime}
  \exp\left\{i\rho_T(g_{\sigma\sigma^\prime})s(i_1^\prime,\ldots,i_{p}^\prime)\alpha(g_{\sigma\sigma^\prime}^{-1}g_0, g_{\sigma\sigma^\prime}^{-1}g_{i_1'},\ldots, g_{\sigma\sigma^\prime}^{-1}g_{i_{p}'})\right\}|g_{i_1^\prime}\rangle_{i_1'}\otimes\cdots\otimes|g_{i_N'}\rangle_{i_N'}.
  \label{eq:Psi-alpha2}
  \end{align}
\end{widetext}
This can be viewed as a wave function constructed from the following cochain $\alpha^\prime$,
\begin{equation}
  \label{eq:alpha2}
  \alpha^\prime(g_0,\ldots,g_{p+1})
  =\rho_T(g_{\sigma\sigma'})\alpha(g_{\sigma\sigma^\prime}^{-1}g_0,\ldots g_{\sigma\sigma^\prime}^{-1}g_N).
\end{equation}
In other words, in order to get a $G$-symmetric decoration, if $\sigma$ is decorated with a cochain $\alpha$, $\sigma^\prime$ must be decorated with the cochain $\alpha'$ defined in Eq.~\eqref{eq:alpha2}.

This result motivates us to define a symmetry action on the $G$-valued cochains: for any $g\in G$ and a $G$-valued $p$-cochain $\alpha$, we define $g\cdot\alpha$ as the following $p$-cochain:
\begin{equation}
  \label{eq:g-dot-alpha}
  (g\cdot\alpha)(g_0,\ldots,g_p)=\rho_T(g)\alpha(g^{-1}g_0,\ldots,g^{-1}g_p).
\end{equation}
Using this notation, the constraint in Eq.~\eqref{eq:alpha2} can be written in a concise way, as $\alpha^\prime = g_{\sigma\sigma^\prime}\cdot\alpha$.
This group action is used in Secs.~\ref{sec:blocks} and \ref{sec:connector} in the main text, and the explicit form in Eq.~\eqref{eq:g-dot-alpha} is used in Sec.~\ref{sec:full}.

We end this appendix by listing some obvious but interesting properties of this symmetry action:
\begin{enumerate}
  \item A cochain is $H$-invariant if for all $h\in H$, $h\cdot\alpha = \alpha$.
  This can be checked directly by comparing Eqs.~\eqref{eq:homogeneous} and \eqref{eq:g-dot-alpha}.
  \item If a cochain $\alpha$ is $H$-invariant, then the cochain $g\cdot\alpha$ is $gHg^{-1}$-invariant.
  \item The group action commutes with the coboundary operator. Therefore, it maps cocycles to cocycles, and coboundaries to coboundaries.
  \item Combining properties 2 and 3, we see that the group action $g\cdot\alpha$ induces an isomorphism between $H^p[H, \uone_T]$ and $H^p[gHg^{-1}, \uone_T]$.
  This isomorphism is used in Sec.~\ref{sec:blocks}.
  We notice that, in this way, this isomorphism is expressed using Eq.~\eqref{eq:g-dot-alpha} in terms of $G$-valued cochains.
  Alternatively, it can also be expressed using $H$-valued and $gHg^{-1}$-valued cochains.
  Using this notation, we get the following isomorphism between $\alpha\in H^p[H, \uone_T]$ and $\alpha^\prime\in H^p[gHg^{-1}, \uone_T]$:
  \begin{equation}
    \label{eq:isoH}
    \alpha(g_0,\ldots,g_p) = \alpha^\prime(gg_0g^{-1},\ldots,gg_pg^{-1}).
  \end{equation}
  This looks different from the $g$-action in Eq.~\eqref{eq:g-dot-alpha}, but they actually gives the same isomorphism between the two cohomology groups, which can be calculated either using $G$-valued or $H$-valued cochains.
  The derivation of Eq.~\eqref{eq:isoH} can be found in Ref.~\cite{Brown}.
\end{enumerate}

Now let us explain why Eq.~\eqref{eq:isoH} gives same isomorphism between the two cohomology groups with Eq.~\eqref{eq:g-dot-alpha}.
First, we show that the $H$-valued cochains and $G$-valued cochains form the same cohomology group.
On one hand, by limiting the group elements of a $G$-valued cochain in $H$, we can map each $G$-valued cochain to a $H$-valued cochain.
Apparently, (i) the resulted $H$-valued cohain is homogeneous under $H$ and (ii) this map commutes with the boundary map $d$.
On the other hand, we can map each $H$-valued cochain back to a $G$-valued cochain.
We choose a representative for each right coset $H \backslash G$.
For a given $g_i \in G$, we denote the representative of the belonging coset as $\bar{g}_i$.
Then we define the map $\rho:\ G\to H$ as $ \rho(g_i) = g_i \bar{g}_i^{-1}$.
$\rho$ introduces a map from $H$-valued cohains to $G$-valued cochains
\begin{equation}
(\rho \alpha) (g_0, \cdots, g_p) = \alpha(\rho(g_0),\cdots, \rho(g_p)).
\end{equation}
It is direct to verify that (i) $\rho \alpha$ is homogeneous under $H$, and (ii) the boundary map $d$ commutes with $\rho$.
Therefore the cohomology groups formed by $\rho \alpha$ and $\alpha$ are isomorphic.
We also introduce the map $\rho^\prime:\ G \to g H g^{-1}$ as $\rho^\prime(g_i) = g_i (g \bar{g}_i)^{-1}$, by choosing the representative of the right coset $g H g^{-1} \backslash G$ $g_i$ belonging to as $g\bar{g}_i$.
Similarly, $\rho^\prime$ introduces a map from $gHg^{-1}$-valued cochains, $\alpha^\prime$, to $G$-valued cochains, $\rho^\prime \alpha^\prime$, and, the cohomology groups formed by $\alpha^\prime$ and $\rho^\prime \alpha^\prime$ are isomorphic.
Then it is direct to show that Eq.~\eqref{eq:isoH} implies
\begin{equation}
(\rho^\prime \alpha^\prime) (g_0,\cdots,g_p) = (\rho \alpha) (g^{-1}g_0, \cdots, g^{-1}g_p),
\end{equation}
which is equivalent with Eq.~\eqref{eq:g-dot-alpha}.

\section{Transfer maps}
\label{app:transfer}

In this appendix, we introduce the transfer map between the cohomology groups of two groups $H$ and $G$, where $H$ is a subgroup of $G$.
This transfer map can be used to compute the $d_1$ map in Secs.~\ref{sec:gluing}.
A detailed introduction to the transfer map can be found in Sec.~3.9 of Ref.~\cite{Brown}.

To describe the transfer map, it is convenient to use the $G$-valued cochains.
Consider a $p$-cochain $\alpha$ valued in $G$ but invariant in $H$.
Since it is not invariant under $G$, acting with an element $g\in G$ will generate a different cochain $g\cdot\alpha$, as defined in Appendix~\ref{app:gaction}.
Hence, we use the following summation to obtain a $G$-invariant cochain, which we denote by $\overline\tr$,
\begin{equation}
  \label{eq:transfer-app}
  \overline\tr^G_H\alpha = \sum_{g\in G/H}g\cdot\alpha.
\end{equation}
It is easy to check that the $\overline\tr$ map commutes with the coboundary operator $d$.
Therefore, it induces a homomorphism from $H^p[H,\uone_T]$ to $H^p[G,\uone_T]$, which is known as the transfer map and is denote by $\tr^G_H$.
The explicit form of $\tr^G_H$ is the following,
\begin{equation}
  \label{eq:transfer-app2}
  \tr^G_H[\alpha] = \sum_{g\in G/H}g\cdot[\alpha].
\end{equation}

Using the transfer map in \eqref{eq:transfer-app2}, one can rewrite the $d_1$ map in Eq.~\eqref{eq:defd1} in the following compact form,
\begin{equation}
  \label{eq:d1-transf}
  (d_1[\psi])|_\tau
  = \sum_{\sigma\in Y_p/G}
  \sum_{g\in G_\tau\setminus G/G_\sigma}
  \left<\partial\tau|g\sigma\right>
  \tr^{G_\tau}_{gG_\sigma g^{-1}}\left(g\cdot[\psi|_\sigma]\right).
\end{equation}
Similarly, the $\overline{tr}$ map in \eqref{eq:transfer-app} can be used to express Eq.~\eqref{eq:partial} in the following form,
\begin{equation}
  \label{eq:partial-transf}
  (\partial\psi)|_\tau
  = \sum_{\sigma\in Y_p/G}
  \sum_{g\in G_\tau\setminus G/G_\sigma}
  \left<\partial\tau|g\sigma\right>
  \overline\tr^{G_\tau}_{gG_\sigma g^{-1}}\left(g\cdot\psi|_\sigma\right).
\end{equation}
\section{Spectral sequence and equivariant cohomology}
\label{app:ss}

In this appendix, we prove that for bosonic SPTs, the TCS classification we compute using the algorithm presented in this work, especially in Sec.~\ref{sec:full}, exactly reproduces the result of Ref.~\cite{ThorngrenElse2018}.
We first briefly review the results of Ref.~\cite{ThorngrenElse2018}, which state that the classification is given by an equivariant cohomology.
Following Chapter 7 of Ref.~\cite{Brown}, we point out that the equivariant cohomology can be expressed as the cohomology group of a double cochain complex, which allows it to be computed by constructing a spectral sequence.
We then construct the equivariant cohomology using the dual complex of $Y$ as the topological space, and prove that the spectral sequence obtained in this way exactly agrees with the TCS computation presented in Sec.~\ref{sec:full}.
Mathematical details of our proof can be found in Ref.~\cite{Brown}, especially Chapter 7.

\citet{ThorngrenElse2018} had shown that the classification of $d$-dimensional TCS protected by symmetry group $G$ (with the exception of so-called ``beyond-group-cohomology'' states) can be computed using the following group-cohomology formula,
\begin{equation}
  \label{eq:appss:TE}
  H^{d+1}[G, \uone_{PT}],
\end{equation}
where $\uone_{PT}$ denotes the nontrivial action $G$ has on the $\uone$ coefficients.
Furthermore, this group cohomology can be computed as an equivariant cohomology, using a $G$-complex $X$~\cite{ThorngrenElse2018},
\begin{equation}
  \label{eq:H=HX}
  H^{d+1}[G, \uone_{PT}]\simeq H^{d+1}_G[X, \uone_{PT}],
\end{equation}
provided that $X$ is topologically trivial, $X\simeq\text{pt}$.
A formal prove of this can be found in Proposition (7.3) of Ref.~\cite{Brown}.

\begin{figure}
  \includegraphics{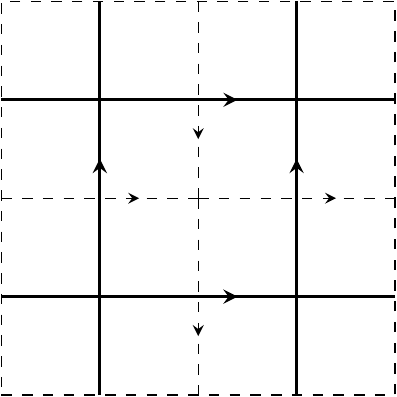}
  \caption{The duality between the two $G$-complexes $X$ and $Y$. The 1-cells in $Y$ and $X$ are represented by solid and dashed lines, respectively. The orientations on the 1-cells in $X$ and $Y$ are related by the right-hand rule.}
  \label{fig:dualXY}
\end{figure}

Here, we take $X$ to be the dual complex of the $G$-complex $Y$ we constructed in Sec.~\ref{sec:simplified}.
As illustrated in Fig.~\ref{fig:dualXY}, the $p$-cells of $Y$ is in one-to-one correspondence to the $(d-p)$-cells of $X$, $Y_p\simeq X_{d-p}$, where $d$ denotes the dimensionality of $Y$.
We denote the cells in $Y$ using greek letters without bars.
In particular, for a $p$-cell $\bar\sigma\in Y_p$, we denote its dual $(d-p)$-cell as $\sigma\in X_{d-p}$.


As shown in Fig.~~\ref{fig:dualXY}, the orientation of cells on $Y$ naturally induces an orientation of cells on $X$.
For example, in 2D, the orientation on 1-cells in $Y$ and dual 1-cells in $X$ are related by the right-hand-rule.
In particular, the relation between the orientations on $X$ and $Y$ depends on the orientation of the space $\mathbb R^n$.
Hence, this relation is reversed under improper symmetry operations in $SG$, such as mirror reflections and glide symmetries.
Recall that the orientation on $Y$ is invariant under $G_\sigma$ because the actions are pointwise.
This implies that its dual $\bar\sigma$ transforms as the following under $G_\sigma$:
\begin{equation}
\label{eq:gsigmaX}
g\sigma = \rho_P(g)\sigma, \forall g\in G_\sigma.
\end{equation}

Since $X$ is the same as $\mathbb R^d$, which is contractible, Eq.~\eqref{eq:H=HX} implies that the SPT classification can be computed using the equivariant cohomology $H^{d+1}_G[X,\uone_{PT}]$.
Following Sec.~7.7 of Ref.~\cite{Brown}, this equivariant cohomology can be expressed as the cohomology group of a double cochain complex, which can then be computed using a spectral sequence.
As shown in Eq.~(7.10) in Ref.~\cite{Brown}, the $E_1$ page of this spectral sequence is given by
\begin{equation}
\label{eq:E1X}
E^{pq}_1
=\bigoplus_{\bar\sigma\in X_p/G}
H^q[G_\sigma,\uone_T].
\end{equation}
Here, the coefficient module in the r.h.s becomes $\uone_T$ instead of $\uone_{PT}$ in Eq.~\eqref{eq:H=HX}, because the symmetry action in Eq.~\eqref{eq:gsigmaX} cancels the $P$-action in $\uone_{PT}$.
Using the duality $X_p\simeq Y_{d-p}$ and the fact that $H^q[G_\sigma, \uone_T]$ classifies the SPT phases $\Phi^{q-1}(G_\sigma)$, we see that $E^{pq}_1$ is the same as the first-page module $E^{q-1}_{d-p,1}$ in Eq.~\eqref{eq:E1}.
Moreover, the two coboundary operations in the double cochain complex is given by $d_1=d$, the coboundary operation of the cochains, and the dual of the $\partial$ map defined in Eq.~\eqref{eq:partial}.
Hence, more generally, the $E^q_{p,r}$ modules computed in Sec.~\ref{sec:full} are the same as the $r$-th page modules $E^{d-p,q+1}_r$ in the spectral sequence \eqref{eq:E1X}.
Therefore, using the result of Eq.~(7.10) in Ref.~\cite{Brown}, we can prove that the TCS classification computed in the main text exactly reproduces the group-cohomology classification in Ref.~\cite{ThorngrenElse2018},
\begin{equation}
  \label{eq:specseq}
  E^q_{p,r}\Rightarrow H^{q-p+d+1}[G,\uone_{PT}].
\end{equation}

\section{Proof of trivial $d_2$ maps and trivial group extensions when $G=SG\times G_0$.}
\label{app:trivial}

In this appendix, we prove that for 2D and 3D bosonic TCSs, the $d_2$ maps and group-extension problems are all trivial.
Consequently, the simplified formula \eqref{eq:sum1} in Sec.~\ref{sec:simplified} produces the correct results for these cases.

We begin by pointing out some obvious cases where the two problems are trivial.
First, recall that the $d_2$ map reduces the spatial dimension by $2$: it maps a TCS pattern on $p$-cells to an anomaly pattern on $(p-2)$-cells.
Therefore, the $d_2$ map can only be nontrivial if $p\geq2$.
Second, recall that the group-extension problem arises when $n$ copies of a $d_b=p$ TCS pattern becomes trivial on $p$-cells but nontrivial on lower-dimensional cells.
Since we don't view $d_b=0$ TCSs as nontrivial states in \eqref{eq:sum1}, the group-extension problem may only exist for $p\geq2$ as well.

Last but not least, we argue that these two problems do not arise for the top building-block dimension $d_b=d$.
This is because by construction, the $d$-cells are the AUs of the space group (or the wallpaper group in 2D), where the local symmetry group is just $G_\sigma=G_0$.
Therefore, the local SPT states we can decorate on the $d$-cells are simply $d$-dimensional SPT states protected by $G_0$ alone, which are described by $\Phi^d(G_0)$.
If every AU is filled with the same element $[\alpha]\in\Phi^d(G_0)$,
the entire system is nothing but the 3D strong SPT phase $[\alpha]$.
When $SG$ has no improper symmetries (like mirror reflections and glide planes), it is well known that $\alpha$ is always compatible with $SG$.
When $SG$ has improper symmetries, $\alpha$ is compatible with $SG$ if and only if $[\alpha]\sim-[\alpha]$.
In either cases, whether the assembly $[\alpha]$ satisfies the no-open-edge conditions can be figured out from $d_1$ map, and the higher-page maps and group-extensions are all trivial.

In summary, ruling out all these obvious cases, we see that the $d_2$ maps and the group-extension problems can only be nontrivial for the case of $p=2$ in 3D.

We now focus on this case, and consider an assembly $[\psi]$ in $E^2_{2,\infty}$, representing 3D TCS states constructed by decorating 2D SPT states on 2-cells.
The building blocks $\psi_2$ contains 3-cocycles $\psi_2|_\sigma$ attached to 2-cells.
In order to compute the subleading terms, we need to express these 3-cocycles using $G$-valued cochains.
The main result of this section is to show that, when $G$ is a direct product of a 3D space group $SG$ and an onsite symmetry group $G_0$, there is a canonical way to lift these 3-cocycles of $G_\sigma$ to $G$-valued cochains, which happen to be not only invariant in $G_\sigma$, but also invariant in $G$.
As a result, one can choose a vanishing subleading term $\psi_1=0$.
This implies that the map $d_2$ vanishes.
It also implies that the group extension of combining $E^2_{2,\infty}$ and $E^2_{2,\infty}$ is trivial.

Consider a 3-cocycle $\psi_2|\sigma$ decorated on a 2-cell $\sigma$.
As discussed in Appendix~\ref{app:spt-review}, it can be lifted to a $G$-valued cochain $\phi_\sigma^\ast\psi|_\sigma$, using a map $\phi_\sigma:G\rightarrow G_\sigma$,
\begin{equation}
\phi_\sigma^\ast\psi|_\sigma(g_0,g_1,g_2,g_3)
=\psi_\sigma(\phi_\sigma(g_0),\phi_\sigma(g_1),\phi_\sigma(g_2),\phi_\sigma(g_3)).
\end{equation}
In particular, we construct the following group homomorphism $\phi_\sigma:G\rightarrow G_\sigma$.
Since $G=SG\times G_0$, we can also decompose $G_\sigma$ as $G_\sigma = SG_\sigma\times G_0$, where $SG_\sigma = G_\sigma\cap SG$ denotes the subgroup of $SG$ that keeps $\sigma$ invariant.
Notice that for two-cells in 3D, there are only two possibilities of $SG_\sigma$: $SG_\sigma=\mathbb Z_2$ if $\sigma$ is a mirror plane, and $SG_\sigma=\mathbb Z_1$ (the trivial group) if it is not.
An element $g\in G$ can be expressed as $g=g_1g_2$, where $g_1\in SG$ and $g_2\in G_0$, respectively.
For a mirror plane, we construct the following map,
\begin{equation}
\phi_\sigma(g)=\phi_\sigma(g_1g_2)=\begin{cases}
g_2, &\rho_M(g_1)=1;\\
m_\sigma g_2, &\rho_M(g_1)=-1.
\end{cases}
\end{equation}
Here, $m_\sigma$ denotes the nontrivial element of $SG_\sigma=\mathbb Z_2$.
For a non-mirror plane, we simply define $\phi_\sigma = g_2$.

It is straightforward to check that the map $\phi_\sigma$ defined above is a group homomorphism, $\phi_\sigma(g_1g_2)=\phi_\sigma(g_1)\phi_\sigma(g_2)$.
Furthermore, it satisfies the condition that its restriction to $G_\sigma$ is identity.
Using this properties, we can show that $\phi_\sigma^\ast\psi|_\sigma$ is actually invariant under $G$ actions:
\begin{align*}
\phi_\sigma^\ast\psi|_\sigma(gg_0,\ldots,gg_3)
&=\psi|_\sigma(\phi_\sigma(g)\phi_\sigma(g_0),\ldots,\phi_\sigma(g)\phi_\sigma(g_3))\\
&=\rho_T(\phi_\sigma(g))\psi|_\sigma(\phi_\sigma(g_0),\ldots,\phi_\sigma(g_3)).
\end{align*}
Therefore, this map defines a canonical way to embed $H^3[G_\sigma,\uone_T]$ into $H^3[G,\uone_T]$ as a subset.
This means that when calculating the $\partial$ map in Eq.~\eqref{eq:partial}, one can ignore the $g$-action and just simply adds the cochains up.
One can then explicitly check that, in such simple cases, if $[\partial\psi_2]$, there is a choice of $\partial\psi_2$ such that $\partial_1=0$.
As a result, the subtleties of $d_2$ map and group-extension won't arise.

\section{2-cells and 1-cells in space group $P\bar4n2$}\label{app:P-4n2}
Here we give the definitions of the 2-cells and 1-cells shown in Fig. \ref{fig:nonLC}.
For simplicity, we assume the lattice constant as $1$.
The 2-cells and 1-cells in one unit cell are defined by their vertices as
\begin{enumerate}
\item $\sigma_1$: $(00\frac14)$-$(00\frac34)$-$(\frac120\frac34)$-$(\frac120\frac14)$.
\item $\sigma_2$: $(0\frac12\frac14)$-$(0\frac12\frac34)$-$(\frac12\frac12\frac34)$-$(\frac12\frac12\frac14)$.
\item $\sigma_3$: $(00\frac14)$-$(0\frac12\frac14)$-$(\frac12\frac12\frac14)$-$(\frac120\frac14)$.
\item $\sigma_4$: $(00\frac34)$-$(0\frac12\frac34)$-$(\frac12\frac12\frac34)$-$(\frac120\frac34)$.
\item $\tau_1$: $(0\frac12\frac14)$-$(0\frac12\frac34)$.
\item $\tau_2$: $(00\frac34)$-$(\frac120\frac34)$.
\item $\tau_3$: $(00\frac14)$-$(00\frac34)$.
\item $\tau_4$: $(00\frac34)$-$(0\frac12\frac34)$.
\item $\tau_5$: $(\frac12\frac12\frac14)$-$(\frac12\frac12\frac34)$.
\end{enumerate}
The equivalent cells can be generated by applying the space group operations on these cells.
The generators of $P\bar4n2$ are
\begin{enumerate}
\item Translation $t_x$: $(x,y,z)\to(x+1,y,z)$.
\item Translation $t_y$: $(x,y,z)\to(x,y+1,z)$.
\item Translation $t_z$: $(x,y,z)\to(x,y,z+1)$.
\item Roto-reflection $S_4$: $(x,y,z)\to (-y,x,-z)$.
\item Glide mirror $n$: $(x,y,z)\to (-x+\frac12,y+\frac12,z+\frac12)$.
\end{enumerate}

Since the building block used in Section \ref{sec:examples} is $\mathbb{Z}_2$ 2D SPT state, we only need to match the boundary anomalies up to mod 2.
Thus, in Eq. \eqref{eq:sigma1_boundary} to \eqref{eq:sigma4_boundary}, we have omitted the boundary terms with even multiplicities.

\section{Full classification for bosonic TCS}
\label{app:wgsg}
\renewcommand\arraystretch{1.2}

\begin{longtable}{ c C{0.225\linewidth}C{0.225\linewidth}C{0.225\linewidth} }
\caption[Classifications of 2D SPT phases.]{Classifications of two-dimensional SPT phases. \label{tab:2D}}\\
\hline
$G_0$  & $E^2_{2,\infty}$ & $E^1_{1,\infty}$ & $E^0_{0,\infty}$ \\
\hline\hline\hline
\endfirsthead
\multicolumn{ 4 }{c} {{\tablename\ \thetable{} (Continued.)}}\\
\hline
$G_0$  & $E^2_{2,\infty}$ & $E^1_{1,\infty}$ & $E^0_{0,\infty}$ \\
\hline\hline\hline
\endhead
\hline \\
\endfoot
\hline \hline \hline
\endlastfoot
\hline
\noalign{\vskip0.03cm}
\multicolumn{ 4 }{c}{Wallpaper group \#1 $p1$  } \\
\hline
$\varnothing$ & $ $ & $ $ & $ $ \\
$Z_2^T$ & $ $ & $\mathbb{Z}_{2}^{2} $ & $ $ \\
$Z_2$ & $\mathbb{Z}_{2} $ & $ $ & $\mathbb{Z}_{2} $ \\
$Z_2 \times Z_2^T$ & $\mathbb{Z}_{2}^{2} $ & $\mathbb{Z}_{2}^{4} $ & $\mathbb{Z}_{2} $ \\
$Z_4$ & $\mathbb{Z}_{4} $ & $ $ & $\mathbb{Z}_{4} $ \\
$Z_6$ & $\mathbb{Z}_{6} $ & $ $ & $\mathbb{Z}_{6} $ \\
$Z_2\times Z_2$ & $\mathbb{Z}_{2}^{3} $ & $\mathbb{Z}_{2}^{2} $ & $\mathbb{Z}_{2}^{2} $ \\
$Z_4\times Z_4$ & $\mathbb{Z}_{4}^{3} $ & $\mathbb{Z}_{4}^{2} $ & $\mathbb{Z}_{4}^{2} $ \\
$Z_6\times Z_6$ & $\mathbb{Z}_{6}^{3} $ & $\mathbb{Z}_{6}^{2} $ & $\mathbb{Z}_{6}^{2} $ \\
$U(1)$ & $\mathbb{Z} $ & $ $ & $\mathbb{Z} $ \\
$SU(2)$ & $\mathbb{Z} $ & $ $ & $ $ \\
$SO(3)$ & $\mathbb{Z} $ & $\mathbb{Z}_{2}^{2} $ & $ $ \\
\hline
\noalign{\vskip0.03cm}
\multicolumn{ 4 }{c}{Wallpaper group \#2 $p2$  } \\
\hline
$\varnothing$ & $ $ & $ $ & $\mathbb{Z}_{2}^{4} $ \\
$Z_2^T$ & $ $ & $\mathbb{Z}_{2}^{3} $ & $\mathbb{Z}_{2}^{4} $ \\
$Z_2$ & $\mathbb{Z}_{2} $ & $ $ & $\mathbb{Z}_{2}^{8} $ \\
$Z_2 \times Z_2^T$ & $\mathbb{Z}_{2}^{2} $ & $\mathbb{Z}_{2}^{6} $ & $\mathbb{Z}_{2}^{8} $ \\
$Z_4$ & $\mathbb{Z}_{4} $ & $ $ & $\mathbb{Z}_{2}^{7} \times\mathbb{Z}_{4} $ \\
$Z_6$ & $\mathbb{Z}_{6} $ & $ $ & $\mathbb{Z}_{2}^{7} \times\mathbb{Z}_{6} $ \\
$Z_2\times Z_2$ & $\mathbb{Z}_{2}^{3} $ & $\mathbb{Z}_{2}^{3} $ & $\mathbb{Z}_{2}^{12} $ \\
$Z_4\times Z_4$ & $\mathbb{Z}_{4}^{3} $ & $\mathbb{Z}_{2}^{3} $ & $\mathbb{Z}_{2}^{10} \times\mathbb{Z}_{4}^{2} $ \\
$Z_6\times Z_6$ & $\mathbb{Z}_{6}^{3} $ & $\mathbb{Z}_{2}^{3} $ & $\mathbb{Z}_{2}^{10} \times\mathbb{Z}_{6}^{2} $ \\
$U(1)$ & $\mathbb{Z} $ & $ $ & $\mathbb{Z} \times\mathbb{Z}_{2}^{7} $ \\
$SU(2)$ & $\mathbb{Z} $ & $ $ & $\mathbb{Z}_{2}^{4} $ \\
$SO(3)$ & $\mathbb{Z} $ & $\mathbb{Z}_{2}^{3} $ & $\mathbb{Z}_{2}^{4} $ \\
\hline
\noalign{\vskip0.03cm}
\multicolumn{ 4 }{c}{Wallpaper group \#3 $pm$  } \\
\hline
$\varnothing$ & $ $ & $ $ & $\mathbb{Z}_{2}^{2} $ \\
$Z_2^T$ & $ $ & $\mathbb{Z}_{2}^{5} $ & $\mathbb{Z}_{2}^{2} $ \\
$Z_2$ & $\mathbb{Z}_{2} $ & $\mathbb{Z}_{2}^{2} $ & $\mathbb{Z}_{2}^{4} $ \\
$Z_2 \times Z_2^T$ & $\mathbb{Z}_{2}^{2} $ & $\mathbb{Z}_{2}^{10} $ & $\mathbb{Z}_{2}^{4} $ \\
$Z_4$ & $\mathbb{Z}_{2} $ & $\mathbb{Z}_{2}^{2} $ & $\mathbb{Z}_{2}^{3} \times\mathbb{Z}_{4} $ \\
$Z_6$ & $\mathbb{Z}_{2} $ & $\mathbb{Z}_{2}^{2} $ & $\mathbb{Z}_{2}^{3} \times\mathbb{Z}_{6} $ \\
$Z_2\times Z_2$ & $\mathbb{Z}_{2}^{3} $ & $\mathbb{Z}_{2}^{7} $ & $\mathbb{Z}_{2}^{6} $ \\
$Z_4\times Z_4$ & $\mathbb{Z}_{2}^{3} $ & $\mathbb{Z}_{2}^{6} \times\mathbb{Z}_{4} $ & $\mathbb{Z}_{2}^{4} \times\mathbb{Z}_{4}^{2} $ \\
$Z_6\times Z_6$ & $\mathbb{Z}_{2}^{3} $ & $\mathbb{Z}_{2}^{6} \times\mathbb{Z}_{6} $ & $\mathbb{Z}_{2}^{4} \times\mathbb{Z}_{6}^{2} $ \\
$U(1)$ & $ $ & $ $ & $\mathbb{Z} \times\mathbb{Z}_{2}^{3} $ \\
$SU(2)$ & $ $ & $ $ & $\mathbb{Z}_{2}^{2} $ \\
$SO(3)$ & $ $ & $\mathbb{Z}_{2}^{3} $ & $\mathbb{Z}_{2}^{2} $ \\
\hline
\noalign{\vskip0.03cm}
\multicolumn{ 4 }{c}{Wallpaper group \#4 $pg$  } \\
\hline
$\varnothing$ & $ $ & $ $ & $ $ \\
$Z_2^T$ & $ $ & $\mathbb{Z}_{2}^{2} $ & $ $ \\
$Z_2$ & $\mathbb{Z}_{2} $ & $ $ & $\mathbb{Z}_{2} $ \\
$Z_2 \times Z_2^T$ & $\mathbb{Z}_{2}^{2} $ & $\mathbb{Z}_{2}^{4} $ & $\mathbb{Z}_{2} $ \\
$Z_4$ & $\mathbb{Z}_{2} $ & $ $ & $\mathbb{Z}_{4} $ \\
$Z_6$ & $\mathbb{Z}_{2} $ & $ $ & $\mathbb{Z}_{6} $ \\
$Z_2\times Z_2$ & $\mathbb{Z}_{2}^{3} $ & $\mathbb{Z}_{2}^{2} $ & $\mathbb{Z}_{2}^{2} $ \\
$Z_4\times Z_4$ & $\mathbb{Z}_{2}^{3} $ & $\mathbb{Z}_{2} \times\mathbb{Z}_{4} $ & $\mathbb{Z}_{4}^{2} $ \\
$Z_6\times Z_6$ & $\mathbb{Z}_{2}^{3} $ & $\mathbb{Z}_{2} \times\mathbb{Z}_{6} $ & $\mathbb{Z}_{6}^{2} $ \\
$U(1)$ & $ $ & $ $ & $\mathbb{Z} $ \\
$SU(2)$ & $ $ & $ $ & $ $ \\
$SO(3)$ & $ $ & $\mathbb{Z}_{2}^{2} $ & $ $ \\
\hline
\noalign{\vskip0.03cm}
\multicolumn{ 4 }{c}{Wallpaper group \#5 $cm$  } \\
\hline
$\varnothing$ & $ $ & $ $ & $\mathbb{Z}_{2} $ \\
$Z_2^T$ & $ $ & $\mathbb{Z}_{2}^{3} $ & $\mathbb{Z}_{2} $ \\
$Z_2$ & $\mathbb{Z}_{2} $ & $\mathbb{Z}_{2} $ & $\mathbb{Z}_{2}^{2} $ \\
$Z_2 \times Z_2^T$ & $\mathbb{Z}_{2}^{2} $ & $\mathbb{Z}_{2}^{6} $ & $\mathbb{Z}_{2}^{2} $ \\
$Z_4$ & $\mathbb{Z}_{2} $ & $\mathbb{Z}_{2} $ & $\mathbb{Z}_{2} \times\mathbb{Z}_{4} $ \\
$Z_6$ & $\mathbb{Z}_{2} $ & $\mathbb{Z}_{2} $ & $\mathbb{Z}_{2} \times\mathbb{Z}_{6} $ \\
$Z_2\times Z_2$ & $\mathbb{Z}_{2}^{3} $ & $\mathbb{Z}_{2}^{4} $ & $\mathbb{Z}_{2}^{3} $ \\
$Z_4\times Z_4$ & $\mathbb{Z}_{2}^{3} $ & $\mathbb{Z}_{2}^{3} \times\mathbb{Z}_{4} $ & $\mathbb{Z}_{2} \times\mathbb{Z}_{4}^{2} $ \\
$Z_6\times Z_6$ & $\mathbb{Z}_{2}^{3} $ & $\mathbb{Z}_{2}^{3} \times\mathbb{Z}_{6} $ & $\mathbb{Z}_{2} \times\mathbb{Z}_{6}^{2} $ \\
$U(1)$ & $ $ & $ $ & $\mathbb{Z} \times\mathbb{Z}_{2} $ \\
$SU(2)$ & $ $ & $ $ & $\mathbb{Z}_{2} $ \\
$SO(3)$ & $ $ & $\mathbb{Z}_{2}^{2} $ & $\mathbb{Z}_{2} $ \\
\hline
\noalign{\vskip0.03cm}
\multicolumn{ 4 }{c}{Wallpaper group \#6 $p2mm$  } \\
\hline
$\varnothing$ & $ $ & $ $ & $\mathbb{Z}_{2}^{8} $ \\
$Z_2^T$ & $ $ & $\mathbb{Z}_{2}^{8} $ & $\mathbb{Z}_{2}^{8} $ \\
$Z_2$ & $\mathbb{Z}_{2} $ & $\mathbb{Z}_{2}^{4} $ & $\mathbb{Z}_{2}^{12} $ \\
$Z_2 \times Z_2^T$ & $\mathbb{Z}_{2}^{2} $ & $\mathbb{Z}_{2}^{16} $ & $\mathbb{Z}_{2}^{12} $ \\
$Z_4$ & $\mathbb{Z}_{2} $ & $\mathbb{Z}_{2}^{4} $ & $\mathbb{Z}_{2}^{11} \times\mathbb{Z}_{4} $ \\
$Z_6$ & $\mathbb{Z}_{2} $ & $\mathbb{Z}_{2}^{4} $ & $\mathbb{Z}_{2}^{11} \times\mathbb{Z}_{6} $ \\
$Z_2\times Z_2$ & $\mathbb{Z}_{2}^{3} $ & $\mathbb{Z}_{2}^{12} $ & $\mathbb{Z}_{2}^{16} $ \\
$Z_4\times Z_4$ & $\mathbb{Z}_{2}^{3} $ & $\mathbb{Z}_{2}^{12} $ & $\mathbb{Z}_{2}^{14} \times\mathbb{Z}_{4}^{2} $ \\
$Z_6\times Z_6$ & $\mathbb{Z}_{2}^{3} $ & $\mathbb{Z}_{2}^{12} $ & $\mathbb{Z}_{2}^{14} \times\mathbb{Z}_{6}^{2} $ \\
$U(1)$ & $ $ & $ $ & $\mathbb{Z} \times\mathbb{Z}_{2}^{11} $ \\
$SU(2)$ & $ $ & $ $ & $\mathbb{Z}_{2}^{8} $ \\
$SO(3)$ & $ $ & $\mathbb{Z}_{2}^{4} $ & $\mathbb{Z}_{2}^{8} $ \\
\hline
\noalign{\vskip0.03cm}
\multicolumn{ 4 }{c}{Wallpaper group \#7 $p2mg$  } \\
\hline
$\varnothing$ & $ $ & $ $ & $\mathbb{Z}_{2}^{3} $ \\
$Z_2^T$ & $ $ & $\mathbb{Z}_{2}^{4} $ & $\mathbb{Z}_{2}^{3} $ \\
$Z_2$ & $\mathbb{Z}_{2} $ & $\mathbb{Z}_{2} $ & $\mathbb{Z}_{2}^{6} $ \\
$Z_2 \times Z_2^T$ & $\mathbb{Z}_{2}^{2} $ & $\mathbb{Z}_{2}^{8} $ & $\mathbb{Z}_{2}^{6} $ \\
$Z_4$ & $\mathbb{Z}_{2} $ & $\mathbb{Z}_{2} $ & $\mathbb{Z}_{2}^{5} \times\mathbb{Z}_{4} $ \\
$Z_6$ & $\mathbb{Z}_{2} $ & $\mathbb{Z}_{2} $ & $\mathbb{Z}_{2}^{5} \times\mathbb{Z}_{6} $ \\
$Z_2\times Z_2$ & $\mathbb{Z}_{2}^{3} $ & $\mathbb{Z}_{2}^{5} $ & $\mathbb{Z}_{2}^{9} $ \\
$Z_4\times Z_4$ & $\mathbb{Z}_{2}^{3} $ & $\mathbb{Z}_{2}^{5} $ & $\mathbb{Z}_{2}^{7} \times\mathbb{Z}_{4}^{2} $ \\
$Z_6\times Z_6$ & $\mathbb{Z}_{2}^{3} $ & $\mathbb{Z}_{2}^{5} $ & $\mathbb{Z}_{2}^{7} \times\mathbb{Z}_{6}^{2} $ \\
$U(1)$ & $ $ & $ $ & $\mathbb{Z} \times\mathbb{Z}_{2}^{5} $ \\
$SU(2)$ & $ $ & $ $ & $\mathbb{Z}_{2}^{3} $ \\
$SO(3)$ & $ $ & $\mathbb{Z}_{2}^{3} $ & $\mathbb{Z}_{2}^{3} $ \\
\hline
\noalign{\vskip0.03cm}
\multicolumn{ 4 }{c}{Wallpaper group \#8 $p2gg$  } \\
\hline
$\varnothing$ & $ $ & $ $ & $\mathbb{Z}_{2}^{2} $ \\
$Z_2^T$ & $ $ & $\mathbb{Z}_{2}^{2} $ & $\mathbb{Z}_{2}^{2} $ \\
$Z_2$ & $\mathbb{Z}_{2} $ & $ $ & $\mathbb{Z}_{2}^{4} $ \\
$Z_2 \times Z_2^T$ & $\mathbb{Z}_{2}^{2} $ & $\mathbb{Z}_{2}^{4} $ & $\mathbb{Z}_{2}^{4} $ \\
$Z_4$ & $\mathbb{Z}_{2} $ & $ $ & $\mathbb{Z}_{2}^{3} \times\mathbb{Z}_{4} $ \\
$Z_6$ & $\mathbb{Z}_{2} $ & $ $ & $\mathbb{Z}_{2}^{3} \times\mathbb{Z}_{6} $ \\
$Z_2\times Z_2$ & $\mathbb{Z}_{2}^{3} $ & $\mathbb{Z}_{2}^{2} $ & $\mathbb{Z}_{2}^{6} $ \\
$Z_4\times Z_4$ & $\mathbb{Z}_{2}^{3} $ & $\mathbb{Z}_{2}^{2} $ & $\mathbb{Z}_{2}^{4} \times\mathbb{Z}_{4}^{2} $ \\
$Z_6\times Z_6$ & $\mathbb{Z}_{2}^{3} $ & $\mathbb{Z}_{2}^{2} $ & $\mathbb{Z}_{2}^{4} \times\mathbb{Z}_{6}^{2} $ \\
$U(1)$ & $ $ & $ $ & $\mathbb{Z} \times\mathbb{Z}_{2}^{3} $ \\
$SU(2)$ & $ $ & $ $ & $\mathbb{Z}_{2}^{2} $ \\
$SO(3)$ & $ $ & $\mathbb{Z}_{2}^{2} $ & $\mathbb{Z}_{2}^{2} $ \\
\hline
\noalign{\vskip0.03cm}
\multicolumn{ 4 }{c}{Wallpaper group \#9 $c2mm$  } \\
\hline
$\varnothing$ & $ $ & $ $ & $\mathbb{Z}_{2}^{5} $ \\
$Z_2^T$ & $ $ & $\mathbb{Z}_{2}^{5} $ & $\mathbb{Z}_{2}^{5} $ \\
$Z_2$ & $\mathbb{Z}_{2} $ & $\mathbb{Z}_{2}^{2} $ & $\mathbb{Z}_{2}^{8} $ \\
$Z_2 \times Z_2^T$ & $\mathbb{Z}_{2}^{2} $ & $\mathbb{Z}_{2}^{10} $ & $\mathbb{Z}_{2}^{8} $ \\
$Z_4$ & $\mathbb{Z}_{2} $ & $\mathbb{Z}_{2}^{2} $ & $\mathbb{Z}_{2}^{7} \times\mathbb{Z}_{4} $ \\
$Z_6$ & $\mathbb{Z}_{2} $ & $\mathbb{Z}_{2}^{2} $ & $\mathbb{Z}_{2}^{7} \times\mathbb{Z}_{6} $ \\
$Z_2\times Z_2$ & $\mathbb{Z}_{2}^{3} $ & $\mathbb{Z}_{2}^{7} $ & $\mathbb{Z}_{2}^{11} $ \\
$Z_4\times Z_4$ & $\mathbb{Z}_{2}^{3} $ & $\mathbb{Z}_{2}^{7} $ & $\mathbb{Z}_{2}^{9} \times\mathbb{Z}_{4}^{2} $ \\
$Z_6\times Z_6$ & $\mathbb{Z}_{2}^{3} $ & $\mathbb{Z}_{2}^{7} $ & $\mathbb{Z}_{2}^{9} \times\mathbb{Z}_{6}^{2} $ \\
$U(1)$ & $ $ & $ $ & $\mathbb{Z} \times\mathbb{Z}_{2}^{7} $ \\
$SU(2)$ & $ $ & $ $ & $\mathbb{Z}_{2}^{5} $ \\
$SO(3)$ & $ $ & $\mathbb{Z}_{2}^{3} $ & $\mathbb{Z}_{2}^{5} $ \\
\hline
\noalign{\vskip0.03cm}
\multicolumn{ 4 }{c}{Wallpaper group \#10 $p4$  } \\
\hline
$\varnothing$ & $ $ & $ $ & $\mathbb{Z}_{2} \times\mathbb{Z}_{4}^{2} $ \\
$Z_2^T$ & $ $ & $\mathbb{Z}_{2}^{2} $ & $\mathbb{Z}_{2}^{3} $ \\
$Z_2$ & $\mathbb{Z}_{2} $ & $ $ & $\mathbb{Z}_{2}^{4} \times\mathbb{Z}_{4}^{2} $ \\
$Z_2 \times Z_2^T$ & $\mathbb{Z}_{2}^{2} $ & $\mathbb{Z}_{2}^{4} $ & $\mathbb{Z}_{2}^{6} $ \\
$Z_4$ & $\mathbb{Z}_{4} $ & $ $ & $\mathbb{Z}_{2}^{2} \times\mathbb{Z}_{4}^{4} $ \\
$Z_6$ & $\mathbb{Z}_{6} $ & $ $ & $\mathbb{Z}_{2}^{3} \times\mathbb{Z}_{4}^{2} \times\mathbb{Z}_{6} $ \\
$Z_2\times Z_2$ & $\mathbb{Z}_{2}^{3} $ & $\mathbb{Z}_{2}^{2} $ & $\mathbb{Z}_{2}^{7} \times\mathbb{Z}_{4}^{2} $ \\
$Z_4\times Z_4$ & $\mathbb{Z}_{4}^{3} $ & $\mathbb{Z}_{2} \times\mathbb{Z}_{4} $ & $\mathbb{Z}_{2}^{3} \times\mathbb{Z}_{4}^{6} $ \\
$Z_6\times Z_6$ & $\mathbb{Z}_{6}^{3} $ & $\mathbb{Z}_{2}^{2} $ & $\mathbb{Z}_{2}^{5} \times\mathbb{Z}_{4}^{2} \times\mathbb{Z}_{6}^{2} $ \\
$U(1)$ & $\mathbb{Z} $ & $ $ & $\mathbb{Z} \times\mathbb{Z}_{2}^{2} \times\mathbb{Z}_{4}^{3} $ \\
$SU(2)$ & $\mathbb{Z} $ & $ $ & $\mathbb{Z}_{2} \times\mathbb{Z}_{4}^{2} $ \\
$SO(3)$ & $\mathbb{Z} $ & $\mathbb{Z}_{2}^{2} $ & $\mathbb{Z}_{2} \times\mathbb{Z}_{4}^{2} $ \\
\hline
\noalign{\vskip0.03cm}
\multicolumn{ 4 }{c}{Wallpaper group \#11 $p4mm$  } \\
\hline
$\varnothing$ & $ $ & $ $ & $\mathbb{Z}_{2}^{6} $ \\
$Z_2^T$ & $ $ & $\mathbb{Z}_{2}^{6} $ & $\mathbb{Z}_{2}^{6} $ \\
$Z_2$ & $\mathbb{Z}_{2} $ & $\mathbb{Z}_{2}^{3} $ & $\mathbb{Z}_{2}^{9} $ \\
$Z_2 \times Z_2^T$ & $\mathbb{Z}_{2}^{2} $ & $\mathbb{Z}_{2}^{12} $ & $\mathbb{Z}_{2}^{9} $ \\
$Z_4$ & $\mathbb{Z}_{2} $ & $\mathbb{Z}_{2}^{3} $ & $\mathbb{Z}_{2}^{7} \times\mathbb{Z}_{4}^{2} $ \\
$Z_6$ & $\mathbb{Z}_{2} $ & $\mathbb{Z}_{2}^{3} $ & $\mathbb{Z}_{2}^{8} \times\mathbb{Z}_{6} $ \\
$Z_2\times Z_2$ & $\mathbb{Z}_{2}^{3} $ & $\mathbb{Z}_{2}^{9} $ & $\mathbb{Z}_{2}^{12} $ \\
$Z_4\times Z_4$ & $\mathbb{Z}_{2}^{3} $ & $\mathbb{Z}_{2}^{8} \times\mathbb{Z}_{4} $ & $\mathbb{Z}_{2}^{8} \times\mathbb{Z}_{4}^{4} $ \\
$Z_6\times Z_6$ & $\mathbb{Z}_{2}^{3} $ & $\mathbb{Z}_{2}^{9} $ & $\mathbb{Z}_{2}^{10} \times\mathbb{Z}_{6}^{2} $ \\
$U(1)$ & $ $ & $ $ & $\mathbb{Z} \times\mathbb{Z}_{2}^{7} \times\mathbb{Z}_{4} $ \\
$SU(2)$ & $ $ & $ $ & $\mathbb{Z}_{2}^{6} $ \\
$SO(3)$ & $ $ & $\mathbb{Z}_{2}^{3} $ & $\mathbb{Z}_{2}^{6} $ \\
\hline
\noalign{\vskip0.03cm}
\multicolumn{ 4 }{c}{Wallpaper group \#12 $p4gm$  } \\
\hline
$\varnothing$ & $ $ & $ $ & $\mathbb{Z}_{2}^{2} \times\mathbb{Z}_{4} $ \\
$Z_2^T$ & $ $ & $\mathbb{Z}_{2}^{3} $ & $\mathbb{Z}_{2}^{3} $ \\
$Z_2$ & $\mathbb{Z}_{2} $ & $\mathbb{Z}_{2} $ & $\mathbb{Z}_{2}^{4} \times\mathbb{Z}_{4} $ \\
$Z_2 \times Z_2^T$ & $\mathbb{Z}_{2}^{2} $ & $\mathbb{Z}_{2}^{6} $ & $\mathbb{Z}_{2}^{5} $ \\
$Z_4$ & $\mathbb{Z}_{2} $ & $\mathbb{Z}_{2} $ & $\mathbb{Z}_{2}^{2} \times\mathbb{Z}_{4}^{3} $ \\
$Z_6$ & $\mathbb{Z}_{2} $ & $\mathbb{Z}_{2} $ & $\mathbb{Z}_{2}^{3} \times\mathbb{Z}_{4} \times\mathbb{Z}_{6} $ \\
$Z_2\times Z_2$ & $\mathbb{Z}_{2}^{3} $ & $\mathbb{Z}_{2}^{4} $ & $\mathbb{Z}_{2}^{6} \times\mathbb{Z}_{4} $ \\
$Z_4\times Z_4$ & $\mathbb{Z}_{2}^{3} $ & $\mathbb{Z}_{2}^{3} \times\mathbb{Z}_{4} $ & $\mathbb{Z}_{2}^{2} \times\mathbb{Z}_{4}^{5} $ \\
$Z_6\times Z_6$ & $\mathbb{Z}_{2}^{3} $ & $\mathbb{Z}_{2}^{4} $ & $\mathbb{Z}_{2}^{4} \times\mathbb{Z}_{4} \times\mathbb{Z}_{6}^{2} $ \\
$U(1)$ & $ $ & $ $ & $\mathbb{Z} \times\mathbb{Z}_{2}^{2} \times\mathbb{Z}_{4}^{2} $ \\
$SU(2)$ & $ $ & $ $ & $\mathbb{Z}_{2}^{2} \times\mathbb{Z}_{4} $ \\
$SO(3)$ & $ $ & $\mathbb{Z}_{2}^{2} $ & $\mathbb{Z}_{2}^{2} \times\mathbb{Z}_{4} $ \\
\hline
\noalign{\vskip0.03cm}
\multicolumn{ 4 }{c}{Wallpaper group \#13 $p3$  } \\
\hline
$\varnothing$ & $ $ & $ $ & $\mathbb{Z}_{3}^{3} $ \\
$Z_2^T$ & $ $ & $ $ & $ $ \\
$Z_2$ & $\mathbb{Z}_{2} $ & $ $ & $\mathbb{Z}_{2} \times\mathbb{Z}_{3}^{3} $ \\
$Z_2 \times Z_2^T$ & $\mathbb{Z}_{2}^{2} $ & $ $ & $\mathbb{Z}_{2} $ \\
$Z_4$ & $\mathbb{Z}_{4} $ & $ $ & $\mathbb{Z}_{3}^{3} \times\mathbb{Z}_{4} $ \\
$Z_6$ & $\mathbb{Z}_{6} $ & $ $ & $\mathbb{Z}_{3}^{5} \times\mathbb{Z}_{6} $ \\
$Z_2\times Z_2$ & $\mathbb{Z}_{2}^{3} $ & $ $ & $\mathbb{Z}_{2}^{2} \times\mathbb{Z}_{3}^{3} $ \\
$Z_4\times Z_4$ & $\mathbb{Z}_{4}^{3} $ & $ $ & $\mathbb{Z}_{3}^{3} \times\mathbb{Z}_{4}^{2} $ \\
$Z_6\times Z_6$ & $\mathbb{Z}_{6}^{3} $ & $\mathbb{Z}_{3}^{2} $ & $\mathbb{Z}_{3}^{7} \times\mathbb{Z}_{6}^{2} $ \\
$U(1)$ & $\mathbb{Z} $ & $ $ & $\mathbb{Z} \times\mathbb{Z}_{3}^{5} $ \\
$SU(2)$ & $\mathbb{Z} $ & $ $ & $\mathbb{Z}_{3}^{3} $ \\
$SO(3)$ & $\mathbb{Z} $ & $ $ & $\mathbb{Z}_{3}^{3} $ \\
\hline
\noalign{\vskip0.03cm}
\multicolumn{ 4 }{c}{Wallpaper group \#14 $p3m1$  } \\
\hline
$\varnothing$ & $ $ & $ $ & $\mathbb{Z}_{2} $ \\
$Z_2^T$ & $ $ & $\mathbb{Z}_{2}^{2} $ & $\mathbb{Z}_{2} $ \\
$Z_2$ & $\mathbb{Z}_{2} $ & $\mathbb{Z}_{2} $ & $\mathbb{Z}_{2}^{2} $ \\
$Z_2 \times Z_2^T$ & $\mathbb{Z}_{2}^{2} $ & $\mathbb{Z}_{2}^{4} $ & $\mathbb{Z}_{2}^{2} $ \\
$Z_4$ & $\mathbb{Z}_{2} $ & $\mathbb{Z}_{2} $ & $\mathbb{Z}_{2} \times\mathbb{Z}_{4} $ \\
$Z_6$ & $\mathbb{Z}_{2} $ & $\mathbb{Z}_{2} $ & $\mathbb{Z}_{2} \times\mathbb{Z}_{3}^{2} \times\mathbb{Z}_{6} $ \\
$Z_2\times Z_2$ & $\mathbb{Z}_{2}^{3} $ & $\mathbb{Z}_{2}^{3} $ & $\mathbb{Z}_{2}^{3} $ \\
$Z_4\times Z_4$ & $\mathbb{Z}_{2}^{3} $ & $\mathbb{Z}_{2}^{3} $ & $\mathbb{Z}_{2} \times\mathbb{Z}_{4}^{2} $ \\
$Z_6\times Z_6$ & $\mathbb{Z}_{2}^{3} $ & $\mathbb{Z}_{2}^{2} \times\mathbb{Z}_{3} \times\mathbb{Z}_{6} $ & $\mathbb{Z}_{2} \times\mathbb{Z}_{3}^{4} \times\mathbb{Z}_{6}^{2} $ \\
$U(1)$ & $ $ & $ $ & $\mathbb{Z} \times\mathbb{Z}_{2} \times\mathbb{Z}_{3}^{2} $ \\
$SU(2)$ & $ $ & $ $ & $\mathbb{Z}_{2} $ \\
$SO(3)$ & $ $ & $\mathbb{Z}_{2} $ & $\mathbb{Z}_{2} $ \\
\hline
\noalign{\vskip0.03cm}
\multicolumn{ 4 }{c}{Wallpaper group \#15 $p31m$  } \\
\hline
$\varnothing$ & $ $ & $ $ & $\mathbb{Z}_{2} \times\mathbb{Z}_{3} $ \\
$Z_2^T$ & $ $ & $\mathbb{Z}_{2}^{2} $ & $\mathbb{Z}_{2} $ \\
$Z_2$ & $\mathbb{Z}_{2} $ & $\mathbb{Z}_{2} $ & $\mathbb{Z}_{2}^{2} \times\mathbb{Z}_{3} $ \\
$Z_2 \times Z_2^T$ & $\mathbb{Z}_{2}^{2} $ & $\mathbb{Z}_{2}^{4} $ & $\mathbb{Z}_{2}^{2} $ \\
$Z_4$ & $\mathbb{Z}_{2} $ & $\mathbb{Z}_{2} $ & $\mathbb{Z}_{2} \times\mathbb{Z}_{3} \times\mathbb{Z}_{4} $ \\
$Z_6$ & $\mathbb{Z}_{2} $ & $\mathbb{Z}_{2} $ & $\mathbb{Z}_{2} \times\mathbb{Z}_{3}^{2} \times\mathbb{Z}_{6} $ \\
$Z_2\times Z_2$ & $\mathbb{Z}_{2}^{3} $ & $\mathbb{Z}_{2}^{3} $ & $\mathbb{Z}_{2}^{3} \times\mathbb{Z}_{3} $ \\
$Z_4\times Z_4$ & $\mathbb{Z}_{2}^{3} $ & $\mathbb{Z}_{2}^{3} $ & $\mathbb{Z}_{2} \times\mathbb{Z}_{3} \times\mathbb{Z}_{4}^{2} $ \\
$Z_6\times Z_6$ & $\mathbb{Z}_{2}^{3} $ & $\mathbb{Z}_{2}^{2} \times\mathbb{Z}_{6} $ & $\mathbb{Z}_{2} \times\mathbb{Z}_{3}^{3} \times\mathbb{Z}_{6}^{2} $ \\
$U(1)$ & $ $ & $ $ & $\mathbb{Z} \times\mathbb{Z}_{2} \times\mathbb{Z}_{3}^{2} $ \\
$SU(2)$ & $ $ & $ $ & $\mathbb{Z}_{2} \times\mathbb{Z}_{3} $ \\
$SO(3)$ & $ $ & $\mathbb{Z}_{2} $ & $\mathbb{Z}_{2} \times\mathbb{Z}_{3} $ \\
\hline
\noalign{\vskip0.03cm}
\multicolumn{ 4 }{c}{Wallpaper group \#16 $p6$  } \\
\hline
$\varnothing$ & $ $ & $ $ & $\mathbb{Z}_{2} \times\mathbb{Z}_{3} \times\mathbb{Z}_{6} $ \\
$Z_2^T$ & $ $ & $\mathbb{Z}_{2} $ & $\mathbb{Z}_{2}^{2} $ \\
$Z_2$ & $\mathbb{Z}_{2} $ & $ $ & $\mathbb{Z}_{2}^{3} \times\mathbb{Z}_{3} \times\mathbb{Z}_{6} $ \\
$Z_2 \times Z_2^T$ & $\mathbb{Z}_{2}^{2} $ & $\mathbb{Z}_{2}^{2} $ & $\mathbb{Z}_{2}^{4} $ \\
$Z_4$ & $\mathbb{Z}_{4} $ & $ $ & $\mathbb{Z}_{2}^{2} \times\mathbb{Z}_{3} \times\mathbb{Z}_{4} \times\mathbb{Z}_{6} $ \\
$Z_6$ & $\mathbb{Z}_{6} $ & $ $ & $\mathbb{Z}_{2} \times\mathbb{Z}_{3} \times\mathbb{Z}_{6}^{3} $ \\
$Z_2\times Z_2$ & $\mathbb{Z}_{2}^{3} $ & $\mathbb{Z}_{2} $ & $\mathbb{Z}_{2}^{5} \times\mathbb{Z}_{3} \times\mathbb{Z}_{6} $ \\
$Z_4\times Z_4$ & $\mathbb{Z}_{4}^{3} $ & $\mathbb{Z}_{2} $ & $\mathbb{Z}_{2}^{3} \times\mathbb{Z}_{3} \times\mathbb{Z}_{4}^{2} \times\mathbb{Z}_{6} $ \\
$Z_6\times Z_6$ & $\mathbb{Z}_{6}^{3} $ & $\mathbb{Z}_{6} $ & $\mathbb{Z}_{2} \times\mathbb{Z}_{3} \times\mathbb{Z}_{6}^{5} $ \\
$U(1)$ & $\mathbb{Z} $ & $ $ & $\mathbb{Z} \times\mathbb{Z}_{2} \times\mathbb{Z}_{3} \times\mathbb{Z}_{6}^{2} $ \\
$SU(2)$ & $\mathbb{Z} $ & $ $ & $\mathbb{Z}_{2} \times\mathbb{Z}_{3} \times\mathbb{Z}_{6} $ \\
$SO(3)$ & $\mathbb{Z} $ & $\mathbb{Z}_{2} $ & $\mathbb{Z}_{2} \times\mathbb{Z}_{3} \times\mathbb{Z}_{6} $ \\
\hline
\noalign{\vskip0.03cm}
\multicolumn{ 4 }{c}{Wallpaper group \#17 $p6mm$  } \\
\hline
$\varnothing$ & $ $ & $ $ & $\mathbb{Z}_{2}^{4} $ \\
$Z_2^T$ & $ $ & $\mathbb{Z}_{2}^{4} $ & $\mathbb{Z}_{2}^{4} $ \\
$Z_2$ & $\mathbb{Z}_{2} $ & $\mathbb{Z}_{2}^{2} $ & $\mathbb{Z}_{2}^{6} $ \\
$Z_2 \times Z_2^T$ & $\mathbb{Z}_{2}^{2} $ & $\mathbb{Z}_{2}^{8} $ & $\mathbb{Z}_{2}^{6} $ \\
$Z_4$ & $\mathbb{Z}_{2} $ & $\mathbb{Z}_{2}^{2} $ & $\mathbb{Z}_{2}^{5} \times\mathbb{Z}_{4} $ \\
$Z_6$ & $\mathbb{Z}_{2} $ & $\mathbb{Z}_{2}^{2} $ & $\mathbb{Z}_{2}^{4} \times\mathbb{Z}_{6}^{2} $ \\
$Z_2\times Z_2$ & $\mathbb{Z}_{2}^{3} $ & $\mathbb{Z}_{2}^{6} $ & $\mathbb{Z}_{2}^{8} $ \\
$Z_4\times Z_4$ & $\mathbb{Z}_{2}^{3} $ & $\mathbb{Z}_{2}^{6} $ & $\mathbb{Z}_{2}^{6} \times\mathbb{Z}_{4}^{2} $ \\
$Z_6\times Z_6$ & $\mathbb{Z}_{2}^{3} $ & $\mathbb{Z}_{2}^{5} \times\mathbb{Z}_{6} $ & $\mathbb{Z}_{2}^{4} \times\mathbb{Z}_{6}^{4} $ \\
$U(1)$ & $ $ & $ $ & $\mathbb{Z} \times\mathbb{Z}_{2}^{4} \times\mathbb{Z}_{6} $ \\
$SU(2)$ & $ $ & $ $ & $\mathbb{Z}_{2}^{4} $ \\
$SO(3)$ & $ $ & $\mathbb{Z}_{2}^{2} $ & $\mathbb{Z}_{2}^{4} $ \\
\end{longtable}

\begin{longtable}{ c C{0.12\linewidth}C{0.2\linewidth}C{0.2\linewidth}C{0.225\linewidth} }
\caption[Classifications of 3D SPT phases.]{Classifications of three-dimensional SPT phases. \label{tab:3D}}\\
\hline
$G_0$  & $E^3_{3,\infty}$ & $E^2_{2,\infty}$ & $E^1_{1,\infty}$ & $E^0_{0,\infty}$ \\
\hline\hline\hline
\endfirsthead
\multicolumn{ 5 }{c} {{\tablename\ \thetable{} (Continued.)}}\\
\hline
$G_0$  & $E^3_{3,\infty}$ & $E^2_{2,\infty}$ & $E^1_{1,\infty}$ & $E^0_{0,\infty}$ \\
\hline\hline\hline
\endhead
\hline \\
\endfoot
\hline \hline \hline
\endlastfoot
\hline
\noalign{\vskip0.03cm}
\multicolumn{ 5 }{c}{Space group \#1 $P1$  } \\
\hline
$\varnothing$ & $ $ & $ $ & $ $ & $ $ \\
$Z_2^T$ & $\mathbb{Z}_{2} $ & $ $ & $\mathbb{Z}_{2}^{3} $ & $ $ \\
$Z_2$ & $ $ & $\mathbb{Z}_{2}^{3} $ & $ $ & $\mathbb{Z}_{2} $ \\
$Z_2 \times Z_2^T$ & $\mathbb{Z}_{2}^{3} $ & $\mathbb{Z}_{2}^{6} $ & $\mathbb{Z}_{2}^{6} $ & $\mathbb{Z}_{2} $ \\
$Z_4$ & $ $ & $\mathbb{Z}_{4}^{3} $ & $ $ & $\mathbb{Z}_{4} $ \\
$Z_6$ & $ $ & $\mathbb{Z}_{6}^{3} $ & $ $ & $\mathbb{Z}_{6} $ \\
$Z_2\times Z_2$ & $\mathbb{Z}_{2}^{2} $ & $\mathbb{Z}_{2}^{9} $ & $\mathbb{Z}_{2}^{3} $ & $\mathbb{Z}_{2}^{2} $ \\
\end{longtable}

\bibliography{homology,spt,CF_ref,overlap}

\end{document}